\documentclass[11pt,
a4paper,
oneside,
notitlepage, 
]{article}
\pdfoutput=1
\usepackage{jheppub}
\usepackage{hyperref}
\usepackage{graphicx,xcolor}
\usepackage{sectsty,amssymb,amsmath}
\usepackage{enumerate}
\usepackage{booktabs}
\usepackage{multirow}
\usepackage{slashed}
\usepackage{comment}
\usepackage[utf8]{inputenc} 
\usepackage{subcaption}
\usepackage{siunitx}
\usepackage{here}
\usepackage{cleveref}
\usepackage{tikz}
\usepackage[normalem]{ulem}
\usepackage{pifont}
\usepackage{makecell}
\usepackage{pdflscape}

\allowdisplaybreaks 

\definecolor{vermilion}{RGB}{227, 66, 52}
\definecolor{burntorange}{RGB}{204, 85, 0}
\definecolor{sienna}{RGB}{160, 82, 45}
\definecolor{firebrick}{RGB}{178, 34, 34}
\definecolor{garnet}{RGB}{115, 0, 0}
\definecolor{royalblue}{RGB}{65, 105, 225}
\definecolor{indigo}{RGB}{75, 0, 130}
\definecolor{darkorchid}{RGB}{153, 50, 204}
\definecolor{deepviolet}{RGB}{148, 0, 211}
\definecolor{darkcyan}{RGB}{0, 139, 139}
\definecolor{pinegreen}{RGB}{1, 121, 111}
\definecolor{oceanblue}{RGB}{0, 105, 148}
\definecolor{navy}{RGB}{0, 0, 128}
\definecolor{slatepurple}{RGB}{106, 90, 205}
\definecolor{emerald}{RGB}{4, 120, 87}
\definecolor{rosso}{cmyk}{0,1,1,0.4}
\definecolor{rossos}{cmyk}{0,1,1,0.55}
\definecolor{rossoc}{cmyk}{0,1,1,0.2}
\definecolor{blu}{cmyk}{1,1,0,0.3}
\definecolor{blus}{cmyk}{1,1,0,0.6}
\definecolor{bluc}{cmyk}{1,1,0,0.1}
\definecolor{verde}{cmyk}{0.92,0,0.59,0.25}
\definecolor{verdec}{cmyk}{0.92,0,0.59,0.15}
\definecolor{verdes}{cmyk}{0.92,0,0.59,0.4}
\definecolor{grigio}{cmyk}{0,0,0,0.07}
\definecolor{rosa}{cmyk}{0,0.1,0.1,0.02}
\definecolor{rosino}{cmyk}{0,0.05,0.05,0.02}
\definecolor{rosas}{cmyk}{0,0.3,0.25,0.05}
\definecolor{celeste}{cmyk}{0.1,0,0,0.02}
\definecolor{giallino}{cmyk}{0,0,0.4,0.02}
\definecolor{rosso}{cmyk}{0,1,1,0.4}
\definecolor{rossos}{cmyk}{0,1,1,0.55}
\definecolor{rossoc}{cmyk}{0,1,1,0.2}
\definecolor{blu}{cmyk}{1,1,0,0.3}
\definecolor{bluc}{cmyk}{1,1,0,0.1}
\definecolor{blucc}{cmyk}{0.7,0.5,0,0}
\definecolor{viola}{cmyk}{0,1,0,0.6}
\definecolor{viola2}{cmyk}{0,1,0.2,0.6}
\definecolor{verde}{cmyk}{0.92,0,0.59,0.25}
\definecolor{verdec}{cmyk}{0.92,0,0.59,0.15}
\definecolor{verdes}{cmyk}{0.92,0,0.59,0.4}
\definecolor{verdino}{cmyk}{0.12,0,0.09,0.05}
\definecolor{giallo}{cmyk}{0,0,1,0}
\definecolor{gialloverde}{cmyk}{0.44,0,0.74,0}
\definecolor{grey}{rgb}{0.6,0.6,0.6}
\definecolor{fuchsia}{rgb}{1,0,1}

\usepackage{chngcntr}
\counterwithin{figure}{section}
\counterwithin{table}{section}
\counterwithin{equation}{section}

\def\({\left(}
\def\){\right)}
\def\be{\begin{equation}}
\def\ee{\end{equation}}
\def\bes{\begin{subequations}}
\def\ees{\end{subequations}}
\def\bea{\begin{eqnarray}}
\def\eea{\end{eqnarray}}
\def\bry{\begin{array}}
\def\ery{\end{array}}
\def\bit{\begin{itemize}}
\def\eit{\end{itemize}}
\def\ben{\begin{enumerate}}
\def\een{\end{enumerate}}

\def\dst{\displaystyle}

\def\mres{m_\text{res}}

\notoc 

\begin{document}

\title{Heavy Vector Triplets at a Muon Collider}

\author[a]{Francesca Acanfora,}   
\author[a]{Michael J.~Baker,}    
\author[a]{Timothy Martonhelyi,}   
\author[a]{Andrea Thamm}     

\emailAdd{facanfora@umass.edu}
\emailAdd{mjbaker@umass.edu}   
\emailAdd{tmartonhelyi@umass.edu}    
\emailAdd{athamm@umass.edu}

\affiliation[a]{Department of Physics, University of Massachusetts Amherst, MA 01003, USA}

\date{\today}

\abstract{Heavy spin-one particles are well-motivated new physics candidates that can have their origin in weakly coupled extensions of the Standard Model gauge group or in strongly coupled Composite Higgs models. Due to the variety of production and decay modes, heavy vector triplets are a useful benchmark for the study and comparison of future colliders. Here we perform a detailed collider analysis of a variety of $2 \to 2$ and $2 \to 3$ processes at a proposed future muon collider. We focus on decays into leptons and Standard Model gauge bosons, and find that heavy vector triplets could be probed up to masses of around 12\,TeV for almost any (perturbative) value of the coupling. We compare the direct reach of a muon collider to the LHC and to updated projections for the HL-LHC, HE-LHC and FCC-hh, and include indirect limits from future measurements of electroweak precision observables. We find that a muon collider offers projected sensitivities that are competitive with future hadron colliders, exceeding those of the HE-LHC in the scenarios considered though not reaching the projected sensitivity of the FCC-hh.

}

\maketitle
\tableofcontents

\section{Introduction}
\label{sec:introduction}

The prospect of a muon collider has steadily gained traction over the past three decades, largely because it would combine the key advantages of hadron and lepton colliders within a relatively compact accelerator design \cite{
Delahaye:2019omf,
mu_smash,
toward_muc,
muon_accelerator_program,
Black:2022cth,
MuonCollider:2022xlm,
Narain:2022qud,
P5:2023wyd,
InternationalMuonCollider:2024jyv,
InternationalMuonCollider:2025sys,
Begel:2025ldu}. 
In particular, a muon collider could achieve high partonic centre-of-mass energies while maintaining a comparatively clean collision environment. Although significant technical challenges remain, most notably those arising from the short lifetime of the muon, the International Muon Collider Collaboration aims to address these issues by 2040 \cite{InternationalMuonCollider:2024jyv}. 

A muon collider would provide an excellent environment for precision studies of Standard Model (SM) processes~\cite{
Ruhdorfer:2023uea,
Andreetto:2024rra,
Marzocca:2025inb,
Costantini:2020stv,
Chiesa:2020awd,
Han:2020pif,
deBlas:2022ofj,
Han:2021lnp,
Forslund:2022xjq,
Dawson:2022zbb,
Liu:2023yrb} 
as well as for the exploration of physics beyond the SM~\cite{
Glioti:2025zpn,
Airen:2026szh,
Han:2020uak,
Ruhdorfer:2019utl,
Li:2023tbx,
Capdevilla:2021rwo,
Capdevilla:2021kcf,
Buttazzo:2020ibd,
Huang:2021biu,
Liu:2021akf,
Cesarotti:2022ttv,
Capdevilla:2021fmj,
Asadi:2021gah,
Gu:2020ldn}. To fully assess its potential, it is essential to benchmark its sensitivity against other proposed future collider facilities within a well-defined framework. In this work, we focus on heavy vector resonances as a representative and well-motivated class of new physics scenarios. Heavy vector resonances have been widely studied at the LHC and are particularly appealing targets at a muon collider.  The full centre-of-mass energy can be exploited for on-shell production, while the relatively small SM backgrounds mean that many processes are sensitive to off-shell effects. 

Heavy vector bosons are a common prediction in many extensions of the Standard Model. They can arise, for instance, as massive gauge bosons associated with extended gauge symmetries  
\cite{
Barger:1980ix,
Hewett:1988xc,
Cvetic:1995zs,
Rizzo:2006wq,
Agashe:2007hh,
Langacker:2008yv,
Salvioni:2010p2769,
Accomando:2013ve,
Agashe:2009bj,
Schmaltz:2010p2610,
Grojean:2011vu,
Langacker:1989xa,
Frank:2010p2250,
Accomando:2011up,
Accomando:2011gt,
Dobrescu:2021vak}
typically featuring relatively weak couplings to SM particles. Alternatively, in strongly coupled scenarios, heavy vectors may appear as strongly coupled $\rho$-like resonances of the underlying strong dynamics 
\cite{
Chanowitz:1993fc,
Langacker:2008yv,
Barbieri:2008cc,
Barbieri:2009p33,
Agashe:2009dg,
Agashe:2009ve,
Cata:2009iy,
Barbieri:2010mn,
CarcamoHernandez:2010wpm,
CarcamoHernandez:2010qxf,
Accomando:2011gt,
Falkowski:2011ua,
Contino:2011np,
Chanowitz:2011ew,
Bellazzini:2012tv,
Accomando:2012us,
Greco:2014aza,
Low:2015uha,
Accomando:2016mvz,
Liu:2018hum,
Liu:2019bua,
DeCurtis:2021fdm,
Liu:2023jta}.
Under the assumption that they are the lightest new states in the spectrum, heavy vectors can be efficiently described within simplified model frameworks \cite{Pappadopulo:2014qza,Baker:2024xwh}, allowing for a largely model-independent characterization of their interactions without committing to a specific ultraviolet completion. Here, we will make use of the simplified model framework but confine our analysis to two benchmark models.

Heavy vector resonances have previously been studied at a muon collider. The focus has been on searching for $Z'$ bosons transforming as singlets under the SM $su(2)_L$ in leptonic final states \cite{Hosseini:2022urq,Dasgupta:2023zrh,Cheung:2025uaz} and composite resonances produced in associated production decaying into bosonic final states \cite{Liu:2023jta}. In this work, we focus on a colourless spin-one particle that transforms as a triplet under the SM $su(2)_L$ and carries zero hypercharge. Such Heavy Vector Triplets (HVTs) couple to all SM fermions, gauge bosons and the Higgs boson. Since their coupling structure does not inherently favour any specific production environment, they are a particularly well-suited benchmark for comparing future collider facilities. HVTs can be produced through resonant, associated and vector boson fusion production at both hadron colliders \cite{Pappadopulo:2014qza,Baker:2022zxv} and a muon collider. Depending on the explicit benchmark, the HVT can decay into leptons, quarks, gauge bosons or Higgs bosons, leading to a rich and experimentally accessible phenomenology. To capture the range of possible behaviours, we consider two representative benchmarks: a weakly coupled scenario based on an explicit $su(3)_C \times su(2)_L \times su(2)_R \times u(1)_X$ model \cite{Barger:1980ix}, and a strongly coupled scenario motivated by composite Higgs models \cite{Contino:2011np,Contino:2013gna}.

In this work, we quantify the sensitivity of a $3$ and $10$\,TeV muon collider assuming both 1 and 10\,ab$^{-1}$ of integrated luminosity. We consider the $2 \to 2$ processes $\mu^- \mu^+ \to \mu^- \mu^+, e^- e^+$ and $W^- W^+$ as well as the $2 \to 3$ channels $\mu^- \mu^+ \to \ell^- \ell^+ \gamma, W^- W^+\gamma$ and $W^- W^+Z$. We pay particular attention to interference between the SM and the new physics contributions, which we find can be substantial. Neglecting the interference can lead to a significant over or underestimate of the collider sensitivity. We further compare our projections to current limits from the LHC and to the expected reach of the high-luminosity LHC (HL-LHC) \cite{ZurbanoFernandez:2020cco}, high-energy LHC (HE-LHC) \cite{FCC:2018bvk,CidVidal:2018eel} and the future circular collider FCC-hh \cite{FCC:2018vvp}. Sensitivity projections for the HE-LHC and FCC-hh in one of our benchmark scenarios were previously presented in Ref.~\cite{Thamm:2015zwa}. Here, we update these results using current LHC constraints and place them in direct comparison with the projected reach of a muon collider.

This paper is structured as follows: in \cref{sec:background} we discuss the expected design of a muon collider and how this informs the analysis strategy for heavy vector triplets. We describe the weakly coupled and strongly coupled benchmark models we use in this study, and the key production and decay channels at a muon collider. We also provide a detailed discussion of our statistical analysis. \Cref{sec:projections} contains cross-sections, differential cross-sections and sensitivity projections for the most promising $2 \to 2$ and $2 \to 3$ processes. We highlight the significant impact of interference of the new physics with the SM background. In \cref{sec:comparison}, we compare the projected sensitivity of a muon collider with the sensitivity of current and future hadron colliders (the LHC, HL-LHC, HE-LHC and FCC-hh).

\section{Background}
\label{sec:background}

\subsection{A Muon Collider}

Recently there has been considerable interest in the possibility of a muon collider~\cite{Delahaye:2019omf,mu_smash,toward_muc,muon_accelerator_program,Black:2022cth,MuonCollider:2022xlm,Narain:2022qud,P5:2023wyd,InternationalMuonCollider:2024jyv,InternationalMuonCollider:2025sys,Begel:2025ldu}, which combines the best aspects of hadron and lepton colliders. A muon collider could push the energy frontier while also providing a (relatively) clean collision environment. The synchrotron radiation of a muon is much smaller than that of an electron, since power loss goes as $m^{-4}$, and the muon beam can be accelerated to very high energies \cite{mu_smash,Han:2020uid}. Due to the fundamental nature of the muon, the entire energy is available in the parton-level collision, and there is no debris left over from the collision of a composite particle. A muon collider can also provide large luminosities and since muons are colour singlets there is minimal QCD background. For these reasons, such a machine is well suited for both precision Higgs measurements \cite{Dawson:2022zbb,Buttazzo:2020uzc} and for heavy new physics searches \cite{Capdevilla:2021rwo,Han:2021lnp,Han:2022edd,Korshynska:2024suh,Chen:2022msz}.

Recent studies \cite{MuonCollider:2022xlm,Andreetto:2024rra,InternationalMuonCollider:2025sys,MuCoL:2025quu} propose several runs of the muon collider with centre-of-mass energies of $3$ and $10\,$TeV. The integrated luminosity goal is to accumulate $1$ to $10\,$ab$^{-1}$, which would provide good precision for Higgs coupling measurements and extend the mass reach of heavy new physics searches~\cite{Delahaye:2019omf}. These are the benchmark energies and luminosities we will assume throughout this work.

The main technical challenge of a muon collider is related to the short lifetime of the muon, $\tau_\mu = 2.2 \times 10^{-6}\,$s. While time dilation increases the muon lifetime by a factor of $\gamma = E/m$ (up to $0.1\,$s for a beam energy of $5\,$TeV), this is still a very short time to produce, capture, collimate, accelerate and collide the muon beam. The most promising technology is currently based on a proton-driver scheme where a proton beam collides with a fixed target and produces a secondary beam of pions and kaons, which subsequently decay into muons. These muons occupy a large phase space volume and in order to collimate the muons, they must be cooled. Established cooling techniques unfortunately cannot cool the muons before they decay. The Muon Ionization Cooling Experiment (MICE) explores a new technique called ionization cooling, where the muon beam travels through a medium (such as lithium hydride or liquid hydrogen) and cools by ionizing its atoms \cite{MICE:2023vpa}. Subsequently, a radio-frequency accelerating cavity accelerates the muons to produce a beam. The International Muon Collider Collaboration (IMCC) aims to perfect this process by 2040 \cite{InternationalMuonCollider:2024jyv}.

The short muon lifetime also impacts the design of the detectors of a future muon collider \cite{MuonCollider:2022ded}. The muons in the beam will decay and create a significant beam induced background. To shield the detectors from this background, two conical tungsten absorbers with tips almost meeting the interaction point have been proposed, which would be installed along the beam pipe \cite{Ally:2022rgk}. These nozzles are included in the latest detector designs by MAIA (Muon Accelerator Instrumented Apparatus) \cite{MAIA:2025hzm} and MUSIC (MUon System for Interesting Collisions) \cite{Andreetto:2025mrd}. Unfortunately, the nozzles limit the angular coverage of the main detector to
\begin{align}
    \label{eq:base_cuts_eta}
    |\eta|<2.5 \,,
\end{align}
corresponding to $9.4^\circ<\theta <170.6^\circ$.  Throughout this work we will assume that particles can only be detected in this central region. There are proposals to install slimmer absorbers or a forward muon detector behind (or partially inside) the nozzles, which could collect highly energetic muons that survive the tungsten damping and increase the pseudorapidity coverage to $|\eta| < 7$ \cite{Ruhdorfer:2019utl,Ruhdorfer:2024dgz}.  Furthermore, vertex reconstruction and timing resolution could be used to mitigate this problem since the particles from the beam induced background tend to arrive before particles that originate from the hard interaction \cite{Bartosik:2019dzq}. However, in this work we remain conservative and only consider final state particles with $|\eta|<2.5$. 

Beyond these major challenges, there are less significant issues including beam energy spread and beam angular spread, which lead to a small uncertainty in the energy and momentum of the initial state muons. These beam effects are known to impact the kinematics of the final state and can deteriorate the sensitivity of missing energy searches, see e.g., Ref.~\cite{Ruhdorfer:2023uea}. Since we do not consider final states with missing energy, we do not account for these effects in our analyses. 

In addition to the pseudorapidity cut in \cref{eq:base_cuts_eta}, we require
\begin{align}
    \label{eq:base_cuts}
    E_{X} &> 20\,\text{GeV}
    \,,
    \\
    \Delta_R(X,Y) &> 0.4
    \,,
\end{align}
as baseline cuts at generation level in our analyses, for $X,Y \in \{\ell^\pm,\gamma, W^\pm, Z\}$.  The energy threshold mitigates the beam induced background~\cite{Bartosik:2019dzq} while the angular cut ensures that final states do not overlap (which is particularly important for particles which produce jets). Both cuts are current standards and provide a conservative assumption for future hardware capabilities. 

Throughout this work we also assume that the final state particles are visible and can be reconstructed. This amounts to the requirement that
\begin{align}
    p_{\text{fin}} &= (\sqrt{s},\vec{0})
\end{align} 
to a good accuracy, where $p_{\text{fin}}$ is the sum of all final state particle four-momenta. This requirement further reduces the beam induced background and reducible backgrounds with missing four-momenta. It is also advantageous in mitigating the effects of next-to-leading order processes, which can be sizeable at high energies \cite{Bredt:2022dmm,Ma:2024ayr,Frixione:2025guf}.\footnote{Note, for example, that Ref.~\cite{Ma:2024ayr} requires $p_{\text{fin}}^2 \geq 0.64 s$.} However, a more detailed study of next-to-leading order effects on heavy vector triplets at a muon collider is beyond the scope of this work.

Finally, we comment on the physics potential of vector boson fusion processes at a muon collider~\cite{Han:2020uid,Costantini:2020stv}. While they can provide powerful searches for new physics \cite{Buttazzo:2018qqp,Ruhdorfer:2019utl,Liu:2021jyc}, they are most important when the final state muons or neutrinos are very forward (that is, they have large pseudorapidities). Since we focus on small pseudorapidities ($|\eta| < 2.5$) and only visible final states, vector boson fusion processes are expected to be subdominant and a detailed study is left to future work.  Note that vector boson fusion processes can also be described in terms of the muon parton distribution functions, which were recently calculated in Ref.~\cite{Frixione:2023gmf}.  This description can also provide a probe of new physics \cite{Asadi:2026kpt}.

\subsection{Heavy Vector Triplets}

A popular and well-motivated simplified model is the heavy vector triplet (HVT) model, which contains a colour-neutral spin-one $su(2)_L$ triplet with zero hypercharge. We denote this vector by $\mathcal{V}_\mu^a$  where $a=1,2,3$.  Omitting terms containing only SM fields, the most general phenomenological Lagrangian of the HVT coupled to the SM is~\cite{Pappadopulo:2014qza}
\begin{equation}
    \label{eq: HVT Lagrangian}
    \bry{lll}
    \dst{\mathcal{L}}_\mathcal{V} \supset  & \dst-\frac14 D_{[\mu}\mathcal{V}_{\nu ]}^a D^{[\mu}\mathcal{V}^{\nu ]\;a}+\frac{m_\mathcal{V}^{2}}2\mathcal{V}_\mu^a \mathcal{V}^{\mu\;a} \vspace{2mm}\\
    &\dst+\, i\,g_V  c_H \mathcal{V}_\mu^a H^\dagger \tau^a {\overset{{}_{\leftrightarrow}}{D}}^\mu H +\frac{g^2}{g_V} c_q \mathcal{V}_\mu^a \sum_Q\overline{Q_L}\gamma^\mu\tau^a Q_L +\frac{g^2}{g_V} c_\ell \mathcal{V}_\mu^a \sum_L\overline{L_L}\gamma^\mu\tau^a L_L \, \vspace{2mm}\\
    &+ \dfrac{g_V}{2}c_{VVV} \epsilon_{abc} \mathcal{V}_\mu^a \mathcal{V}_\nu^b D^{[\mu}\mathcal{V}^{\nu]\,c} + g_V^2 c_{VVHH} \mathcal{V}_\mu^a\mathcal{V}^{\mu\,a} H^\dagger H - \dfrac{g}{2}c_{VVW} \epsilon_{abc} W^{\mu\nu\,a}\mathcal{V}_\mu^b \mathcal{V}_\nu^c 
    \,,
    \ery
\end{equation}
where 
\begin{align}
    \label{eq: covariant derivatives}
    D_{[\mu}\mathcal{V}^a_{\nu ]} &= D_{\mu}\mathcal{V}^a_{\nu } -D_{\nu}\mathcal{V}^a_{\mu }
    \\
    D_\mu \mathcal{V}_\nu^a &= \partial_\mu \mathcal{V}_\nu^a
    +g\,\epsilon^{abc}W_\mu^b\mathcal{V}_\nu^c
    \,,
\end{align}
$g$ denotes the $su(2)_L$ gauge coupling, $W_\mu^b$ are the $su(2)_L$ gauge bosons, $\tau^a=\sigma^a/2$, $\sigma^a$ are the Pauli matrices, the Higgs current is 
\begin{align}
    i\,H^\dagger \tau^a {\overset{{}_{\leftrightarrow}}{D}}^\mu H
    =
    i\,H^\dagger \tau^a D^\mu H\,-\,i\,(D^\mu H^\dagger) \tau^a  H
    \,,
\end{align}
$Q_L$ are the $\{u_L,d_L\}$, $\{c_L,s_L\}$ and $\{t_L,b_L\}$ $su(2)_L$ doublets, $L_L$ are the electron, muon and tau $su(2)_L$ doublets and $W^{\mu\nu\,a}$ is the $su(2)_L$ field strength tensor.  The coupling $g_V$ is the typical strength of the HVT self-interaction, while the $c_i$ parameters represent model-dependent departures from that typical strength. 

Upon electroweak symmetry breaking (EWSB), the components of the HVT mix with the SM hypercharge and $su(2)_L$ gauge bosons.  Prior to mixing their charge eigenstates are
\begin{equation}
    \label{eq: charge eigenstates}
    \mathcal{V}_\mu^\pm = \dfrac{\mathcal{V}_\mu^1 \mp i \mathcal{V}_\mu^2}{\sqrt{2}}, \qquad \mathcal{V}_\mu^0 = \mathcal{V}_\mu^3
    \,.
\end{equation}
In the absence of the HVT, the SM $B$ and $W^3$ mix to give the photon and the $\hat{Z}$ while $W^{1,2}$ become $\hat{W}^\pm$.  With the HVT, the neutral component $\mathcal{V}^0$ mixes with $\hat{Z}$, giving $V^0$ and $Z$, and the charged component $\mathcal{V}^\pm$ mixes with $\hat{W}^\pm$, giving $V^\pm$ and $W^\pm$.  The mixing angles of the neutral and charged vectors are
\begin{align}
    \tan 2\theta_N 
    &=
    g_V c_H \frac{\hat{v}}{\hat m_V} \frac{ \hat{m}_Z \hat{m}_V }{\hat{m}_V^2 - \hat{m}_Z^2} \,,
    \\
    \tan 2\theta_C
    &=
    g_V c_H \frac{\hat{v}}{\hat m_V} \frac{\hat{m}_W \hat{m}_V}{\hat{m}_V^2 - \hat{m}_W^2} \,,
\end{align}
respectively, where
\begin{align}
    \hat{m}_Z &= \frac{e}{2 \sin \theta_W \cos \theta_W} \hat{v} \,,
    \\
    \hat{m}_W &= \cos \theta_W \hat{m}_Z \,,
    \\
    \hat{m}_V &= \sqrt{m_\mathcal{V}^2 + g_V^2 c_{VVHH} \hat{v}^2} \,,
    \\
    \hat{v} &= \frac{\hat\mu}{\sqrt{\hat\lambda}}
    \,,
\end{align}
$\hat\mu$ and $\hat\lambda$ are the Higgs mass and quartic couplings, and $\theta_W$ is the Weinberg angle (see Ref.~\cite{Pappadopulo:2014qza} for more details). While $\hat{v}$ can differ significantly from $246$\,GeV if there is large mixing, large vector masses ($m_\mathcal{V} \gg 100$\,GeV) lead to small mixing angles and $\hat{v} \approx 246$\,GeV.  While the physical masses of the charged and neutral components turn out to be very similar~\cite{Pappadopulo:2014qza}, throughout this work we use
\begin{align}
    m_V \equiv m_{V^0}
\end{align}
as input and compute the corresponding value of $m_{V^\pm}$.

From the discussion in Ref.~\cite{Pappadopulo:2014qza} we note that the heavy vector two-body decays which are not forbidden or mixing angle suppressed are decays to fermions, $V^0 \to W_L^+ W_L^-$, $V^0 \to Z_L h$, $V^+ \to W_L^+ Z_L$ and $V^+ \to W_L^+ h$.

In \cref{eq: HVT Lagrangian}, the parameter $c_H$ controls the mixing of the HVT with the SM gauge bosons and their decay into the di-boson final states.  Coupling of the HVT to the SM quarks and leptons is controlled by the parameters $c_q$ and $c_\ell$, respectively. We will assume generation-independent couplings throughout this work.  The couplings $c_{VVV}$, $c_{VVHH}$ and $c_{VVW}$ contain more than one HVT field and do not significantly impact the collider phenomenology.

\subsection{Benchmark Models}

While the simplified model introduced above is a powerful framework for presenting experimental limits in a way that is applicable to many particular models, for projections at future colliders it is useful to consider the reach on benchmark models.  We consider two explicit benchmarks, model A and B, which were also considered in Ref.~\cite{Pappadopulo:2014qza}. Model A describes a weakly coupled gauge extension, where the vector triplet emerges from $su(2)_1 \times su(2)_2 \times u(1)_Y$ breaking through a Higgs mechanism to $su(2)_L \times u(1)_Y$ (which can be achieved in a linear $\sigma$-model~\cite{Barger:1980ix}).  Model B is based on a strongly coupled minimal composite Higgs model with a non-linearly realized $SO(5)/SO(4)$ global symmetry~\cite{Contino:2011np}.  Under these explicit models, the $c$ parameters of \cref{eq: HVT Lagrangian} become constants or functions of $g_V$. Their values are shown in \cref{Table: Benchmark parameters}. In model B, the parameter $a_\rho = m_\rho /(g_V f)$ is an $O(1)$ free parameter, which we take as $a_\rho = 1/\sqrt{2}$ in our analysis.\footnote{Note that Refs.~\cite{Pappadopulo:2014qza,Thamm:2015zwa} used $a_\rho = 1$ and Ref.~\cite{Liu:2023jta} used $a_\rho = 1/\sqrt{2}$.}  In these benchmark models, the coupling $g_V$ and the mass $m_V$ are the only remaining free parameters. We define the parameter $k_V = \sqrt{1 - g^2/g_V^2}$ as a useful shorthand, noting that neither model is theoretically well-defined for $g_V \leq g \approx 0.66$.

\begin{table}
    \renewcommand{\arraystretch}{1.1}
    \setlength{\tabcolsep}{0.23cm}
    \begin{center}
    \begin{tabular}{c c c}
        \hline
        \hline
        \rule{0pt}{\normalbaselineskip}
        & {\bf Model A} & {\bf Model B} \\
        & $su(2)_1\times su(2)_2$ & $SO(5)/SO(4)$ \\ 
        \hline
        $g_V$ & $g_2$ & $\pm g_\rho$ \\
        \hline 
        \rule{0pt}{\normalbaselineskip}
        $m_V$ & $\dfrac{|g_V|u}{k_V}$ & $\dfrac{m_\rho}{k_V}$ \\[4mm]
        $c_F$ & $-\dfrac{1}{k_V}$ & $\dfrac{1}{k_V}$ \\[4mm]
        $c_H$ & $-\dfrac{g^2}{g_V^2 k_V}$ & $-\dfrac{1}{k_V}\left(a_\rho^2 - \dfrac{g^2}{g_V^2}\right)$ \\[4mm]
        $c_{VVV}$ & $-\dfrac{1}{k_V}\left( 1 - \dfrac{2g^2}{g_V^2} \right)$ & $\dfrac{1}{k_V}\left( 1 - \dfrac{2g^2}{g_V^2} \right)$ \\[4mm]
        $c_{VVW}$ & 1 & $1$ \\[4mm]
        $c_{VVHH}$ & $\dfrac{g^4}{4g_V^4 k_V^2}$ & $-\dfrac{g^2}{2g_V^2k_V^2}\left( a_\rho^2 - \dfrac{g^2}{2g_V^2} \right)$ \\[4mm]
        \hline
        \hline
    \end{tabular}
    \\[0.1cm]
    \caption{\small Matchings for models A and B. Here $k_V = \sqrt{1 - g^2/g_V^2}$, $u$ is the VEV of the bi-doublet $\Phi$ in model A, and $a_\rho = \dfrac{m_\rho}{g_V f}$ in model B is an $O(1)$ free parameter which we take to be $a_\rho = \frac{1}{\sqrt{2}}$~\cite{Contino:2011np,Liu:2023jta}.}
    \label{Table: Benchmark parameters}
    \end{center}
\end{table}

Note that for (perturbative) unitarity to hold we require the coupling combinations appearing in \cref{eq: HVT Lagrangian} (e.g., $g_V c_H$, $g^2/g_V c_{q,\ell}$) to be less than $\sqrt{4 \pi}$. In model A, $g_V c_H$ and $g^2/g_V c_{q,\ell}$ tend to zero for large $g_V$ and only $g_V c_{VVV}/2$ increases with $g_V$. We find that $g_Vc_{VVV}/2 < \sqrt{4 \pi}$ for $g_V < 7$. To remain well within the unitary regime, we restrict ourselves to the parameter region $g_V < 5$ where $g_V c_{VVV}/2 \lesssim 2$. In model B, the form of $|g_V c_{VVV}/2|$ is the same as model A, and $g_Vc_H = -g_V c_{VVV}/2$ when $a_\rho = 1/\sqrt{2}$. So from both $g_V c_H$ and $g_V c_{VVV}/2$ we have the same restriction that $g_V < 7$. Since strongly coupled models often have larger couplings than weakly coupled models, for model B we do not restrict ourselves to $g_V<5$ but instead indicate the region where unitarity breaks down.

\begin{figure}[t]
    \centering
    \includegraphics[width=0.49\linewidth]{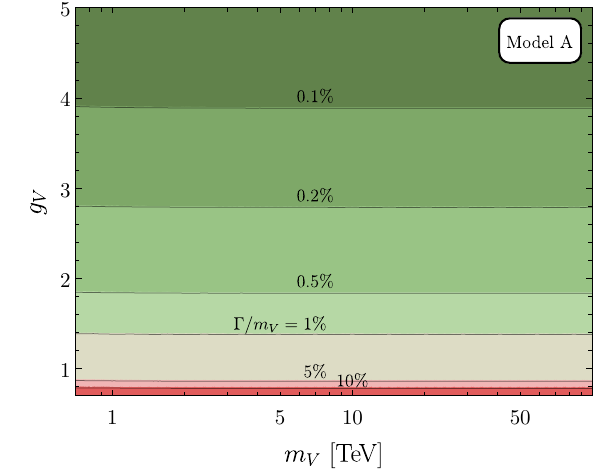}
    \includegraphics[width=0.49\linewidth]{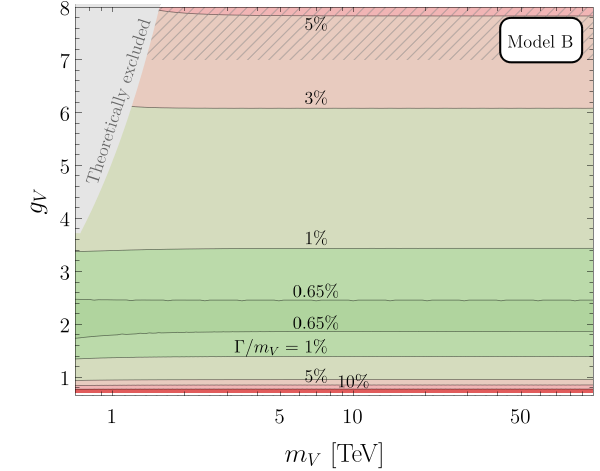}
    \caption{\small The width to mass ratio, $\Gamma_V/m_V$, of $V^0$ as a function of the HVT mass, $m_V$, and the coupling $g_V$ for model A (left) and model B (right).}
    \label{fig:widths}
\end{figure}

In \cref{fig:widths} we show the width to mass ratio of $V^0$ for model A (left) and model B (right).  We see that the width to mass ratio is almost independent of $m_V$, except for small effects due to mixing (which are more evident in model B).  In model A, the width is small for large couplings (less than 0.1\% for $g_V \gtrsim 4$) and larger for small couplings (above 1\% for $g_V \lesssim 1.4$).  The width becomes greater than 10\% for $g_V \lesssim 0.75$.  In model B, however, the different structure of the $c_H$ coupling leads to large widths at both small and large couplings (greater than 3\% for $g_V \lesssim 0.95$ and $g_V \gtrsim 6$) with a minimum of $\Gamma/m_V \approx 0.6\%$ at $g_V \approx 2.2$.

In this work we will not assume the Narrow Width Approximation (NWA). While the NWA can be a useful simplifying tool and a crucial assumption to derive model-independent limits in a simplified model \cite{Pappadopulo:2014qza, Thamm:2015zwa, Baker:2024xwh}, it is not a necessary assumption in the discussion of explicit models. We will see that there are several reasons why we do not take the NWA for the HVT model at a muon collider \cite{Berdine:2007uv}.  Firstly, while the NWA can be useful for $2\to2$ processes at a hadron collider (see, e.g., Ref.~\cite{Pappadopulo:2014qza}), where the parton centre-of-mass energy is not equal to the total centre-of-mass energy, it cannot be used for $2\to2$ processes at a lepton collider (except when $m_V = \sqrt{s}$).  Secondly, for both $2\to2$ and $2\to3$ processes we find that there can be significant interference between resonant and non-resonant processes, which are lost in the NWA.  Thirdly, the NWA is only valid when $m_V\ll \sqrt{s}$.  We will also be interested in $m_V \sim \sqrt{s}$ and larger masses.

\subsection{Key HVT Channels at a Muon Collider}
\label{sec:key-channels}

At a hadron collider, the colliding particles are composite and the parton centre-of-mass energy is variable.  A muon collider, however, has a fixed parton centre-of-mass energy. This limits the sensitivity of $s$-channel mediated $2 \to 2$ processes since the HVT mediator can only be on-shell if its mass coincides with the centre-of-mass energy of the collider. This is the reason why $2\to2$ $t$-channel processes and $2\to3$ processes (often called associated production or radiative return) are particularly interesting at a lepton collider. Although $2\to3$ processes suffer from additional phase space suppression, an additional photon or gauge boson can carry away energy and let a HVT resonance with $m_V < \sqrt{s}$ be on-shell.

As noted above, heavy vector triplets predominantly couple to left-handed fermions, $W_L^+ W_L^-$, $Z_L h$, $W_L^+ Z_L$ and $W_L^+ h$.  This implies that promising production modes at a muon collider are resonant production, production in association with a gauge boson and vector boson fusion. In this work we focus on HVT decay channels into leptons and gauge bosons and first consider the $2 \to 2$ channels
\begin{align}
    \mu^-\mu^+ &\to \mu^-\mu^+\,,\\
    \mu^-\mu^+ &\to e^-e^+\,,\\
    \mu^-\mu^+ &\to W^-W^+
    \,.
\end{align}
We do not consider the $\tau^-\tau^+$ or $q\bar q$ final states since they are difficult to reconstruct and are not expected to be as powerful.  We also do not consider processes with neutrinos in the final state: a single neutrino could be reconstructed via four-momentum conservation but there are no $2\to2$ processes with a single neutrino in the final state due to charge conservation; more than one missed particle can not be individually reconstructed and many different background processes would need to be carefully accounted for.

We then consider the $2 \to 3$ channels 
\begin{align}
    \mu^-\mu^+ &\to \ell^-\ell^+ \gamma\,,\\
    \mu^-\mu^+ &\to W^- W^+ \gamma\,,\\
    \mu^-\mu^+ &\to W^- W^+ Z
    \,,
\end{align}
where $\ell^-$ is an electron or a muon.  We will see that it is useful to separate electrons and muons in the $2\to2$ processes but that it is not so important in $2\to3$ channels.  We do not study the $\ell^- \ell^+ Z$ final state as we expect the reach to be slightly worse than $\ell^-\ell^+\gamma$ due to the $Z$ branching ratios and reconstruction efficiencies.  However, we do study $W^- W^+ Z$ since here extra diagrams appear which are not present in $W^- W^+ \gamma$.  We also do not consider final states containing Higgs bosons or neutrinos. Since the HVT branching ratio into di-boson final states is almost identical to the branching ratio into a Higgs and gauge boson~\cite{Pappadopulo:2014qza}, we expect final states containing the Higgs to be very similar to gauge boson final states. While neutrinos can occur in $2\to3$ processes, the missing energy would reduce the reconstruction efficiency and lead to weaker results.   We will discuss the details of the expected sensitivities of the $2 \to 2$ processes and $2 \to 3$ processes in \cref{sec:2-to-2} and \cref{sec:2-to-3} below.

While $2\to 4$ processes such as $\mu^- \mu^+ \to \mu^-\mu^+ W^- W^+$ may be competitive in certain regions of parameter space (see, e.g., Ref.~\cite{Baker:2022zxv}), the extra phase space suppression combined with the restrictive pseudorapidity requirement on the visible final states, \cref{eq:base_cuts_eta}, will generally weaken the possible constraints in these channels. Promising channels would include visible final states with a wider angular acceptance (e.g., $|\eta| < 7$), and final states with one or two invisible muons \cite{deLima:2025ctj}. A detailed study of these channels is left to future work. 

\subsection{Statistical Analysis}
\label{sec:analysis}

In the following we will compare a new physics hypothesis containing the SM and a heavy vector triplet, $H_\text{HVT}$, with the SM hypotheses, $H_\text{SM}$.  Assuming that the events in a future observation are Poisson distributed the likelihood of observing the events $o$ given the hypothesis, $H$, is~\cite{Cowan:2010js}
\begin{align}
    L(o|H) =  \prod_{i=1}^{n_B}
    \frac{(H^i)^{o_i} e^{-H^i}}{o_i!}
    \,,
\end{align}
where $n_B$ is the number of bins and $o$ and $H$ are lists with $n_B$ entries. 
We will base our test statistic on the log-likelihood ratio,
\begin{align} 
    \label{eq:testStatistic}
    \Lambda^P
    &=
    -2 \ln\left(
    \frac
    {L(o|H_\text{HVT})}
    {L(o|H_\text{SM})}
    \right)
    \\
    &=
    -2 \sum_{i = 1}^{n_B}\left( H_\text{SM}^i -H_\text{HVT}^i + o_i\ln
    \left(
    \frac{H_\text{HVT}^i}
    {H_\text{SM}^i}
    \right)
    \right)
    \,.
\end{align}
In the absence of a measurement we assume that the observation matches the SM expectation (i.e., we will take the Asimov data set).   The test statistic for data consistent with $H_\text{SM}$ is then
\begin{align}
    \Lambda_\text{SM}^P
    &=
    -2 \sum_{i= 1}^{n_B}\left(H_\text{SM}^i -H_\text{HVT}^i + H_\text{SM}^i\ln
    \left(
    \frac{H_\text{HVT}^i}
    {H_\text{SM}^i}
    \right)
    \right) 
    \,.
\end{align}
Using the CL$_\text{s}$ method the confidence level is given by
\begin{align}
    \label{eq:cl}
    \text{C.L.}
    =
    \frac{1-\Phi(\sqrt{\Lambda_\text{SM}^P})}{\Phi(0)} \,,
\end{align}
where $\Phi(x)$  is the cumulative distribution function of the normal distribution with mean 0 and standard deviation 1 (so that $\Phi(0) = 0.5$). If a point in parameter space has $\Lambda_\text{SM} > 3.84$  then it can be excluded with a confidence level of 0.05~\cite{ParticleDataGroup:2024cfk}.

\subsubsection{Statistical Analysis for $2 \to 2$ Processes}
\label{sec:stats-2-to-2}

For the $2 \to 2$ processes we compute the differential cross-sections analytically.  In order to model systematic errors (which we take to be Gaussian distributed) we extend the test statistic for the Asimov data set to \cite{Conway:2011in}
\begin{align}
    \label{eq:interf_log_lik_rat_nuis}
    \Lambda_\text{SM} 
    &= 
    -2 \sum_{i = 1}^{n_B} \left[ H_\text{SM}^i - \beta_i H_\text{HVT}^i + H_\text{SM}^i \ln \left(\frac{\beta_i H_\text{HVT}^i}{H_\text{SM}^i} \right) -\dfrac{(\beta_i-1)^2}{2 \epsilon_i^2}\right]
    \,,
\end{align}
where
\begin{align}
    \label{def:conway_beta}
    \beta_i 
    &=
    \dfrac{1}{2}\left( 1- H_\text{HVT}^i \epsilon_i^2 +\sqrt{4 H_\text{SM} \epsilon_i^2+(\epsilon_i^2 H_\text{HVT}^i-1)^2} \right)
    \,,
\end{align}
and $\epsilon_i$ is the relative error on the predicted number of events under the HVT hypothesis in the $i^\text{th}$ bin, $H_\text{HVT}^i$. We note that $\Lambda_\text{SM} \to \Lambda_\text{SM}^P$ in the limit $\epsilon_i \to 0$.  We will exclude a point in parameter space if it gives $\Lambda_\text{SM} > 3.84$.  We will discuss the systematic errors we use in each channel in the sections below.

\subsubsection{Statistical Analysis for $2 \to 3$ Processes}
\label{sec:stats-2-to-3}

For the $2 \to 3$ processes we compute the differential cross-sections with Monte Carlo techniques, using \texttt{MadGraph}~\cite{Alwall:2014hca}.  We will see that interference effects can be very important, and because of this we will compute $H_\text{SM}$ and $H_\text{HVT}$ independently.  That is, we will compute the background, $b$, and the signal plus background taking into account interference, $s\oplus b$, rather than computing the signal, $s$, and adding it to the background, $s+b$.  This means that in regions of parameter space where the HVT contribution is essentially negligible, $H_\text{SM}$ and $H_\text{HVT}$ can still differ due to Monte Carlo error.  This leads to three separate issues.

First, we may imagine that we can simply incorporate the Monte Carlo error in our statistical analysis.  While \texttt{MadGraph} provides an estimate of the error on the total cross-section, this is not always reliable.  We found for representative points in parameter space that the total cross-section error provided by \texttt{MadGraph} underestimates the actual error we find from running \texttt{MadGraph} multiple times.  Furthermore, the total cross-section error may differ from the error in different regions of phase space. In fact, \texttt{MadGraph} uses optimal sampling so regions of phase space where the matrix element is small are sampled more poorly, resulting in a larger uncertainty. 

To account for this we generate the Monte Carlo data in a way that allows us to estimate the \texttt{MadGraph} error.  For each hypothesis we repeat the Monte Carlo generation ten times and find the differential cross-section for each repetition.  We then compute the standard deviation between the ten runs in every bin separately, which we write as $ \delta(b)_i$ and $ \delta(s\oplus b)_i$ for $H_\text{SM}$ and $H_\text{HVT}$ respectively, and use this as an estimate of the error in each bin.  We use the mean value in each bin as an estimate of the central values.  We stress that the magnitude of the Monte Carlo error is model dependent, process dependent and generation cuts dependent, so we repeat the procedure for every process and for every $(m_V,g_V)$ point.  For $\epsilon_i$ in \cref{eq:interf_log_lik_rat_nuis,def:conway_beta} we then compute the sum in quadrature of the relative error on $b_i$, the relative error on $(s\oplus b)_i$ and a systematic experimental error $\epsilon_\text{syst}$,
\begin{align}
    \label{eq:2_to_3_epsilon_i}
    \epsilon_i^2
    =
    \left(\dfrac{\delta(b)_i}{b_i}\right)^2 + \left(\dfrac{\delta(s\oplus b)_i}{(s\oplus b)_i} \right)^2 + \epsilon_\text{syst}^2
    \,.
\end{align}
Finally we compute the log-likelihood ratio using \cref{eq:interf_log_lik_rat_nuis}.

Second, we partially mitigate the impact of optimal sampling on the error across the phase space by using bias functions. We will see, e.g., in the $\ell^- \ell^+ \gamma$ channel that the differential cross-section can differ by several orders of magnitude for different values of the leptonic invariant mass, $m_{\ell \ell}$ (see, e.g., \cref{fig:mll_dist}).  Without biasing, \texttt{MadGraph} will simulate the region with a large differential cross-section (at large $m_{\ell \ell}$) with better accuracy than the region with smaller cross-section (at small $m_{\ell \ell}$).  This is a good strategy for estimating the total cross-section.  For this reason, when we only need the total cross-section (e.g., in \cref{fig:mm_lla_xsec} and in the 1-bin exclusion in \cref{fig:exclusion_lla}) we use unbiased simulations.  However, when performing a differential analysis on a point in parameter space where the HVT model predicts deviations at small $m_{\ell \ell}$, the large Monte Carlo error will lead to a smaller significance. Fortunately, we can configure \texttt{MadGraph} to sample more often in certain regions of phase space using a bias function, $f_\text{bias}$. It is a function of the four-momenta of all the particles in the scattering, and large values of $f_\text{bias}$ lead to more frequent sampling. For $\ell^- \ell^+ \gamma$ we chose $f_\text{bias}=E_\gamma^2$ (where for this process $E_\gamma = (s-m_{\ell\ell}^2)/(2\sqrt{s})$, so small $m_{\ell \ell}$ corresponds to large $E_\gamma$). This compensates the optimal sampling which prefers small $E_\gamma$.\footnote{This strategy is better than splitting the $m_{\ell \ell}$ range into several different simulations, as even in just a section of the spectrum there is still a sizable difference between low and high $m_{\ell \ell}$.} So unbiased runs have an $E_\gamma$ distribution featuring smaller errors on smaller energies, while biased runs have larger errors on smaller energies. To obtain a distribution featuring small errors across the whole range of $m_{\ell\ell}$, we combine the biased and unbiased distribution bin by bin (as described in \cite{comb_exp_results}).  We perform a similar procedure for $W^- W^+ \gamma$ and $W^- W^+ Z$, see \cref{sec:wwa} for details. 

Third, we assume a future experiment observes the Asimov data set.  This means that, even when there is only a Monte Carlo difference between $H_\text{SM}$ and $H_\text{HVT}$, the observation agrees perfectly with $H_\text{SM}$ and not with $H_\text{HVT}$.  Due to this we find an irreducible contribution to the test statistic $\Lambda_\text{SM}$ even in regions of parameter space where the HVT is decoupled.  Furthermore, this irreducible contribution increases as the number of bins in the differential distribution is increased.  That is, if we increase the number of bins used we find that even decoupled regions of parameter space can randomly have $\Lambda_\text{SM} > 3.84$.  This is not physical, it is just a result of the statistical procedure and the fact that we are taking interference into account so we cannot simply sum signal and background.  To ensure that this does not significantly affect our final results, we keep the number of bins small enough so that this irreducible contribution to $\Lambda_\text{SM}$ gives $\Lambda_\text{SM} < 3.84/2 = 1.92$ for decoupled points in parameter space.  We will see nonetheless that this issue means that our exclusion curves are not smooth for the $2 \to 3$ processes, due to this random irreducible contribution to $\Lambda_\text{SM}$.

\section{Sensitivity Projections}
\label{sec:projections}

In this section, we will discuss the projected sensitivity of a $3$ and $10\,$TeV muon collider with $L = 1$ and $10\,\text{ab}^{-1}$ using the $2\to 2$ and $2\to 3$ processes discussed in \cref{sec:key-channels}.  Unless otherwise stated, we use \texttt{MadGraph} \cite{Alwall:2014hca} to simulate signal, background and interference. We use the Heavy Vector Triplet (Bridge model) UFO model \cite{hepmdb,HVTGitHub} to generate the $H_\text{HVT}$ hypotheses. Note that we do not perform a detector simulation or include showering for this BSM study. To account for these effects, we include systematic errors in our analysis.

\subsection{$2 \to 2$ Processes}
\label{sec:2-to-2}

First we consider the  $\mu^-\mu^+$, $e^-e^+$ and $W^-W^+$ final states. For these $2 \to 2$ processes, we can compute the tree-level cross-section analytically, use known branching ratios and include motivated tagging efficiencies (rather than performing a detector simulation).  We will highlight the impact of interference between the HVT and the SM contributions, which is significant in all three channels. 

In what follows, we use the notation ``HVT$\oplus$SM'' for the combined contribution to the cross-section from a HVT propagator, plus all SM channels, plus interference between the HVT and the SM. Curves that are labeled ``HVT+SM'' correspond to the signal plus background analysis ignoring interference, and ``SM'' represents the Standard Model alone.

\subsubsection{The $\mu^-\mu^+$ Channel} 
\label{sec:mumu}

\begin{figure}
    \centering
    \includegraphics[width=0.2\linewidth]{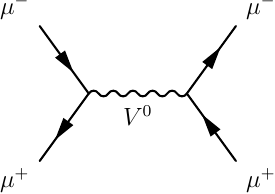}
    \hspace{0.75cm}
    \includegraphics[width=0.2\linewidth]{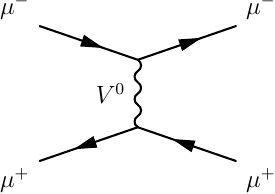}
    \caption{\small
    Feynman diagrams for the signal contributions to $\mu^- \mu^+ \to \mu^-\mu^+$.
    }
    \label{fig:mumu-mumu-feynman-diagrams}
\end{figure}

For the $\mu^- \mu^+$ final state the HVT contributes $s$- and $t$-channel diagrams at leading order, shown in \cref{fig:mumu-mumu-feynman-diagrams}.   Dropping terms of order $\theta_N$,\footnote{In our computation we take the effect of the mixing angles into account.} the matrix element for these two diagrams is
\begin{align}
    -i\mathcal{M}_\text{HVT}(\mu^-\mu^+ \to \mu^-\mu^+) 
    &=
    i\left(\frac{c_\ell\, g^{2}}{2 g_V}\right)^{2} 
    \left[
        \dfrac{\big(\overline{\mu_L} \gamma^\mu \mu_L\big)\big(\overline{\mu_L} \gamma_\mu \mu_L\big)}{s - m_V^{2} + i\, m_V \Gamma_V}
        -
        \dfrac{\big(\overline{\mu_L} \gamma^\mu \mu_L\big)\big(\overline{\mu_L} \gamma_\mu \mu_L\big)}{t - m_V^{2} + i\, m_V \Gamma_V}
    \right] \, .
\end{align}
We see that the matrix element has peaks at $\sqrt{s} = m_V$ and $\sqrt{t} = m_V$. The $s$-channel contribution will dominate on resonance, $m_V \sim \sqrt{s}$.  As the HVT moves further off-shell, the importance of resonant production via the $s$-channel diagram decreases and the $t$-channel contribution becomes more significant. While $t$ can not equal $m_V^2$ (since $-s \leq t \leq 0$), this can still lead to an enhancement at small $|t|$ when $m_V \ll \sqrt{s}$.  We will see that the $t$-channel diagram can give a significant contribution to the cross-section, which would be missed if we only focused on resonant production.    The main background in this channel is the SM $\mu^- \mu^+ \to \mu^-\mu^+$, mediated by an $s$- or $t$-channel $\gamma$ or $Z$.

\begin{figure}
    \centering
    \includegraphics[width=0.49\linewidth]{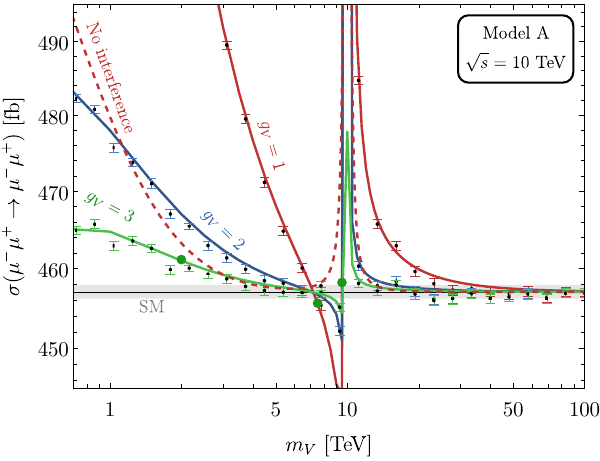}
    \includegraphics[width=0.49\linewidth]{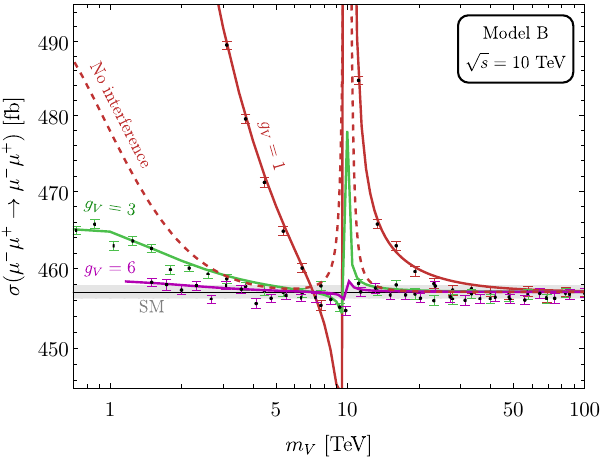}
    \caption{\small The total cross-section, $\sigma(\mu^-\mu^+\to \mu^-\mu^+)$, as a function of the HVT mass, $m_V$, for model A (left) and model B (right) at $\sqrt{s} = 10\,$TeV. The coloured lines show the analytic result for the HVT$\oplus$SM cross-section for different values of $g_V$, and the data points show the corresponding numeric \texttt{MadGraph} results. The black line shows the SM cross-section, and the grey shaded region shows the range of uncertainty of the numeric result. The dashed red line shows the signal plus background for $g_V=1$ neglecting interference effects. The three dark green points show the parameter points discussed in \cref{fig:eta-binning}.}
    \label{fig:analytic-vs-mg5 x-sections}
\end{figure}

\Cref{fig:analytic-vs-mg5 x-sections} shows the total cross-section of $\mu^-\mu^+ \to \mu^-\mu^+$ at $\sqrt{s} = 10\,$TeV as a function of the HVT mass for model A (left) and model B (right). The coloured lines show the analytic HVT$\oplus$SM calculations for $g_V = 1,2,3$ for model A and for $g_V = 1,3,6$ for model B. As a check, the discrete points show the corresponding \texttt{MadGraph} results and the Monte Carlo error. Each point corresponds to $10^5$ Monte Carlo events. The analytic SM $\mu^-\mu^+\to\mu^-\mu^+$ cross-section is shown by the constant black line at $457\,$fb, and the SM \texttt{MadGraph} uncertainty is shown by the grey band.  We kept the $|\eta|< 2.5$ cut consistent across the computations and find good agreement.  With a dashed line we show the analytic HVT+SM cross-section for $g_V = 1$, where interference is neglected.

In \cref{fig:analytic-vs-mg5 x-sections} we first note that the cross-section is quite large, almost 500\,fb.  In both the left and right panel we see the pole at $m_V = \sqrt{s} = 10\,$TeV and a decreasing HVT cross-section for larger HVT masses. For masses below $10\,$TeV, we see a typical dip-hill interference pattern between the signal $s$-channel and the other SM and HVT channels.\footnote{Whether there is a dip and whether the dip comes before or after the hill depends on the complex phase of the $s$-channel couplings, which determines the sign of the interference.  See, e.g., Ref.~\cite{Carena:2016npr}.}  There is a cancellation of the HVT contribution to the total cross-section at $m_V\approx 7\,$TeV in the HVT$\oplus$SM computation, and the $t$-channel contribution dominates at lower masses, leading to a significant increase in the cross-section. The HVT$\oplus$SM cross-section gets closer to the SM as $g_V$ increases, as the $V^0 \mu^- \mu^+$ interaction is proportional to $g^2 c_F/g_V$ in the Lagrangian in \cref{eq: HVT Lagrangian}.  In both models A and B (left and right) the parameter $|c_F|$ approaches $1$ as $g_V$ grows. From \cref{Table: Benchmark parameters} we see that $c_F$ only differs between the models by an overall sign. The only difference between the cross-sections of model A and model B enters through the total width, which includes all the HVT couplings.  Finally, when we neglect interference (dashed red) we see that in both model A and B the HVT+SM cross-section is always larger than the SM alone, in contrast to HVT$\oplus$SM, and that the deviation from the SM is smaller at most masses.  In fact, the interference term is larger than the HVT$^2$ term for most masses.  To obtain accurate exclusion plots it will therefore be important to account for interference.

We now want to discuss both the impact of binning the events in pseudorapidity and of interference.  We first consider the power of using a differential distribution in pseudorapidity.

\begin{figure}
    \centering
    \includegraphics[width=0.49\linewidth]{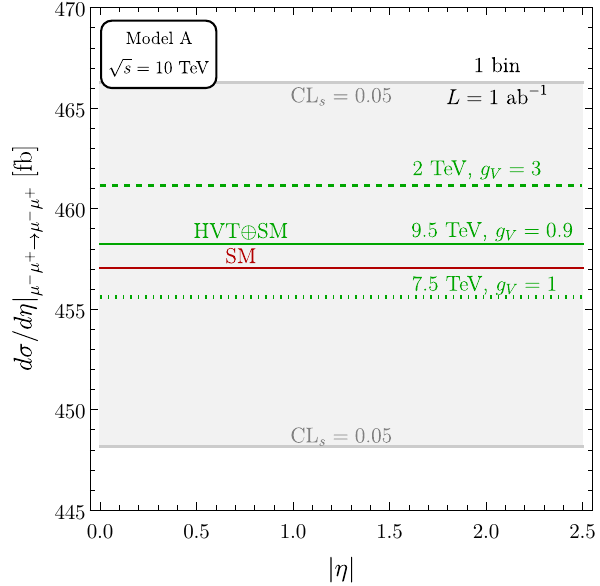}
    \includegraphics[width=0.49\linewidth]{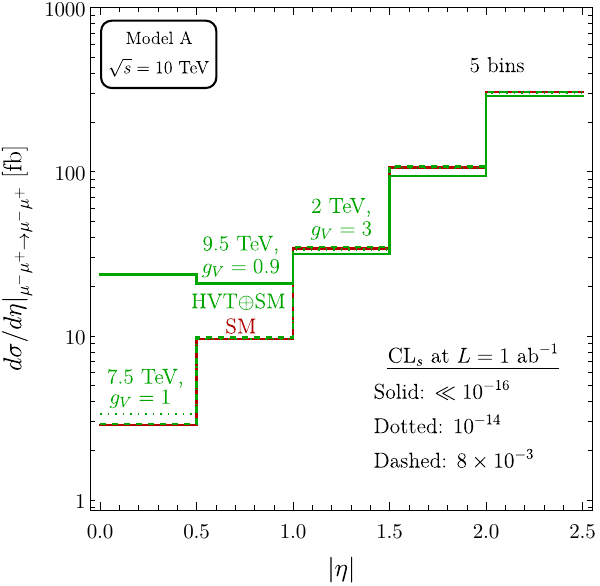}
    \caption{\small Differential cross-section as a function of the absolute pseudorapidity $|\eta|$ in model A for a single bin (left) versus five bins (right) at a $10\,$TeV collider. The red line shows the SM prediction and the green lines show the HVT$\oplus$SM prediction for different masses and couplings. In the left panel, the grey region corresponds to a  CL$_s \geq 0.05$ at $L = 1\,\text{ ab}^{-1}$. In the right panel, the $1\,$ab$^{-1}$ CL for each of the parameter points is shown in the legend.}
    \label{fig:eta-binning}
\end{figure}

\Cref{fig:eta-binning} shows the differential cross-section in model A as a function of the absolute pseudorapidity $|\eta|$ at a 10\,TeV collider. The left panel shows just one bin, equivalent to the total cross-section, and the right panel shows the differential cross-section over five equally spaced bins. The red line corresponds to the distribution under the SM alone and the green lines show the HVT$\oplus$SM prediction for three different parameter points. On the left hand plot, the grey region indicates the region which can not be excluded at $95\%$ CL, using \cref{eq:cl} and \cref{eq:interf_log_lik_rat_nuis}. While we expect the reconstruction efficiency of the final state muons to be high and do not expect a detector simulation to substantially modify our result, we remain conservative by adding a systematic error of $\epsilon_i = 1\%$ to all bins.  We see from the left panel that we have chosen three HVT parameter points that can not be excluded at $95\%$ CL with a 1-bin analysis based on the total cross-section. The same HVT parameter points are also shown in the right panel. We see that heavier HVTs broadly tend to decay to more central muons, which gives a powerful discriminating feature. With $1\,$ab$^{-1}$ all three parameter points can be excluded at $95\%$ CL when binned in five $|\eta|$ bins.  While this may be hard see for the parameter point $m_V = 2\,$TeV, $g_V = 3$, the cross-sections in, e.g., the $0.5<|\eta| < 1.0$ bin are $9.55\,$fb for the SM and $9.75\,$fb for HVT$\oplus$SM, corresponding to 9547 and 9745 events with $L=1\text{ ab}^{-1}$, respectively. The difference of 198 events is then around 2\%, while the statistical error is only around 1\% (note that for our exclusion limit we sum over all bins, as given in \cref{eq:interf_log_lik_rat_nuis}). 

\begin{figure}
    \centering
    \includegraphics[width=0.49\linewidth]{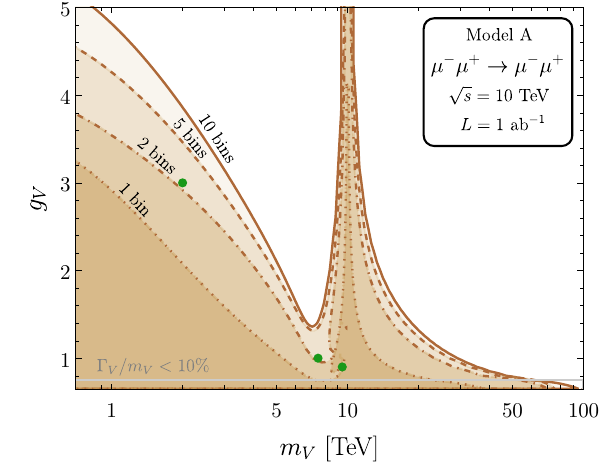}
    \includegraphics[width=0.49\linewidth]{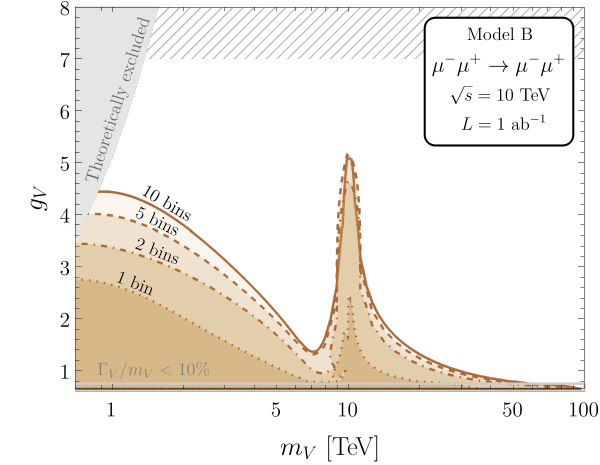}
    \caption{\small 95\% CL sensitivity projections for the $\mu^-\mu^+ \to \mu^-\mu^+$ process for model A (left) and model B (right) in the $(m_V,g_V)$ plane at a 10\,TeV muon collider, for a total cross-section analysis (dotted) and for data binned in $|\eta|$. The limits are shown for an integrated luminosity of 1 ab$^{-1}$. For model B (right), the solid grey region is theoretically excluded, and the hatched region in grey indicates the loss of perturbative unitarity. The three green points in the left panel correspond to the benchmark points discussed in \cref{fig:eta-binning}.}
    \label{fig:ll-different-bins}
\end{figure}

In \cref{fig:ll-different-bins} we show the 95\% CL exclusion limits for this process in the $(m_V,g_V)$ plane for model A (left) and model B (right) using HVT$\oplus$SM at a $10\,$TeV muon collider with an integrated luminosity of $1\,\text{ab}^{-1}$. The dotted, dot-dashed, dashed and solid contours correspond to analyses using one, two, five and ten $|\eta|$ bin(s), respectively. We see that a 1-bin analysis of the total cross-section yields a reasonable exclusion, particularly on resonance and for HVT masses below $5\,$TeV (where the cross-section differs considerably from the SM cross-section, \cref{fig:analytic-vs-mg5 x-sections}). In model A (left), this analysis has sensitivity up to HVT masses of $35\,$TeV. However, increasing the number of $|\eta|$ bins substantially increases the sensitivity at all masses. The 10-bin analysis has sensitivity to HVT masses as large as $100\,$TeV.  While we see the sensitivity increasing with the number of bins, there is diminishing returns and eventually the limit worsens due to limited statistics in each bin.  We do not show this due to the time required for full parameter space scans, but 20 bins only improves marginally on the 10-bin analysis while 50 bins shows no further improvement. The three dots indicate the parameter points discussed in \cref{fig:eta-binning}, where the 1-bin analysis is not sensitive for different reasons.  At $m_V=2$\,TeV and $g_V=3$, there is a small positive interference with the SM. The HVT$\oplus$SM and SM cross-sections are very close and it is only in the $|\eta|$ distribution that we see a very small change in the events in each bin. At $m_V=7.5$\,TeV and $g_V=1$, we see from \cref{fig:analytic-vs-mg5 x-sections} that there is slight negative interference with the SM, which again alters the $|\eta|$ distribution. At $m_V=9.5$\,TeV and $g_V=0.9$, there is an accidental cancellation in the total cross-section between the signal, which is close to resonance, and the negative interference with the SM. Even though the cross-sections are nearly the same, we know in this region there should be a large number of signal events with small $|\eta|$, and indeed we see this in \cref{fig:eta-binning}. In the right-hand panel we show model B. The sensitivity is similar to model A, since the cross-sections are similar. The solid grey region is theoretically excluded, and the hatched grey region shows perturbative unitarity is lost.

\begin{figure}
\centering
\includegraphics[width=0.49\linewidth]{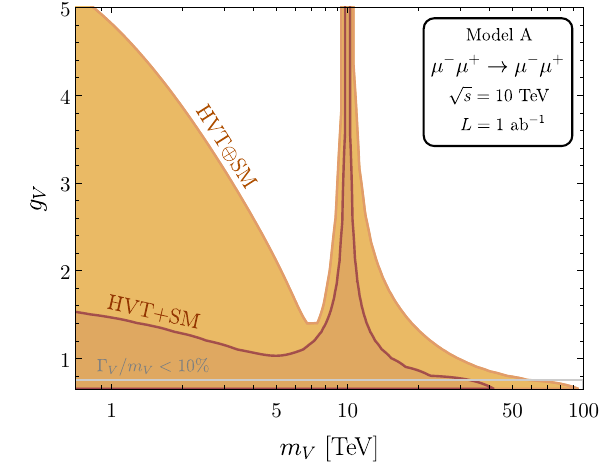}
\includegraphics[width=0.49\linewidth]{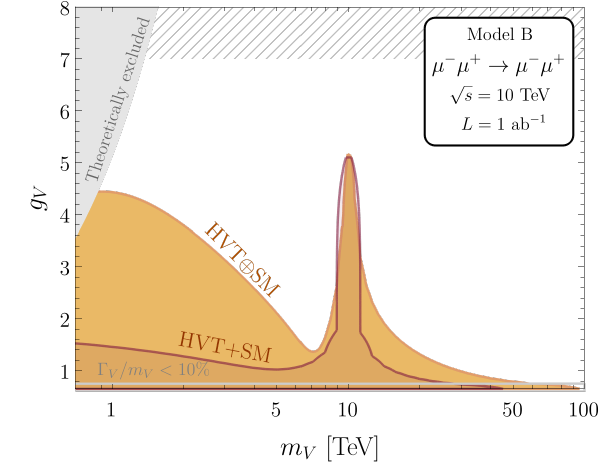}
\caption{\small 95\% CL sensitivity projections for the $\mu^-\mu^+ \to \mu^-\mu^+$ process in the $(m_V,g_V)$ plane at a 10\,TeV muon collider using ten equally spaced bins in $|\eta|<2.5$ with an integrated luminosity of $1\ \text{ab}^{-1}$. The orange regions show the exclusions under the HVT$\oplus$SM analysis, and the darker contours show the HVT+SM exclusions under the assumption of zero interference. Other features are the same as \cref{fig:ll-different-bins}.}
\label{fig:ll-exclusions}
\end{figure}

We now turn to the impact of interference.  In \cref{fig:ll-exclusions} we show the same exclusion as in \cref{fig:ll-different-bins} for the 10-bin analysis. The darker brown HVT curve shows the sensitivity projection of a 10-bin signal plus background analysis if interference were ignored (HVT+SM). We see that the correct inclusion of interference leads to a significantly larger reach in mass and coupling.  This difference grows for smaller HVT masses. In fact, in both panels, the sensitivity below $m_V \approx 7.5$\,TeV is almost entirely due to interference.  In this region the SM-HVT interference contribution is larger than the HVT$^2$ contribution.  Below $m_V \approx3$ to $4$\,TeV, the HVT$^2$ contribution increases due to the $t$-channel enhancement in $\mu^-\mu^+ \to \mu^-\mu^+$, but this effect is not nearly as large as the SM-HVT interference contribution.  It is well known that signal and background interference can play a large role, (see, e.g., Refs.~\cite{Accomando:2011eu,Choudhury:2011cg,Accomando:2013sfa,Pappadopulo:2014qza}) and, in particular, that interference effects become significant outside of a narrow window around the invariant mass peak. Note that, in general, the interference is highly model dependent and excluding interference leads to more model independent projections.

\begin{figure}
    \centering
    \includegraphics[width=0.49\linewidth]{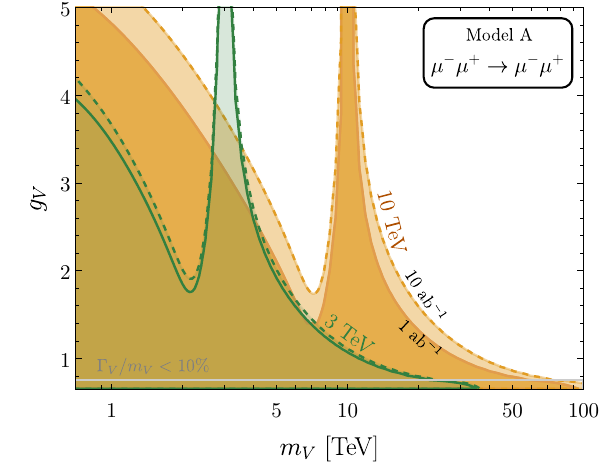}
    \includegraphics[width=0.49\linewidth]{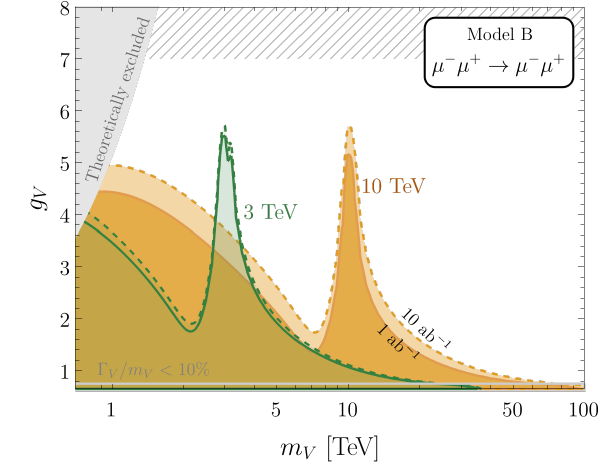}
    \caption{\small 95\% CL sensitivity projections for the $\mu^-\mu^+ \to \mu^-\mu^+$ process in the $(m_V,g_V)$ plane for a 10\,TeV (orange) and a 3\,TeV (green) muon collider. Notation is the same as in previous figures.}
    \label{fig:ll-10-vs-3}
\end{figure}

In \cref{fig:ll-10-vs-3} we show our final 10-bin 95\% CL sensitivity projections for the $\mu^-\mu^+ \to \mu^- \mu^+$ process at a 10\,TeV (orange) and a 3\,TeV (green) muon collider. The solid boundaries show the exclusions for $L=1\text{\,ab}^{-1}$ and the dashed boundaries show $L=10\text{\,ab}^{-1}$. We see that a 10\,TeV muon collider will outperform a 3\,TeV collider for all masses and couplings, apart from at the resonance peak (where the 10\,TeV collider can reach up to $g_V \sim 3.5$). While a $3\,$TeV collider would have a maximum reach of $\sim 35\,$TeV for small coupling values, a $10\,$TeV collider can reach $\sim 100\,$TeV.

\subsubsection{The $e^-e^+$ Channel} 
\label{sec:ee}

\begin{figure}
    \centering
    \includegraphics[width=0.2\linewidth]{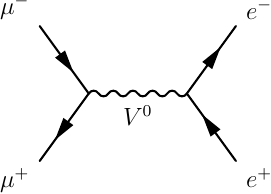}
    \caption{\small
    Feynman diagram for the signal contributions to $\mu^- \mu^+ \to e^-e^+$.
    }
    \label{fig:mumu-ee-feynman-diagrams}
\end{figure}

The $e^- e^+$ channel differs from the $\mu^- \mu^+$ channel as there is only the $s$-channel diagram shown in \cref{fig:mumu-ee-feynman-diagrams}.  Dropping terms of order $\theta_N$, the matrix element is
\begin{align}
    -i\mathcal{M}_\text{HVT}(\mu^-\mu^+ \to e^- e^+) 
    &=
    i\left(\frac{c_\ell\, g^{2}}{2 g_V}\right)^{2} 
        \dfrac{\big(\overline{e_L} \gamma^\mu e_L\big)\big(\overline{\mu_L} \gamma_\mu \mu_L\big)}{s - m_V^{2} + i\, m_V \Gamma_V} \, .
\end{align}
We see that this matrix element has only one pole at $\sqrt{s} = m_V$.  The main background in this channel is the SM $\mu^- \mu^+ \to e^-e^+$, mediated by an $s$-channel $\gamma$ or $Z$.

\begin{figure}
    \centering
    \includegraphics[width=0.49\linewidth]{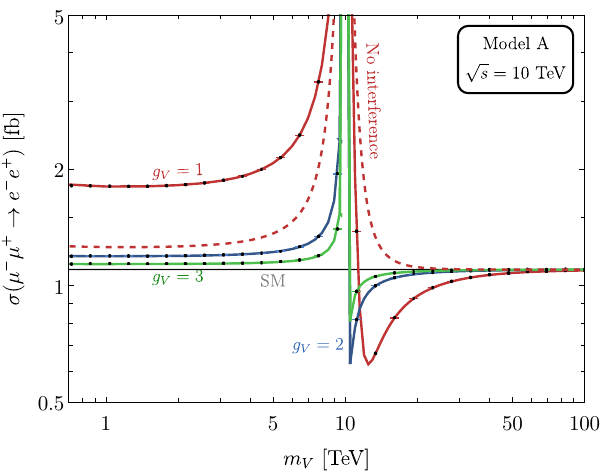}
    \includegraphics[width=0.49\linewidth]{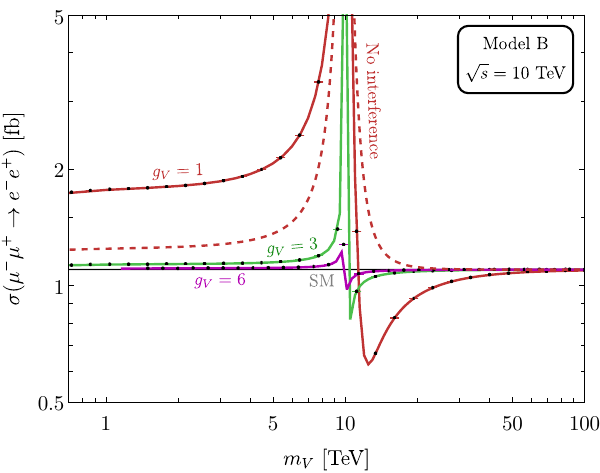}
    \caption{\small The total cross-section, $\sigma(\mu^-\mu^+\to e^-e^+)$, as a function of the HVT mass, $m_V$, for model A (left) and model B (right) at $\sqrt{s} = 10\,$TeV. The coloured lines show the analytic result for the HVT$\oplus$SM cross-section for different values of $g_V$, and the data points show the corresponding numeric \texttt{MadGraph} results. The black line shows the SM cross-section, and the grey shaded region shows the range of uncertainty of the numeric result. The dashed red line shows HVT+SM for $g_V=1$, which neglects interference effects.}
    \label{fig:mumu-ee-x-sections}
\end{figure}

We show the total cross-section as a function of mass for this process in \cref{fig:mumu-ee-x-sections}, for model A (left) and model B (right). We first note that the cross-section is much smaller than for $\mu^-\mu^+$, with a SM cross-section around 1.1\,fb.  In the HVT$\oplus$SM curves in both panels we again see a strong resonance peak, but now the interference shifts the destructive interference to $m_V > 10$\,TeV. In both panels we again see that increasing $g_V$ leads to a smaller HVT contribution.  By considering the no-interference curves we also see in both panels that the HVT-SM contributions are larger than the HVT$^2$ contributions, so correctly modelling the interference is key for setting limits away from the resonance peak.  While in the $\mu^-\mu^+$ channel we saw that the HVT$\oplus$SM cross-section matched the SM cross-section at $m_V \approx 7$\,TeV and at $m_V \approx 10$\,TeV, we see here that they only match at $m_V \approx 10$\,TeV.  We will see that this leads to better sensitivity in the $e^-e^+$ channel at $m_V \approx 7$\,TeV.

\begin{figure}
    \centering
    \includegraphics[width=0.49\linewidth]{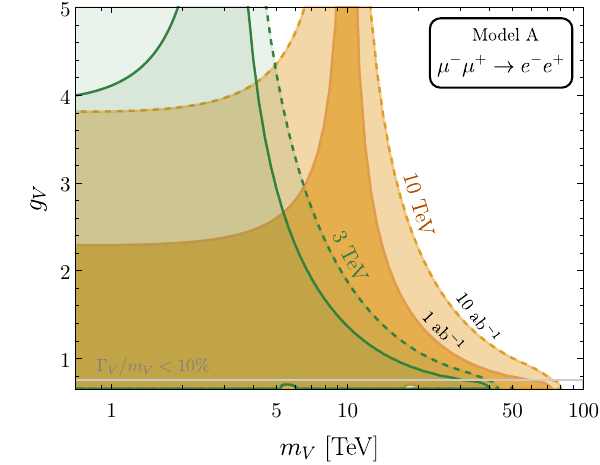}
    \includegraphics[width=0.49\linewidth]{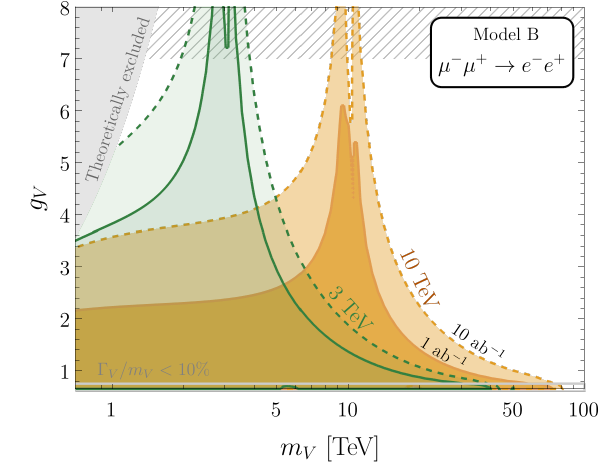}
    \caption{\small 95\% CL sensitivity projections for the $\mu^-\mu^+ \to e^-e^+$ process  in the $(m_V,g_V)$ plane for a 10\,TeV (orange) and a 3\,TeV (green) muon collider.}
    \label{fig:ee-10-vs-3}
\end{figure}

In this case we perform a 10-bin analysis (binned in $|\eta|$) comparing HVT$\oplus$SM with the SM, using the statistical test in \cref{sec:stats-2-to-2}.  We again use a systematic error of $\epsilon_i = 1$\% in each bin.  In \cref{fig:ee-10-vs-3} we show the 95\% CL exclusion limits in the $(m_V,g_V)$ plane for a 10\,TeV (orange) and a 3\,TeV (green) muon collider for 1\,ab$^{-1}$ (solid) and 10\,ab$^{-1}$ (dashed) for model A (left) and model B (right).  Compared to the $\mu^-\mu^+$ channel shown in \cref{fig:ll-10-vs-3}, the $e^-e^+$ channel with 1\,ab$^{-1}$ at a 10\,TeV muon collider is not as powerful for either model at low masses, but performs better in the $4\,\text{TeV} \lesssim m_V \lesssim 8\,\text{TeV}$ region, as the cross-section does not suffer from the cancellation seen in the $\mu^-\mu^+$ channel.  On and above resonance the $\mu^-\mu^+$ and $e^-e^+$ searches are comparable.  Going to 10\,ab$^{-1}$ produces a bigger increase in exclusion in the $e^-e^+$ channel.  This is because the cross-section is smaller, so the greater number of events produced with 10\,ab$^{-1}$ has a stronger impact on the statistical uncertainty.  All the same features are seen at 3\,TeV, but shifted to lower values of $m_V$.

\subsubsection{The $W^-W^+$ Channel} 
\label{sec:ww}

\begin{figure}
    \centering
    \includegraphics[width=0.2\linewidth]{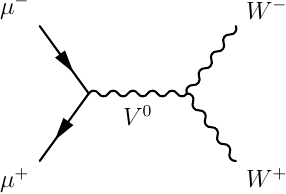}
    \caption{\small
    Feynman diagram for the signal contribution to $\mu^- \mu^+ \to W^-W^+$.
    }
    \label{fig:mumutoww-feynman-diagrams}
\end{figure}

In this section, we discuss $V^0$ production with a subsequent decay into $W^-W^+$, which is mediated by an $s$-channel diagram, \cref{fig:mumutoww-feynman-diagrams}.  The dominant and irreducible SM background is the process $\mu^- \mu^+ \to W^-W^+$ mediated by a photon, $Z$ boson or Higgs boson in the $s$-channel and by a neutrino in the $t$-channel.

We can again compute the amplitude and cross-section analytically. While the analytic expressions in the mass basis are too cumbersome to show explicitly, we highlight a delicate cancellation that occurs in this process.  As in the SM, where the Higgs boson ensures that perturbative unitarity is satisfied in gauge boson scattering as the centre-of-mass energy increases, all the diagrams contributing to $\mu^- \mu^+ \to W^-W^+$ are necessary to obtain an amplitude that does not grow with energy. That is, individual diagrams in $\mu^- \mu^+ \to W^-W^+$ contain energy-growing terms which only cancel in the full sum.

We can show this explicitly in the small $\theta_N,\theta_C$ limit. For simplicity, we also set the muon mass to zero (so we can ignore the Higgs mediated contribution). The matrix element can then be decomposed as
\begin{align}
    |{\cal M}(\mu^- \mu^+ \to W^- W^+)|^2
    =&\,
    |{\cal M}|_\gamma^2 + 
    |{\cal M}|_Z^2 +
    |{\cal M}|_\nu^2 +
    |{\cal M}|_V^2 +
    \notag\\
    &
    |{\cal M}|_{\gamma Z}^2 + 
    |{\cal M}|_{\gamma \nu}^2 + 
    |{\cal M}|_{\gamma V}^2 + 
    |{\cal M}|_{Z \nu}^2 + 
    |{\cal M}|_{Z V}^2 + 
    |{\cal M}|_{\nu V}^2 
    \,,
\end{align}
where the first line contains the SM and HVT squared pieces and the second line contains the interference terms.  Assuming $s,t \gg m_W^2$ and writing the contributions proportional to $t^2/m_W^4$ or $st/m_W^4$ when $s,t \gg m_V$ and at leading order in the small mixing angles $\theta_N$ and $\theta_C$, we find 
\begin{align}
    \label{eq:mumutoWW-unitarity-first}
    |{\cal M}|_V^2 &= \frac{e^6 c_W}{s_W^6}\frac{c_\ell^2}{g_V^2} \frac{s^2t(s+t)}{m_W^4 (s-m_V^2)^2} \theta_C \theta_N 
    \\
    |{\cal M}|_\gamma^2 &=\frac{e^4}{4 s_W^4}\frac{t (s+t)}{m_W^4} \left[{\color{emerald}-8 s_W^4}\right] 
    \\
    |{\cal M}|_Z^2 &=\frac{e^4}{4 s_W^4}\frac{t (s+t)}{m_W^4}\left[{\color{emerald}-8 s_W^4}{\color{royalblue}+4 s_W^2} {\color{deepviolet}-1} {\color{darkorchid}- \frac{2 e c_\ell}{g_V s_W c_W} \theta_N} {\color{darkcyan}-\frac{4 e s_W^3 c_\ell}{g_V c_W} \theta_N} {\color{indigo}+\frac{6 e s_W c_\ell}{g_V c_W} \theta_N} \right] 
    \\
    |{\cal M}|_\nu^2 &= \frac{e^4}{4 s_W^4}\frac{t (s+t)}{m_W^4} \left[{{\color{deepviolet}-1}} {\color{navy}- \frac{4 e c_\ell}{g_V s_W}\theta_C} \right] 
    \\
    |{\cal M}|_{\gamma Z}^2 &=\frac{e^4}{4 s_W^4}\frac{t (s+t)}{m_W^4} \left[ {\color{emerald}16 s_W^4}{\color{royalblue}{-4 s_W^2}} {\color{darkcyan}+\frac{4 e s_W^3 c_\ell}{g_V c_W} \theta_N} {\color{indigo}-\frac{4 e s_W c_\ell}{g_V c_W} \theta_N}\right] 
    \\
    |{\cal M}|_{\gamma \nu}^2 &=\frac{e^4}{4 s_W^4}\frac{t (s+t)}{m_W^4} \left[{\color{royalblue}4 s_W^2} {\color{slatepurple}+\frac{8 e c_\ell s_W}{g_V} \theta_C}\right] 
    \\
    |{\cal M}|_{Z \nu}^2 &=\frac{e^4}{4 s_W^4}\frac{t (s+t)}{m_W^4} \left[{\color{deepviolet}2} {\color{royalblue}-4 s_W^2} {\color{indigo}- \frac{2 e s_W c_\ell }{g_V c_W} \theta_N}  {\color{darkorchid}+ \frac{2 e c_\ell }{g_V c_W s_W} \theta_N} {\color{slatepurple}-\frac{8 e c_\ell s_W}{g_V} \theta_C} {\color{navy}+ \frac{4 e c_\ell}{g_V s_W}\theta_C}\right] 
    \\
    |{\cal M}|_{\gamma V}^2 &=\frac{e^5}{s_W^5}\frac{c_{\ell}}{2 g_V}\frac{s t (s+t)}{m_W^4(s-m_V^2)}\left[{\color{firebrick}4 c_W s_W^2 \theta_C} {\color{burntorange}-2 s_W^2 \theta_N} {\color{sienna}+2s_W^4 \theta_N}\right] 
    \\
    |{\cal M}|_{Z V}^2 &=\frac{e^5}{s_W^5}\frac{c_{\ell}}{2 g_V}\frac{s t (s+t)}{m_W^4(s-m_V^2)} \left[{\color{vermilion}2 c_W \theta_C} {\color{firebrick}-4 c_W s_W^2 \theta_C} {\color{garnet}-\theta_N} {\color{sienna}-2 s_W^4 \theta_N} {\color{burntorange}+3 s_W^2\theta_N}\right] 
    \\
    |{\cal M}|_{\nu V}^2 &= 
    \frac{e^5}{s_W^5}\frac{c_{\ell}}{2 g_V}\frac{s t (s+t)}{m_W^4(s-m_V^2)} \left[{\color{vermilion}-2 c_W \theta_C} {\color{garnet}+ \theta_N} {\color{burntorange}-s_W^2 \theta_N} \right] 
    \,,
    \label{eq:mumutoWW-unitarity-last}
\end{align}
where $s_W=\sin \theta_W$, $c_W=\cos \theta_W$ and $\theta_{W}$ is the Weinberg angle.  The colour coding in \crefrange{eq:mumutoWW-unitarity-first}{eq:mumutoWW-unitarity-last} highlights the terms that sum to zero. We see that the terms growing with the energy cancel at order $\theta_{N,C}$ and unitarity is restored. Note that $|{\cal M}|_V^2$ is proportional to $\theta_C\theta_N$ and cancels against subleading contributions from other terms.  We also note that terms involving the new heavy vector are proportional to $\sin \theta_{C,N}$ indicating that this contribution would vanish in the limit of zero mixing. It is thus only the mixing between the HVT and the SM $Z$ and $W$ bosons which transfers a relevant contribution of the terms required for this cancellation to the HVT diagrams.  Although we refrain from writing it out, the same cancellation occurs for terms proportional to $s/m_W^2$ and $t/m_W^2$ and at higher orders in $\theta_{N,C}$. For these terms a more intricate cancellation occurs between all parts of the matrix element (as opposed to the cancellation within the SM or interference sectors as seen above). In the end the matrix element does not grow with the centre-of-mass energy of the collider and
\begin{align}
    |{\cal M}(\mu^- \mu^+ \to W^- W^+)|^2
    \lesssim&\,
    \mathcal{O}(1)
\end{align}
for all energies.  This discussion highlights the importance of the interference terms. If we ignored them and simply generated our signal from the $s$-channel diagram mediated by the HVT we would obtain a cross-section that is proportional to $t(s+t)/m_W^4$ and vastly overestimates the actual cross-section. This would lead to overly optimistic sensitivity projections, as we will show below. Even for a small mixing of the HVT with the SM gauge bosons, at large enough energies all the interference terms need to be taken into account.

\begin{figure}
    \centering
    \includegraphics[width=0.49\linewidth]{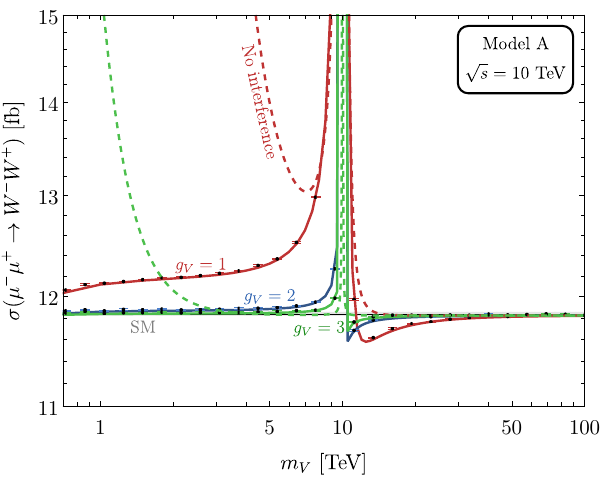}
    \includegraphics[width=0.49\linewidth]{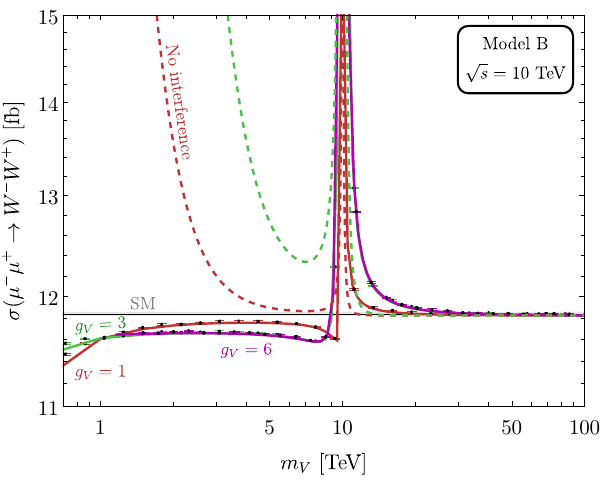}
    \caption{\small The total cross-section, $\sigma(\mu^-\mu^+\to W^-W^+)$, as a function of the HVT mass, $m_V$, for model A (left) and model B (right) at $\sqrt{s} = 10\,$TeV. The coloured lines show the analytic result for the HVT$\oplus$SM cross-section for different values of $g_V$, and the data points show the corresponding \texttt{MadGraph} results. The grey line shows the SM cross-section.}
    \label{fig:analytic-vs-mg5-ww-x-sections}
\end{figure}

\Cref{fig:analytic-vs-mg5-ww-x-sections} shows the $10\,$TeV cross-section $\sigma(\mu^- \mu^+ \to W^- W^+)$ for $g_V = 1,2,3$ for model A (left) and $g_V = 1,3,6$ for model B (right). The solid lines show the results of our analytic computation while the dots depict the corresponding \texttt{MadGraph} results. Note that \texttt{MadGraph}'s errors on the cross-section are included in the plot and are too small to be visible. The grey line shows the SM cross-section. In both panels we see resonance features and a smaller HVT contribution away from the peak.  We see that the cross-section in model A decreases for increasing $g_V$. This is because $|\mathcal{M}|_V^2 $ is proportional to $1/g_V^4$, when we use couplings given in \cref{Table: Benchmark parameters}.  A smaller $g_V$ also leads to a larger width, which leads to a broader resonance peak.   In model B, $|\mathcal{M}|_V^2 $ goes as $(g_V^2/2 - g^2)^2/(g_V^2 - g^2)^2$ which vanishes at $g_V = 0.92$ and tends to $1/4$ for $g_V \gg g$. This explains why the cross-section for $g_V = 1$ is similar to the SM and why the cross-section is similar for $g_V = 6$ and for $g_V = 3$.  The cross-section in model B also increases as $g_V \to g$, which we do not show.  We note that the interference leads to a larger cross-section than the SM below $m_V = 10$\,TeV in model A, but a smaller cross-section in model B.  This is because the interference terms are proportional to $c_F$ (see, e.g., \cref{eq:mumutoWW-unitarity-last}) and the sign of $c_F$ differs between these models (see \cref{Table: Benchmark parameters}).  In both panels we see that if we neglect interference (that is, when we plot $|{\cal M}|_V^2 + |{\cal M}|_\text{SM}^2$), the cross-sections become large when $m_V \ll \sqrt{s}$.  Although it is impractical to show the cancellation of the energy growing terms proportional to $\theta_C \theta_N$ in \crefrange{eq:mumutoWW-unitarity-first}{eq:mumutoWW-unitarity-last}, we see that the term in \cref{eq:mumutoWW-unitarity-first} must be cancelled by interference terms to give an accurate prediction of the cross-section.

\begin{figure}
    \centering
    \includegraphics[width=0.49\linewidth]{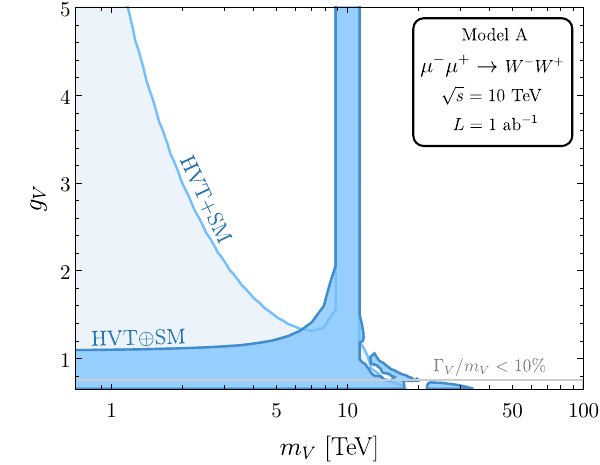}
    \includegraphics[width=0.49\linewidth]{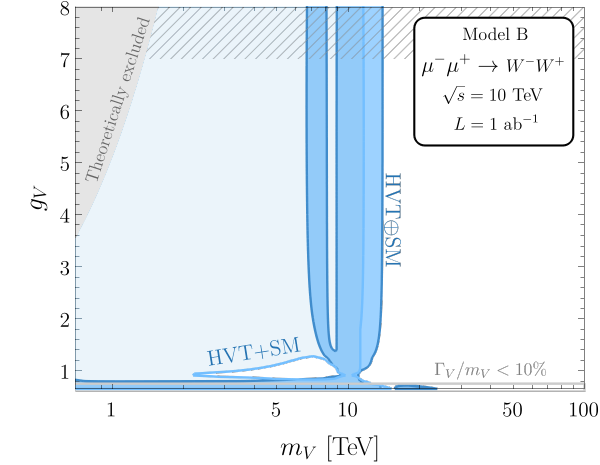}
    \caption{\small 95\% CL sensitivity projections for $\mu^-\mu^+ \to W^-W^+ \to$ hadrons at integrated luminosity $1\ \text{ab}^{-1}$ with the 10-bin HVT$\oplus$SM analysis. The HVT analysis without interference is shown by the lighter region.  The solid grey region is theoretically excluded, and the hatched region in grey indicates the loss of perturbative unitarity.}
    \label{fig:WW-exclusions}
\end{figure}

In \cref{fig:WW-exclusions} we show the projected 95\% CL exclusion limits for $\mu^- \mu^+ \to W^- W^+$ decaying fully hadronically at a 10\,TeV collider with 1 ab$^{-1}$ using an analysis with ten bins evenly spread over $|\eta|<2.5$.  We take BR$(W \to \text{hadrons}) = 67\%$ and account for a $73\%$ tagging efficiency of the hadronic $W$ bosons, following recent CMS results \cite{CMS-DP-2023-065}. We include a $\epsilon_i = 5\%$ systematic error to account for uncertainties in the detector and in reconstruction.  As in \cref{fig:ll-exclusions}, we show the full HVT$\oplus$SM analysis (including interference) with dark blue shaded regions, and the HVT+SM analysis (excluding interference) with light blue regions. In the HVT$\oplus$SM analysis, we see for model A that values of $g_V \lesssim 1.1$ can be excluded for $m_V \lesssim 7.5\,$TeV with $1\,$ab$^{-1}$. The resonant region at $m_V \approx 10\,$TeV has a very high sensitivity, excluding all perturbative values of $g_V$. The sensitivity decreases for larger masses, and masses up to $\sim 30\,$TeV can be reached. As expected, the sensitivity projections for HVT$\oplus$SM in model A are weaker than for the $\mu^- \mu^+$ channel, shown in \cref{fig:ll-exclusions}. The difference is particularly stark for $m_V < 7.5\,$TeV due to the absence of a $t$-channel diagram. In model B, with 1\,ab$^{-1}$ this channel is only sensitive around the resonance mass $m_V \approx 10$\,TeV.  We see in \cref{fig:analytic-vs-mg5-ww-x-sections} that the HVT$\oplus$SM cross-section is very similar to the SM cross-section for mass $1\,\text{TeV}\lesssim m_V \lesssim 6$\,TeV.  There is a loss of sensitivity at $m_V \approx 8.5$\,TeV, as the HVT cross-section matches the SM cross-section here.  The sensitivity loss at $g_V = 0.92$ is due to the fact that the HVT cross-section vanishes at this point and remains small when mixing effects are taken into account.  Finally, we see that in both model A and model B neglecting the interference leads us to significantly overestimate the collider sensitivity. This is due to the cancellation between the signal and background diagrams described above. 

\begin{figure}
    \centering
    \includegraphics[width=0.49\linewidth]{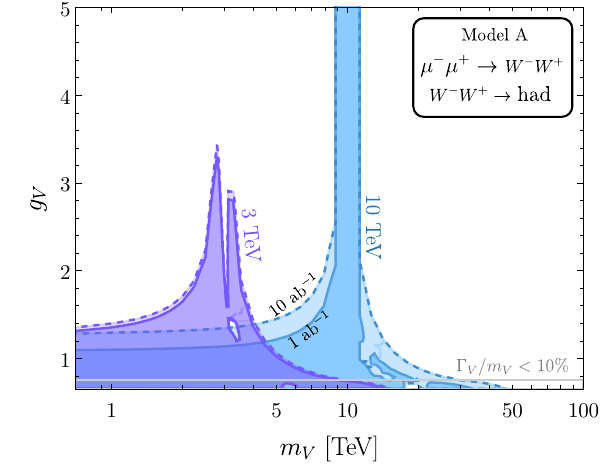}
    \includegraphics[width=0.49\linewidth]{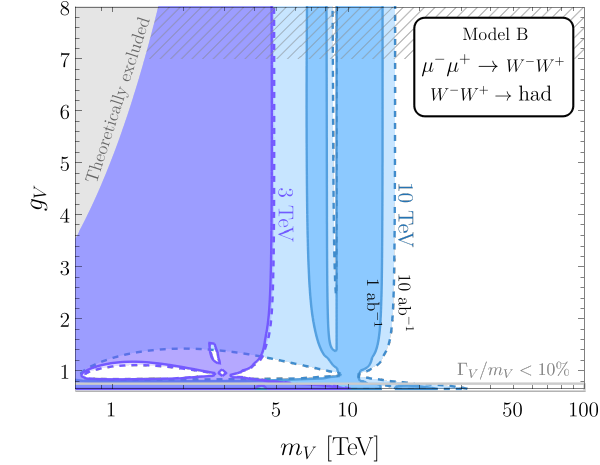}
    \caption{\small 95\% CL sensitivity projections for $\mu^-\mu^+ \to W^-W^+$ in the $(m_V,g_V)$ plane for a 10\,TeV (blue) and a 3\,TeV (purple) muon collider. Notation is the same as in previous figures.}
    \label{fig:WW-10-vs-3}
\end{figure}

\Cref{fig:WW-10-vs-3} shows our sensitivity projections for $3$ and $10\,$TeV centre-of-mass energies in purple and blue, respectively, with 1 and 10\,ab$^{-1}$. Firstly, with 10\,ab$^{-1}$ the reach of a 10\,TeV collider is slightly better on Model A and much better on Model B, probing almost the whole parameter space with $m_V \lesssim 14$\,TeV.  This occurs because the analysis becomes sensitive to the small deviation from the SM seen in \cref{fig:analytic-vs-mg5-ww-x-sections}.  Also note that in model B, this channel is sensitive to large $g_V$, since in this model $c_H$ is unsuppressed when $g_V$ is large.  The sensitivity at a $3\,$TeV collider exhibits the same features as at $10\,$TeV. In model A, we see that the resonant peak moves to $m_V = 3\,$TeV. For smaller masses, the sensitivity in $g_V$ is enhanced compared to a $10\,$TeV analysis. At $3\,$TeV, values of $g_V \sim 1.4$ can be reached for $m_V < 3\,$TeV, compared to $g_V \sim 1.1$ with a $10\,$TeV centre-of-mass energy, with $1\,$ab$^{-1}$. Maximum masses of $m_V \sim 15\,$TeV can be accessed with a $3\,$TeV collider compared to $m_V \sim 50\,$TeV at $10\,$TeV. In model B, a $3\,$TeV machine shows weaker performance than a $10\,$TeV collider with a similar reach in $g_V$ but a mass reach that is reduced to $m_V \lesssim 5$\,TeV.  With a 3\,TeV collider, 1 and 10\,ab$^{-1}$ can probe almost the same regions of parameter space of model B.

\subsection{$2 \to 3$ Processes}
\label{sec:2-to-3}

In this section we discuss the projected sensitivity of the $2 \to 3$ processes, with the final states $\ell^- \ell^+ \gamma$, $W^- W^+ \gamma$ and $W^- W^+ Z$.  At a fixed-energy lepton collider, $2 \to 3$ processes could have an advantage over $2 \to 2$ processes.  Since the extra particle takes away energy, the new heavy particles can be produced on-shell even when $m_V \not\approx \sqrt{s}$. However, this comes at the price of an extra particle in the final state phase space volume and an additional power of $e^2$ (for a final state photon or $Z$ boson) compared to the cross-section of a $2 \to 2$ process. This is not as bad as it may seem though, as this applies to signal, interference and background in the same way.

As in \cref{sec:2-to-2}, we again find that it is important to correctly account for interference.  This is true in the $\ell^-\ell^+ \gamma$ channel, where the HVT can appear both in resonant and non-resonant propagators, and for $W^-W^+\gamma$ and $W^-W^+Z$, to provide accurate unitarity cancellations (analogous to the $W^-W^+$ channel discussed in \cref{sec:ww}).   Due to the large number of possible topologies, we use \texttt{MadGraph} to compute the signal, background and interference. We use \texttt{MadGraph 3.6.3} for the simulation of $\ell^- \ell^+ \gamma$ and $W^- W^+ Z$, and \texttt{MadGraph 3.5.7} for $W^- W^+ \gamma$ (in each case we use the corresponding version of \texttt{MadGraph} when computing the SM background) \cite{Alwall:2014hca}.  The \texttt{MadGraph} simulation introduces a Monte Carlo error that we account for as discussed in \cref{sec:stats-2-to-3}. To isolate the resonance peaks occurring in some diagrams we will bin events in the invariant mass where the resonance is expected, so the invariant mass of final state leptons, $m_{\ell\ell}$, the invariant mass of final $W$ bosons for the $W^-W^+\gamma$ channel, and the best among $m_{W^\pm Z}$ and $m_{WW}$ for the $W^-W^+Z$ channel. To have enough Monte Carlo events in each bin, we do not consider any other kinematic variables.

\subsubsection{The $\ell^-\ell^+ \gamma$ Channel} 
\label{sec:mumua}

In this section we derive sensitivity projections for the process $\mu^- \mu^+ \to \ell^-\ell^+ \gamma$, where $\ell = \{e, \mu\}$. We do not separate muons and electrons as we did for $2 \to 2$ processes since the dominant muon cross-section does not feature a cancellation as it did in the $2 \to 2$ case.  We consider the irreducible SM background $\mu^- \mu^+ \to \ell^-\ell^+ \gamma$ and assume that all other backgrounds can be sufficiently reduced with appropriate triggers and kinematic cuts. 

\begin{figure}
    \centering
    \includegraphics[height=0.17\linewidth]{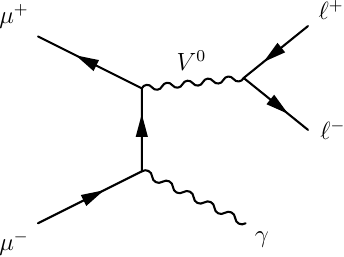}
    \hspace{1cm}
    \includegraphics[height=0.17\linewidth]{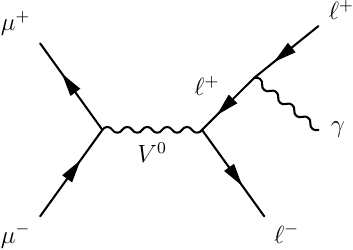}
    \hspace{1cm}
    \includegraphics[height=0.17\linewidth]{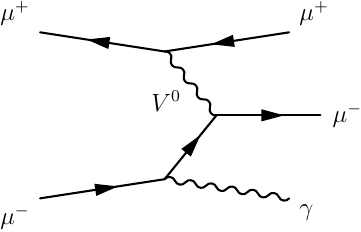}
    \caption{\small
    Representative Feynman diagrams for $\mu^- \mu^+ \to \ell^-\ell^+\gamma$ and $\mu^- \mu^+ \to \mu^-\mu^+\gamma$.
    }
    \label{fig:fey_diag_lla}
\end{figure}

The Feynman diagrams involving the HVT are identical to the SM diagrams featuring an internal $Z$ boson, but where the $Z$ boson is traded for a $V^0$. Counting both $e$ and $\mu$ final states there are 36 possible SM diagrams and 12 diagrams involving a $V^0$.\footnote{Here and in the following consider charge conjugation to give distinct diagrams (following the \texttt{MadGraph} convention).} Of these, we show three representative topologies in \cref{fig:fey_diag_lla}. We see that in some diagrams the leptons will have an invariant mass peaked at $m_{\ell\ell} = m_V$, others will have this feature only when the photon is relatively soft, and others will not show a resonance feature.  While we will not show the results of neglecting interference, they are similar to that seen in \cref{sec:mumu} and interference effects are generally large.

\begin{figure}
    \centering
    \includegraphics[width=0.49\linewidth]{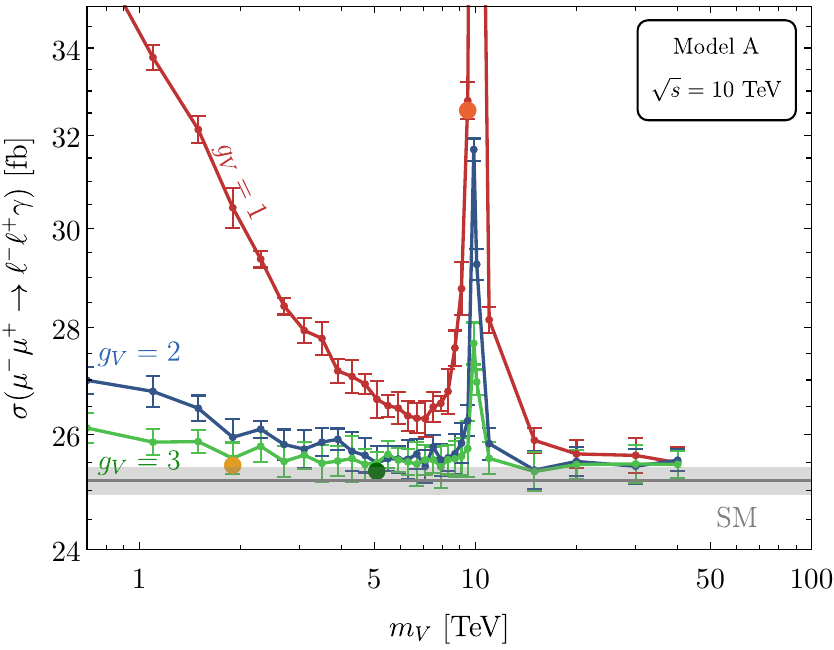}
    \includegraphics[width=0.49\linewidth]{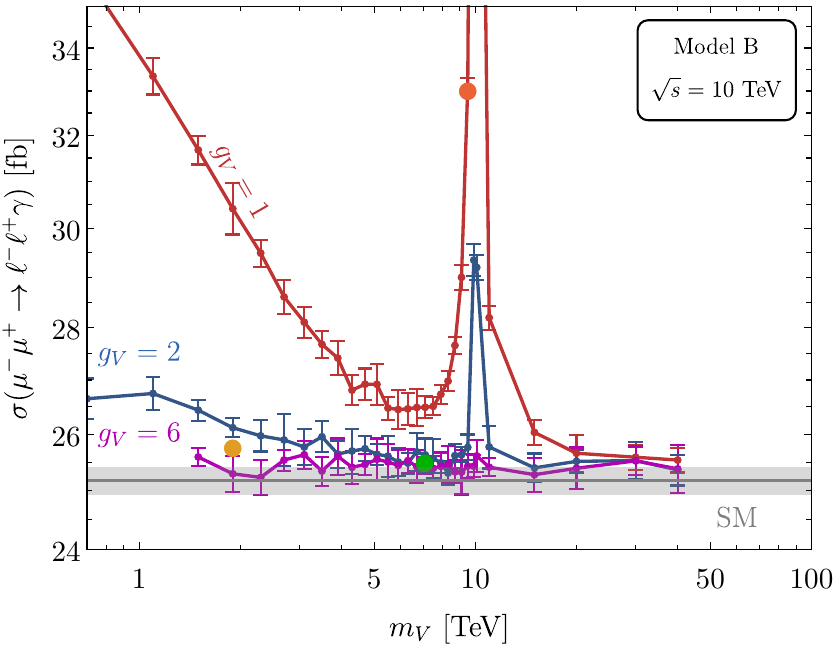}
    \caption{\small The total cross-section $\sigma(\mu^-\mu^+\to \ell^-\ell^+ \gamma)$ as a function of the HVT mass, $m_V$, for model A (left) and model B (right) at $\sqrt{s} = 10\,$TeV. The coloured lines show the cross-sections for different values of $g_V$. The three coloured points correspond to the benchmark points presented in \cref{fig:mll_dist}.  The grey shaded region shows the range of uncertainty on the SM cross-section of the numeric result.}
    \label{fig:mm_lla_xsec}
\end{figure}

We show the total cross-section for $\mu^- \mu^+ \to \ell^-\ell^+ \gamma$ in \cref{fig:mm_lla_xsec}, for $g_V = 1,2,3$ in model A (left) and $g_V = 1,3,6$ in model B (right) as a function of $m_V$. Overall the cross-section is around an order of magnitude smaller than for the $\mu^-\mu^+$ channel, but not as small as may be expected from the suppression factors mentioned above.   This is due to the possibility of having $V^0$ on-shell and to the larger number of diagrams contributing to this process.  For both model A and model B we see that there is resonant enhancement at $m_V \approx 10\,$TeV, which corresponds to a resonant propagator.  In both cases there is a dip for $m_V = 5$ to $9\,$TeV and a sizable enhancement for masses $m_V < 5\,$TeV, which is a result of the topologies that are $t$-channel-like when the photon is soft (e.g., the final diagram in \cref{fig:fey_diag_lla}). Note, however, that the cross-section is always larger than the SM cross-section, so while interference is important the overall sign of the new physics contribution is positive.  Finally, we see that the HVT contribution decreases for increasing $g_V$, as expected following the discussion of the $\mu^-\mu^+$ channel.

\begin{figure}
    \centering
    \includegraphics[width=0.49\linewidth]{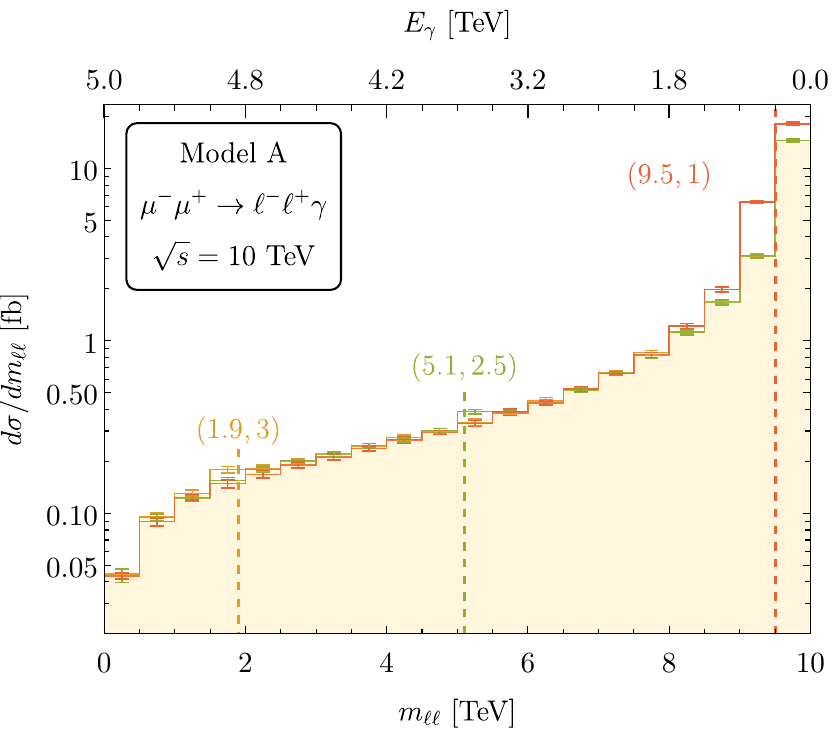}
    \includegraphics[width=0.49\linewidth]{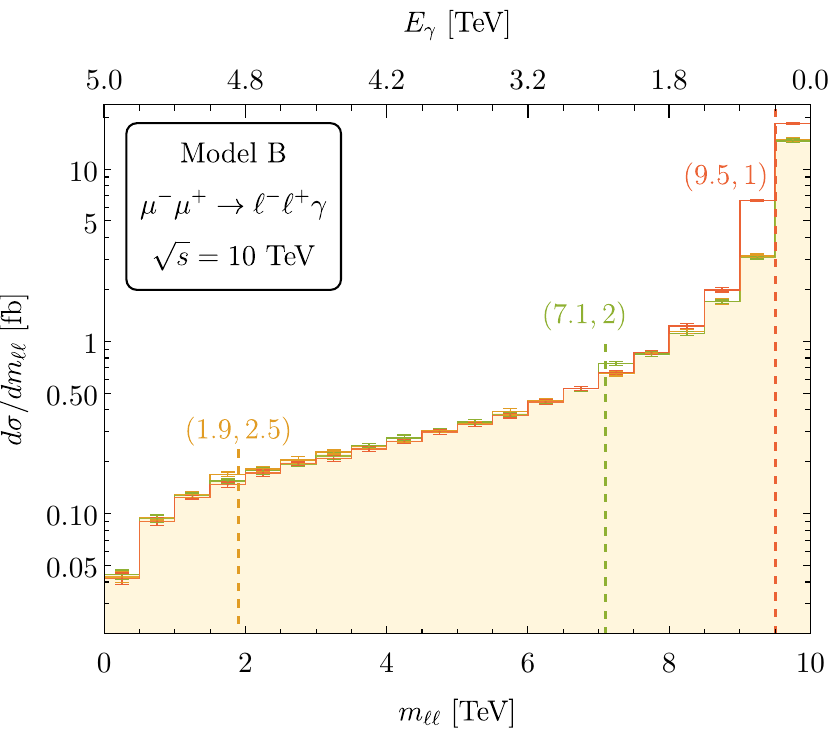}
    \\
    \includegraphics[width=0.49\linewidth]{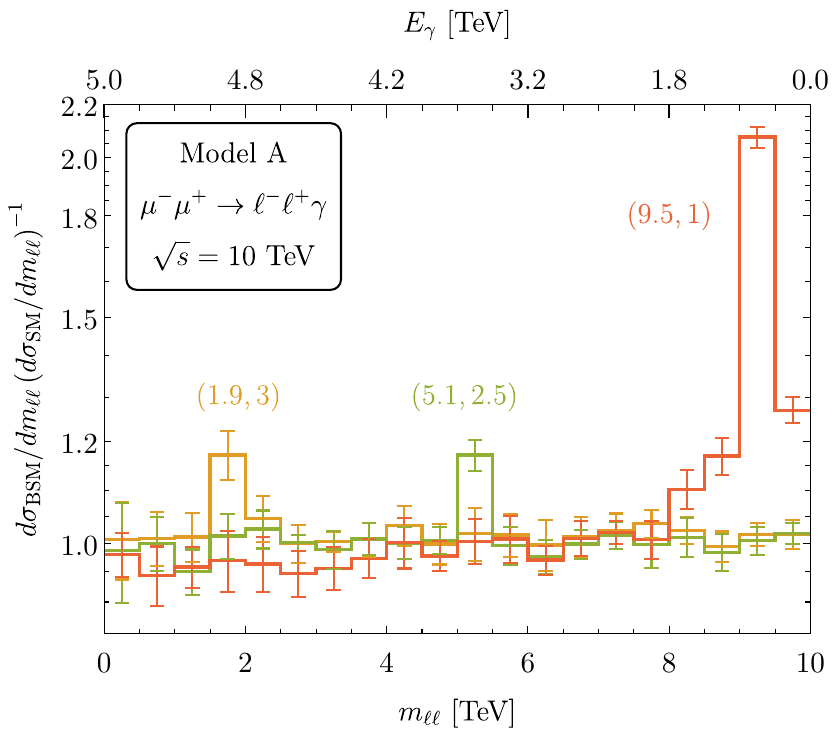}
    \includegraphics[width=0.49\linewidth]{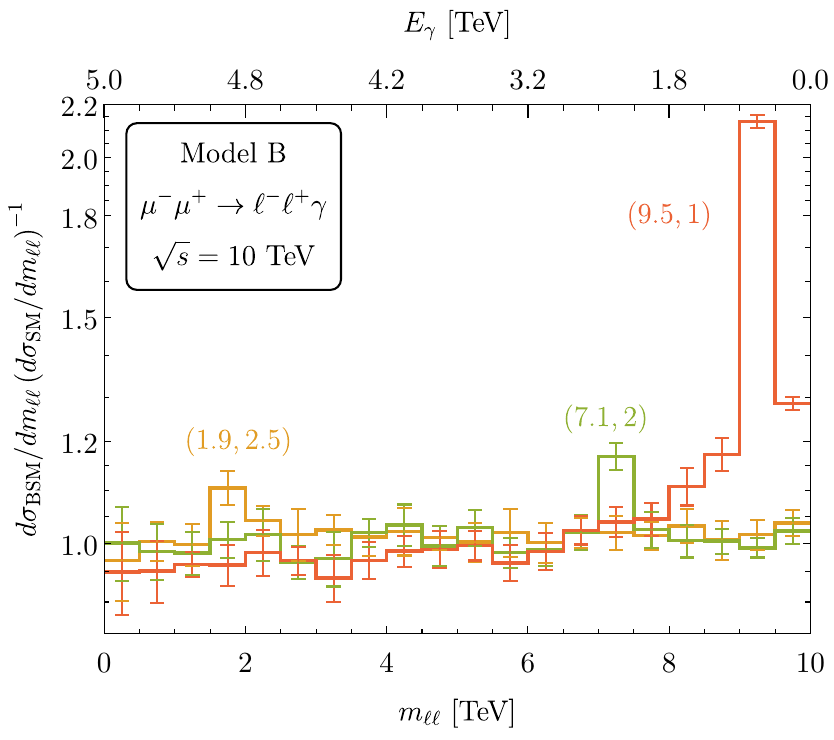}
    \caption{\small Top: The differential cross-section for the $\ell^-\ell^+\gamma$ final state as a function of the invariant mass of the two final state leptons $m_{\ell\ell}$ (bottom axis) and recoil energy (top axis) for model A (left) and model B (right).  Bottom: The ratio of the HVT$\oplus$SM differential cross-section to the SM cross-section, with relative errors.}
    \label{fig:mll_dist}
\end{figure}

In \cref{fig:mll_dist} (top) we show the differential cross-section for the SM and for HVT$\oplus$SM for three example points in the HVT parameter space, denoted by $(m_V, g_V)$, as a function of the invariant mass of the two leptons.  We see that there are resonance peaks when $m_{\ell\ell} \approx m_V$, even when $m_{\ell\ell} \not\approx \sqrt{s}$.   This is particularly clear in the bottom plots, which show the deviations from the SM expectation.  As discussed in \cref{sec:stats-2-to-3}, we see from the top plots that the distribution is sharply peaked at $m_{\ell\ell} \approx \sqrt{s}$.  To obtain a sufficient number of Monte Carlo samples at low $m_{\ell\ell}$ (which is important for points in parameter space where $m_V \ll \sqrt{s}$) we use the bias function 
\begin{align}
    f_\text{bias}^{\ell^-\ell^+\gamma}=E_\gamma^2
    \,.
\end{align}
For this process $E_\gamma = (s-m_{\ell\ell}^2)/(2\sqrt{s})$, so small $m_{\ell \ell}$ corresponds to large $E_\gamma$ (note that we plot $E_\gamma$ at the top of the panels in \cref{{fig:mll_dist}} and $m_{\ell\ell}$ at the bottom).  Even though the cross-section is much smaller at small $m_{\ell\ell}$, we see from the bottom panels that the relative errors are roughly constant across the $m_{\ell\ell}$ range.

\begin{figure}
    \centering 
    \includegraphics[width=0.49\linewidth]{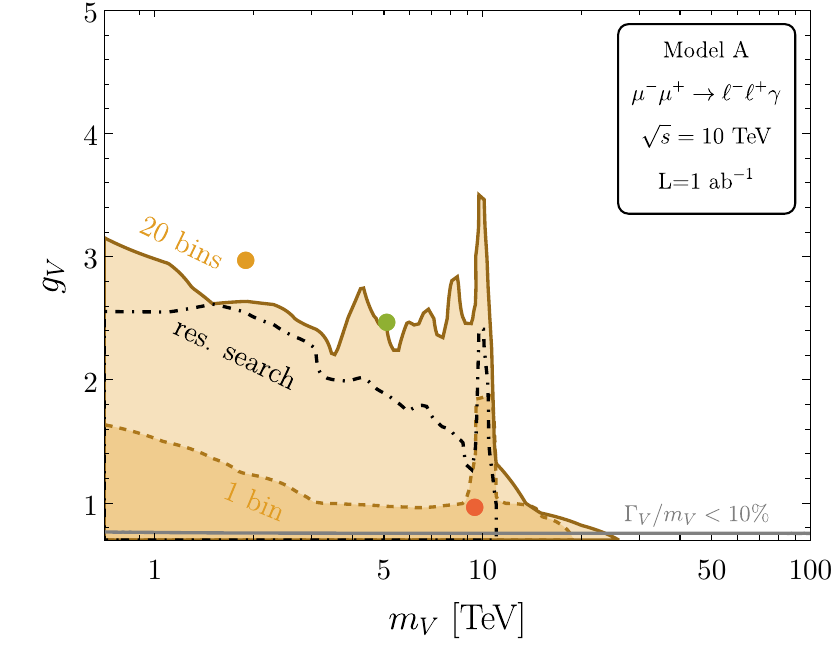}  
    \includegraphics[width=0.49\linewidth]{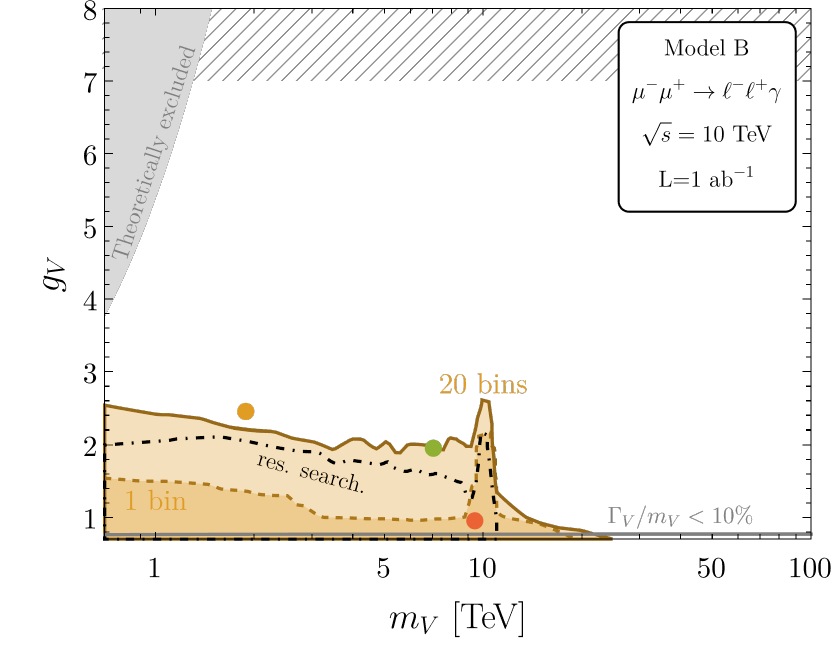}
    \caption{\small 95\% CL sensitivity projections for the $\mu^-\mu^+ \to \ell^-\ell^+ \gamma$ process for model A (left) and model B (right) in the $(m_V,g_V)$ plane at a $10\,$TeV muon collider for data binned in $m_{\ell \ell}$ (solid), unbinned data (dashed) and for a resonance search (black). The limits are shown for an integrated luminosity of 1\,ab$^{-1}$. The three coloured points correspond to the benchmark points discussed in \cref{fig:mll_dist}. For model B (right), the solid grey region is theoretically excluded and the hatched region in grey indicates the loss of perturbative unitarity.}
    \label{fig:exclusion_lla}
\end{figure}

We then compute the projected sensitivity using the procedure discussed in \cref{sec:stats-2-to-3}, where now we bin in $m_{\ell\ell}$.  We use a systematic error of $\epsilon_\text{syst} = 5$\%, to approximately take into account reconstruction uncertainties.  We show the projected sensitivity in \cref{fig:exclusion_lla} for model A (left) and model B (right) for $\sqrt{s} = 10\,$TeV and $ L = 1\,$ab$^{-1}$. The solid brown line shows the sensitivity of a 20-bin analysis. We also show the sensitivity of an unbinned analysis using only the value of the total cross-section with a dashed orange line. We see that the binned analysis is significantly better for all masses, particularly those below the resonance mass. Above the resonance mass the range of $g_V$ probed is similar, but the binned analysis can reach up to around 30\,TeV, while the total cross-section analysis can only reach around 20\,TeV.  We also show the projected sensitivity for a more typical resonance search, where we require
\begin{align}
    m_{\ell \ell} \in [m_V-2\Delta,m_V+2\Delta]
    \,.
\end{align} 
To be able to compare our later results with those derived in Ref.~\cite{Liu:2023jta} for di-boson final states, we use the same width-dependent $\Delta$,
\begin{align}
    \label{eq:Delta}
    \Delta &=
    \begin{cases}
        0.05\,m_V & \Gamma_V < 0.05\,m_V \\
        \Gamma_V & 0.05\,m_V < \Gamma_V < 1.5\,\text{TeV} \\
        1.5\,\text{TeV} & \Gamma_V > 1.5\,\text{TeV}\,.
    \end{cases}
\end{align}
We see that the sensitivity is weaker than the analysis that considers all invariant mass bins. We also see that the sensitivity of this analysis decreases as $m_V$ increases, until the resonance region is reached. The resonance search is not sensitive above the centre-of-mass energy of the collider.

\begin{figure}
    \centering
    \includegraphics[width=0.49\linewidth]{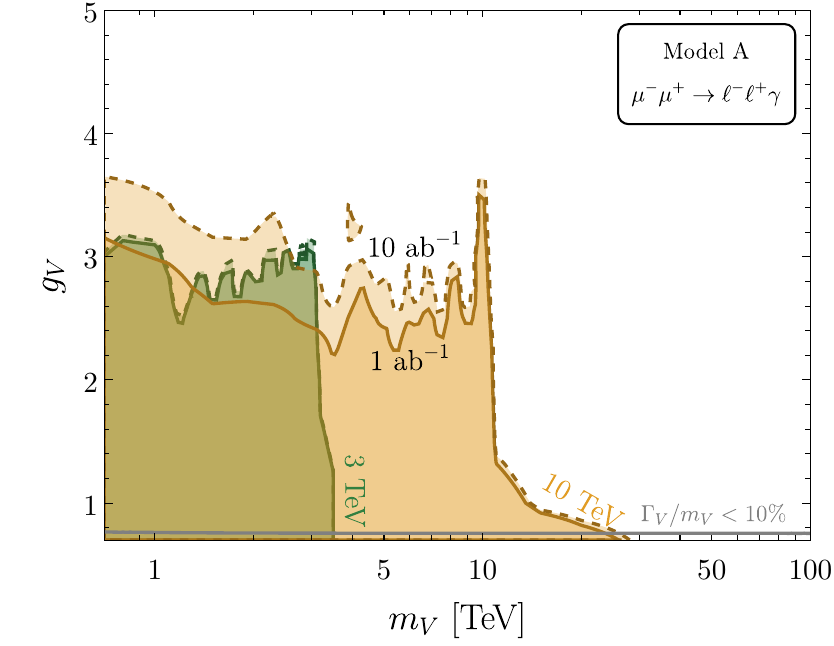}
    \includegraphics[width=0.49\linewidth]{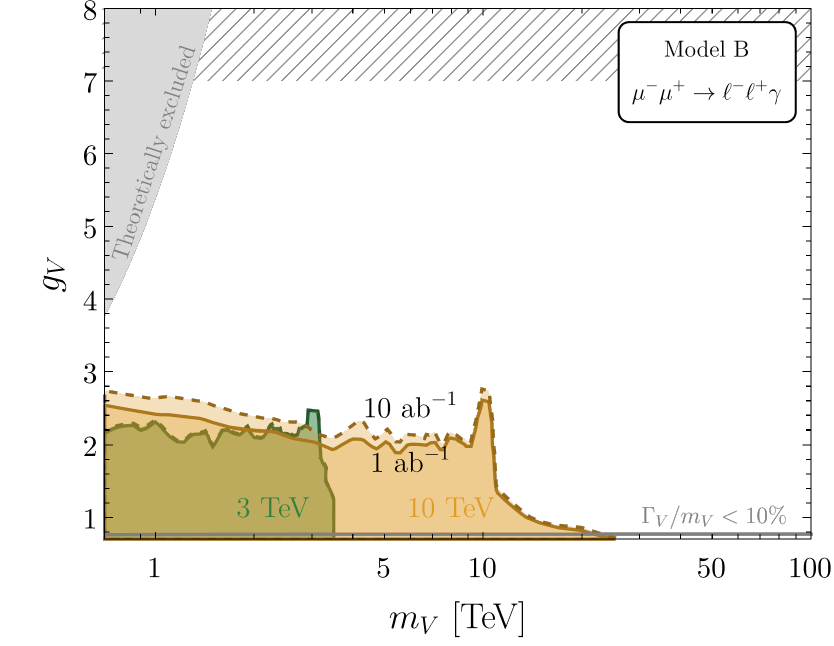}
    \caption{\small 95\% CL sensitivity projections for the $\mu^-\mu^+ \to \ell^-\ell^+\gamma$ process in the $(m_V,g_V)$ plane for a 10\,TeV (orange) and a 3\,TeV (green) muon collider for $1\,$ab$^{-1}$ (solid) and $10\,$ab$^{-1}$ (dashed). Notation is the same as in previous figures.}
    \label{fig:lly-10-vs-3}
\end{figure}

In \cref{fig:lly-10-vs-3} we see sensitivity projections for $\sqrt{s} = 3$ and $10\,$TeV in orange and green, respectively, for an integrated luminosity of $1$ and $10\,$ab$^{-1}$, in solid and dashed. For both model A and B we see that the mass reach increases from $3.5$ to $20\,$TeV with a higher centre-of-mass energy. The effect of an increased luminosity is less significant. At a $3\,$TeV collider, the increase in luminosity leads to a negligible improvement in sensitivity. For model A, at a $10\,$TeV collider, increasing the luminosity from $1$ to $10\,$ab$^{-1}$ improves the reach on $g_V$ from $\sim 3$ to $\sim 3.5$ at small $m_V$. The model B reach is not as large as in model A due to a larger width.

\subsubsection{The $W^-W^+\gamma$ Channel}
\label{sec:wwa}

\begin{figure}
    \centering
    \includegraphics[width=0.23\linewidth]{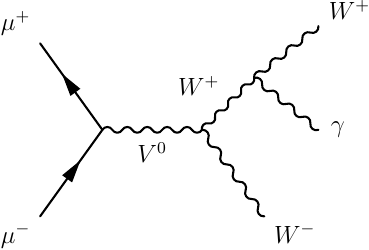}
    \hspace{0.05cm}
    \includegraphics[width=0.23\linewidth]{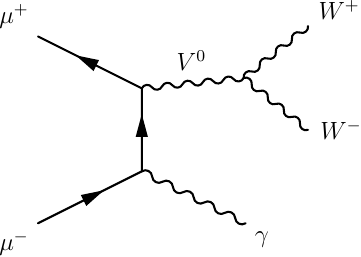}
    \hspace{0.05cm}
    \includegraphics[width=0.23\linewidth]{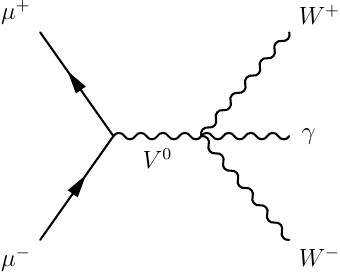}
    \caption{\small
    Representative Feynman diagrams for signal contributions to $\mu^- \mu^+ \to W^-W^+\gamma $.
    }
    \label{fig:fey_diag_wwa}
\end{figure}

In this section we compute sensitivity projections for the process $\mu^- \mu^+ \to W^- W^+ \gamma$. For the SM processes there are 18 distinct tree-level diagrams and in principle another 15 diagrams involving the  HVT. However, in both model A and B the coupling of the $V^\pm W^\mp \gamma$ vertex is $i c_C s_C g s_W(1-c_{VVW})=0$, since $c_{VVW}=1$. This leaves five non-zero signal topologies in this channel.  We show three of these in \cref{fig:fey_diag_wwa}.  The other two are like the left and middle diagrams but with the photon attached to the $W^-$ and $\mu^+$ legs, respectively.

In the first diagram of \cref{fig:fey_diag_wwa} the product of propagators is
\begin{equation}
    -i\mathcal{M} \propto
    \dfrac{1}{s-m_V^2+i m_V \Gamma_V} \dfrac{1}{s-2\sqrt{s}E_{W^-}+i m_W \Gamma_W}
    \,,
\end{equation}
where $E_{W^-}$ is the energy of the $W^-$ boson.  We see that the first propagator is enhanced when $m_V \approx \sqrt{s}$ and the second propagator is enhanced when $E_{W^-} \approx \sqrt{s}/2$.  Also note that the $\mu^-\mu^+V^0$ coupling can have the opposite sign to the $\mu^- \mu^+ \gamma$ and $\mu^- \mu^+ Z$ couplings. The interference can therefore be negative and induce cancellations between the HVT and SM diagrams. 

The second diagram also has two peaks, coming from the $t$-channel muon propagator and the HVT propagator,
\begin{align}
    -i\mathcal{M} \propto
    \dfrac{1}{t} \dfrac{1}{m_{WW}^2-m_V^2+im_V \Gamma_V} 
    \,.
\end{align}
The momentum exchange $t = (p_{\mu^-}-p_{\gamma})^2 = -2 E_\gamma (\sqrt{s}/2) (1 - \cos \theta)$ is minimised for photons that are close to the beam line. Since we impose $|\eta_\gamma| < 2.5$, photons from this topology are likely to have an absolute pseudorapidity close to $2.5$. The variable $|t|$ is also minimised by $E_\gamma \sim 0$.  In this case the invariant mass of the $W^-W^+$ pair is $m_{WW}\approx \sqrt{s}$. The $s$-channel propagator is maximised when $m_{WW} \sim m_V$. So there can be an enhancement in both propagators if $m_V \sim \sqrt{s}$. For smaller HVT masses both propagators cannot be enhanced at the same time. This diagram with an on-shell HVT is commonly referred to as associated production.

The third diagram in \cref{fig:fey_diag_wwa} includes a contact interaction and contains an $s$-channel propagator, which enhances this process when $m_V \approx \sqrt{s}$. 

From these considerations we see that it is important to include all diagrams rather than just a subset. Furthermore, as in \cref{sec:ww}, it is important to compute interference effects between the SM and HVT diagrams, to ensure that the matrix element does not grow with energy. We therefore simulate all signal diagrams in a gauge invariant amplitude and allow them to interfere with themselves and the SM background, where both unitarity cancellations and destructive interference (between signal and background and among different signal channels) can play an important role. As a consequence, there are no simple cuts which favour the signal and reduce the background. The most promising differential distribution is the resonant mass of the SM bosons produced in the HVT decay. A cut on the invariant mass would predominantly select the associated production diagrams. However, we will see below that this weakens the constraints compared to using the full resonant mass distribution.  Since it is computationally expensive to simulate all diagrams including interference, we only consider a differential distribution in the resonant mass distribution (and not also in $|\eta|$, as we did previously).

\begin{figure}
    \centering
    \includegraphics[width=0.49\linewidth]{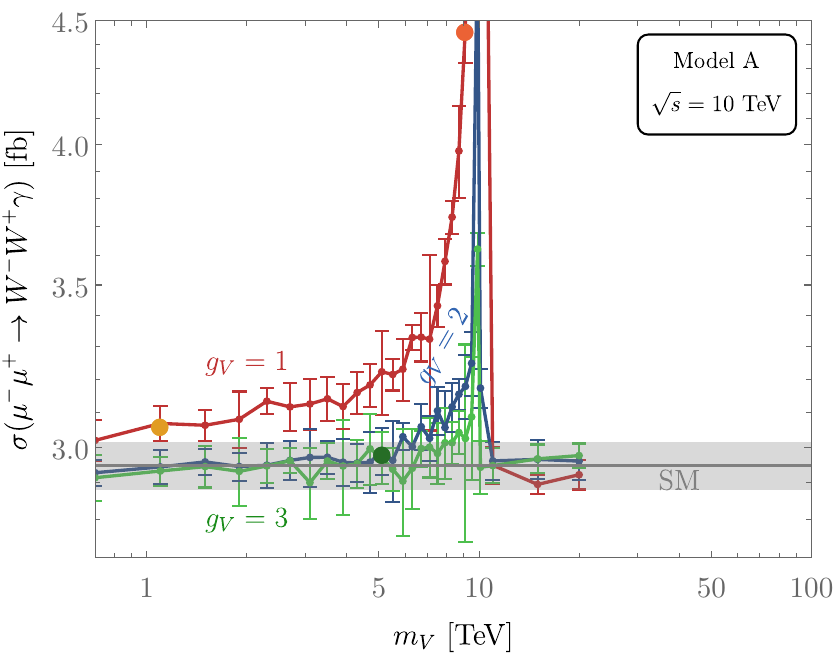}
    \includegraphics[width=0.49\linewidth]{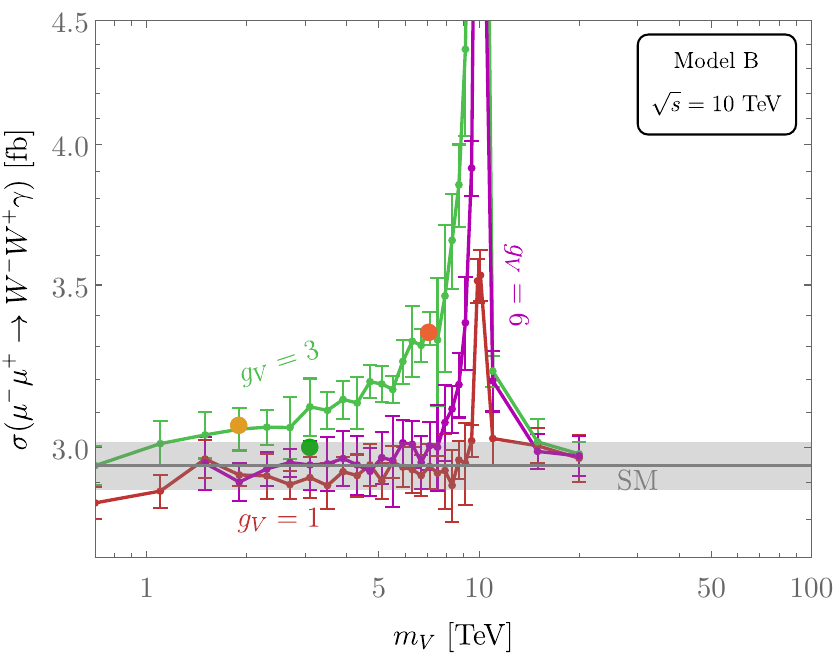}
    \caption{\small The total cross-section $\sigma(\mu^-\mu^+\to W^-W^+ \gamma)$ as a function of the HVT mass, $m_V$, for model A (left) and model B (right) at $\sqrt{s} = 10\,$TeV. The coloured lines show the cross-sections for different values of $g_V$ taken from \texttt{MadGraph} simulations, along with an error which we compute as discussed in \cref{sec:stats-2-to-3}. The three coloured points correspond to the benchmark points presented in \cref{fig:wwa_mres_dist}. The grey line shows the SM cross-section, the grey shaded region shows the range of uncertainty.}
    \label{fig:wwa_cross_sections}
\end{figure}

In \cref{fig:wwa_cross_sections}, we show the total cross-sections $\sigma(\mu^-\mu^+\to W^-W^+ \gamma)$ as a function of the HVT mass, $m_V$, for model A (left) and model B (right) at $\sqrt{s} = 10\,$TeV. The coloured lines show the cross-section for $g_V = 1,2,3$ in model A and for $g_V = 1,3,6$ in model B. We see that the general behaviour of the cross-sections for $\sigma(\mu^-\mu^+\to W^-W^+ \gamma)$ is very similar to $\sigma(\mu^-\mu^+\to W^-W^+)$, shown in \cref{fig:analytic-vs-mg5-ww-x-sections}. Here the cross-section is around 3\,fb, which is smaller than the 12\,fb seen for $W^-W^+$, but not much smaller as may be expected from an extra coupling and phase space suppression.  The cross-section is significantly enhanced for $m_V \sim \sqrt{s}$, and in model A it is larger for smaller $g_V$.  In model B we see that the cross-section increases as $g_V$ increases from $g_V = 1$ to $g_V = 3$, but then decreases as we go to $g_V = 6$.  In general the cross-section decreases as $g_V$ increases, but as discussed in \cref{sec:ww} there is a cancellation at $g_V = 0.92$, explaining the small cross-section for $g_V = 1$. Furthermore, the cross-sections for $g_V = 3$ and $6$ are now significantly different, in contrast to the $\mu^- \mu^+ \to W^- W^+$ cross-section shown in \cref{fig:analytic-vs-mg5-ww-x-sections}. In the $2\to 2$ process, the cross-section can only peak when the HVT is on-shell for $m_V\sim \sqrt{s}$. For different HVT masses, the cross-section is close the SM. In the $2\to 3$ process, the propagating HVT can be on-shell for $m_V \le \sqrt{s}$ leading to an enhanced cross-section for these HVT masses, as we see for $g_V = 3$. For $g_V = 6$ this effect is again reduced due to the larger total width.

The Feynman diagrams suggest that the resonant mass of the $W$ bosons produced in the HVT decay provides a useful discriminator. In a realistic setting where the four-momenta of the $W$ bosons are not well reconstructed, it may be better to use the recoiling energy of the photon, $E_\gamma$, which is in one-to-one correspondence since
\begin{align} \label{eq:mWW}
    m_{WW}^2 = s -2\sqrt{s}E_\gamma
    \,.
\end{align}

\begin{figure}
\centering  
    \includegraphics[width=0.49\linewidth]{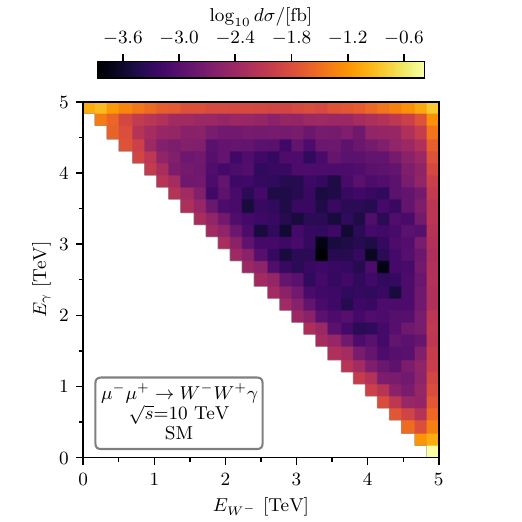}
    \includegraphics[width=0.49\linewidth]{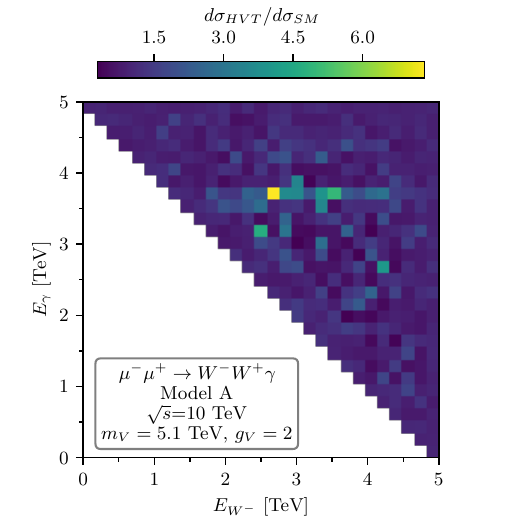}
    \caption{\small The SM cross-section (left) and the ratio of an example HVT cross-section to the SM cross-section (right) as a function of the final state energies of $W^-$ and $\gamma$ for a 10\,TeV centre-of-mass energy.}
    \label{fig:2d_energy_dis_wwa}
\end{figure}

In \cref{fig:2d_energy_dis_wwa} we show the cross-section for the SM background (left) and the ratio of the cross-section of the model A HVT parameter point ($m_V = 5.1$\,TeV, $g_V = 2$) to the SM cross-section (right) as a function of the energies of the final state particles. Note that we only need to consider the $(E_{W^-}, E_\gamma)$~plane as, by charge symmetry, the  distribution in the $(E_{W^+}, E_\gamma)$~plane is identical. In the left panel we see the well-known kinematic effect that the cross-section is largest at the phase space boundaries, especially in the corners \cite{Byers:1964ryc,Kajantie:1968jtt,Burns:2009zi}. In the right panel we see that the cross-section with an HVT can differ the most from the SM for intermediate values of $E_\gamma$ and $E_{W^-}$. In fact, the largest enhancement of the HVT cross-section in the figure is at $E_\gamma \lesssim 3.7$\,TeV and is largely independent of $E_{W^-}$. Using \cref{eq:mWW} this corresponds to $m_{WW} \sim 5.1\,$TeV, where an on-shell HVT decays to $W^-W^+$. This shows that it is important to correctly simulate intermediate values of $E_\gamma$ and $E_{W^-}$. However, as discussed in \cref{sec:mumua}, \texttt{MadGraph} samples most often in regions of large cross-section. More central values of $E_\gamma$ and $E_{W^-}$ would then be relatively poorly sampled and suffer from a large Monte Carlo error. To mitigate this we use the bias function
\begin{align} \label{eq:biasfunction}
    f_\text{bias}
    =
    E_{W^-}\left(\frac{\sqrt{s}}{2}-E_{W^-}\right)
    +
    E_{W^+}\left(\frac{\sqrt{s}}{2}-E_{W^+}\right)
    +
    E_\gamma\left(\frac{\sqrt{s}}{2}-E_\gamma\right)
    \,,
\end{align}
which boosts the sampling in the central region and diminishes it near the boundary. We find that with this bias function the error bars stay small for a wide range of resonant masses and for different $m_V$ and $g_V$.

\begin{figure}
    \centering
    \includegraphics[width=0.49\linewidth]{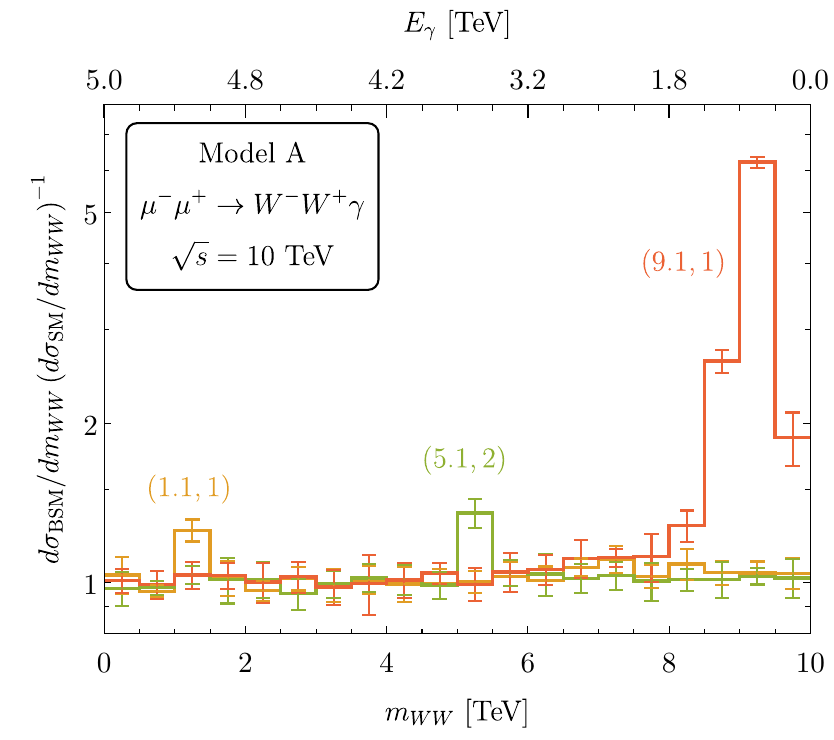}
    \includegraphics[width=0.49\linewidth]{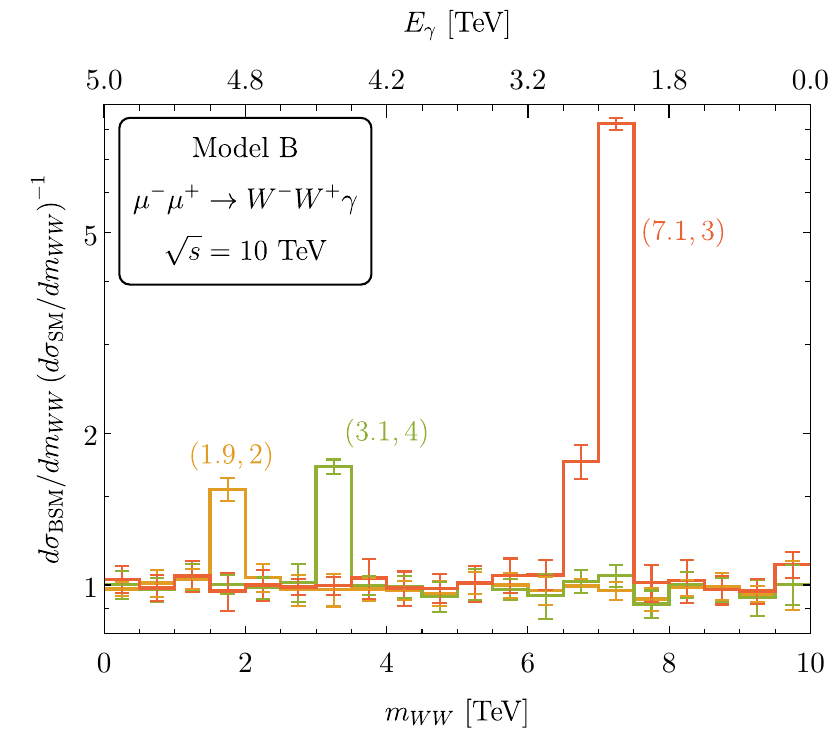}
    \caption{\small The ratio of example HVT cross-sections to the SM cross-section as a function of the resonant mass (bottom axis) and recoil energy (top axis) for model A (left) and model B (right). The parameter points are specified by their values $(m_V, g_V)$. The error bars show the Monte Carlo error after employing our bias function in \cref{eq:biasfunction}.}
    \label{fig:wwa_mres_dist}
\end{figure}

In \cref{fig:wwa_mres_dist} we show the ratio of the differential HVT and SM cross-sections, for three example HVT parameter points.  In the left panel we show three model A parameter points specified by their values $(m_V, g_V)$.  The point with $m_V = 5.1$\,TeV and $g_V = 2$ is the same point shown in the right panel of \cref{fig:2d_energy_dis_wwa}.  We see that there is a large excess at $E_\gamma \approx 3.7$\,TeV and $m_{WW} \approx 5$\,TeV, as expected. We also show two other parameter points, with small and large masses, which feature deviations from the SM expectation near the boundaries.  We see that thanks to our bias function the error bars are of a similar size for all values of $m_{WW}$.  In the right panel, we see similar features, with a central resonance for $m_{WW} \approx m_V$.  In some cases we see that if we select only the resonant peak, we would miss deviations from the SM in other regions of phase space.

\begin{figure}
    \centering
    \includegraphics[width=0.49\linewidth]{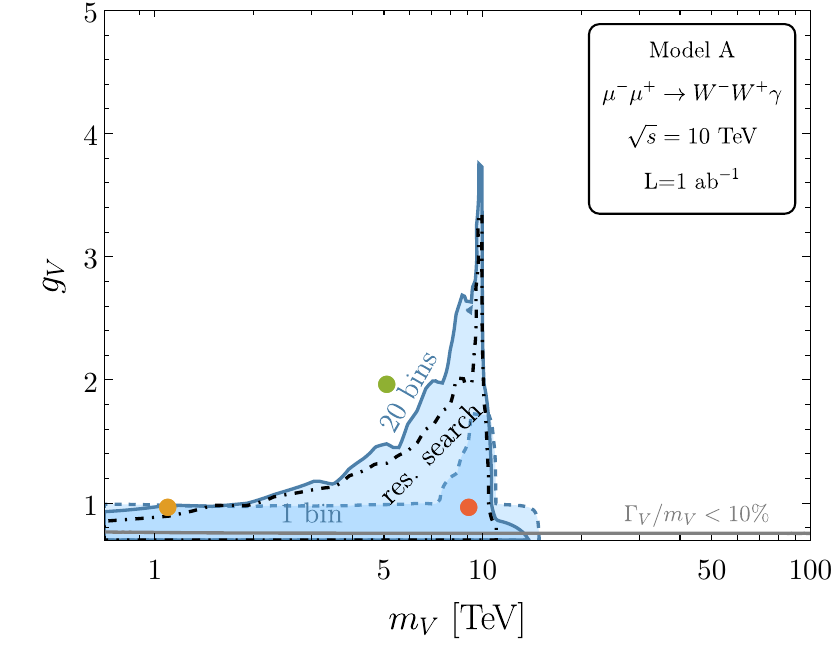}  
    \includegraphics[width=0.49\linewidth]{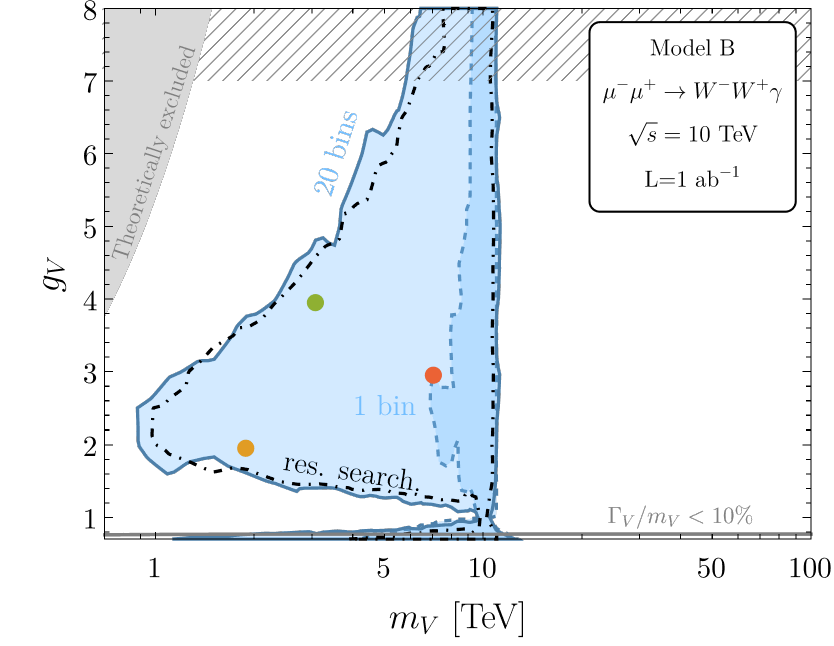} 
    \caption{\small 95\% CL sensitivity projections for three different analyses of $\mu^-\mu^+ \to W^-W^+ \gamma$ with hadronically decaying $W$s for model A (left) and model B (right) in the $(m_V,g_V)$ plane at a $10\,$TeV muon collider for data binned in $m_{WW}$ (solid), unbinned data (dashed) and for a resonance search (black). The limits are shown for an integrated luminosity of 1\,ab$^{-1}$. The three coloured points correspond to the benchmark points presented in \cref{fig:wwa_mres_dist}. For model B (right), the solid grey region is theoretically excluded, and the hatched region in grey indicates the loss of perturbative unitarity.}
    
    \label{fig:exclusion_wwa}
\end{figure}

\Cref{fig:exclusion_wwa} shows the expected 95\% confidence limits of a 10\,TeV muon collider with 1\,ab$^{-1}$ for model A (left) and model B (right).    We take BR$(W \to \text{hadrons}) = 67\%$ and account for a $73\%$ tagging efficiency of the hadronic $W$ bosons~\cite{CMS-DP-2023-065}.  We show three different analysis strategies based on the statistical procedure outlined in \cref{sec:stats-2-to-3} with a systematic error of $\epsilon_{\text{syst}} = 5\%$.  We perform a total cross-section analysis, an analysis using 20 equally spaced bins in $m_{WW}$, and an analysis that targets the resonant production of $W^- W^+$.  In this third analysis we require
\begin{align}
    m_{WW} \in [m_V-2\Delta,m_V+2\Delta]
    \,,
\end{align} 
with $\Delta$ as defined in \cref{eq:Delta}.  We see that in both models the best sensitivity is when $m_V \approx 10$\,TeV.  Model B also features good sensitivity for $m_V \gtrsim 1$\,TeV for intermediate values of $g_V$ but this is lost around $g_V = 0.92$, for the same reasons as in the $W^-W^+$ final state.  There is poor sensitivity in both models when $m_V \gg \sqrt{s}$.  In both models and all regions of parameter space, the total cross-section analysis performs the worst while the analysis utilising the full $m_{WW}$ distribution performs the best.  

\begin{figure}
    \centering
    \includegraphics[width=0.49\linewidth]{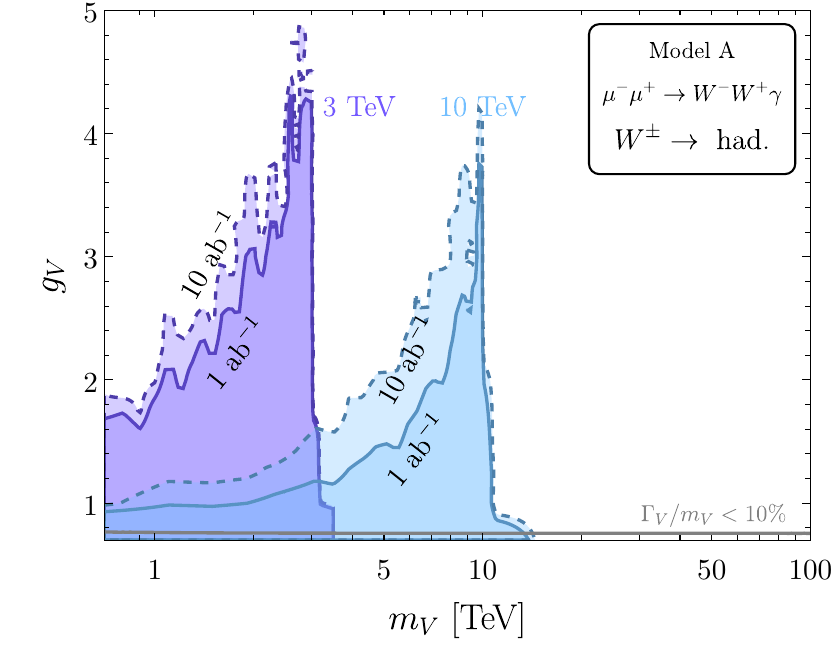}
    \includegraphics[width=0.49\linewidth]{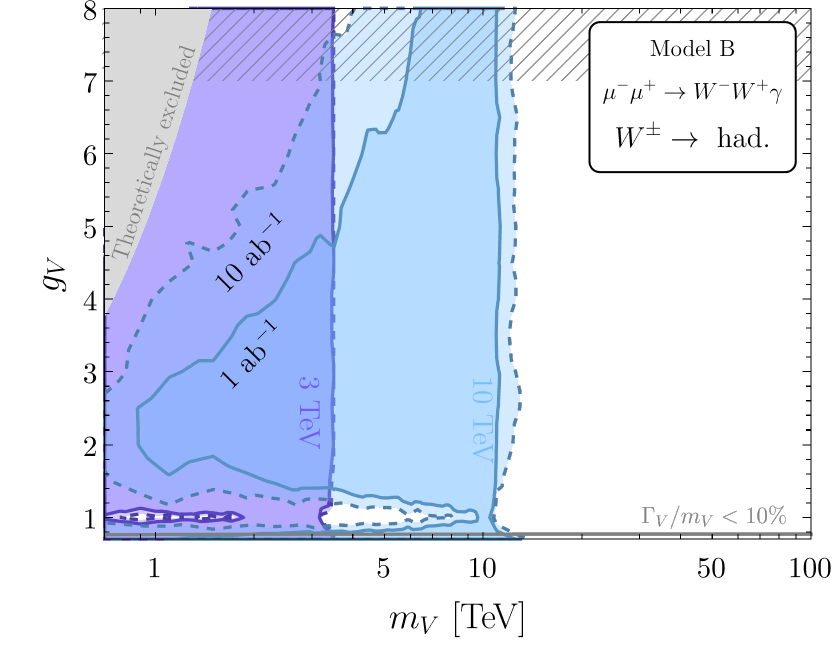}
    \caption{\small 95\% CL sensitivity projections for the $\mu^-\mu^+ \to W^-W^+\gamma$ process in the $(m_V,g_V)$ plane for a 10\,TeV (blue) and a 3\,TeV (purple) muon collider with luminosities of 1 and 10\,ab$^{-1}$. Notation is the same as in previous figures.}
    \label{fig:wwa-10-vs-3}
\end{figure}

In \cref{fig:wwa-10-vs-3} we show sensitivity projections for $3$ and $10\,$TeV collider energies, in purple and blue, and $1$ and $10\,$ab$^{-1}$ with solid and dashed lines, respectively. We see that the increase in collider energies increases the mass reach from $m_V = 3$ to $10\,$TeV. A larger integrated luminosity increases the sensitivity to slightly larger values of $g_V$. As before, this increase is marginal at a $3\,$TeV collider and more pronounced at $10\,$TeV. Model B can be probed significantly better in this channel than model A due to the $g_V$ dependence of the cross-section, which means that it can be sizeable for intermediate values of $g_V$. Since the sensitivity is always largest for $m_V \sim \sqrt{s}$, in this search channel a $3\,$TeV run would be sensitive to large regions of parameter space that a $10\,$TeV run could not access.

\subsubsection{The $W^-W^+Z$ Channel}
\label{sec:wwz}

Finally, we turn to the $W^-W^+Z$ final state.  This channel shares many similarities with the $W^-W^+\gamma$ final state.  At the energies we are considering the $W$ and $Z$ boson masses are almost negligible, so the main difference comes from the fact that we now have a non-zero $V^\pm W^\mp Z$ vertex.

\begin{figure}
    \centering
    \includegraphics[width=0.23\linewidth]{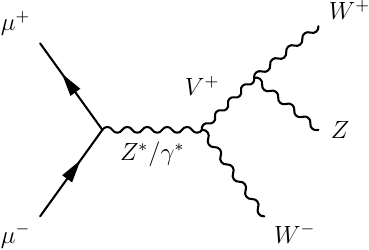}
    \hspace{0.05cm}
    \includegraphics[width=0.23\linewidth]{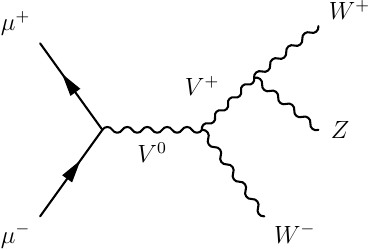}
    \includegraphics[width=0.23\linewidth]{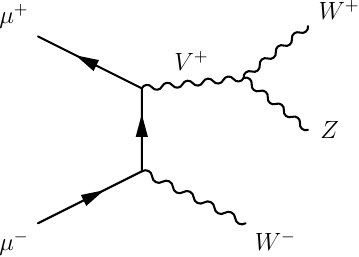}
    \hspace{0.05cm}
    \includegraphics[width=0.23\linewidth]{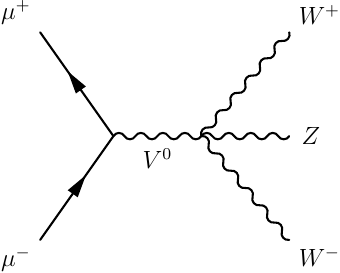}
    \caption{\small
    Representative Feynman diagrams for signal contributions to $\mu^- \mu^+ \to W^-W^+Z$.
    }
    \label{fig:fey_diag_wwz}
\end{figure}

In \cref{fig:fey_diag_wwz} we show some representative Feynman diagrams out of the 17 diagrams involving the HVT, some of which are the same as those shown in \cref{fig:fey_diag_wwa} but with $\gamma \to Z$. There are 21 SM diagrams. Note that the first three diagrams now feature a virtual $V^\pm$ and were not present in $\mu^- \mu^+ \to W^-W^+ \gamma$. This is interesting since searching for both $W^-W^+ \gamma$  and $ W^-W^+ Z$ final states lets us determine whether a charged component of the heavy vector is present and, thus, can differentiate between an $su(2)_L$ triplet or singlet.  As in the $W^-W^+\gamma$ section, diagrams can be obtained by taking the $\mu^-\mu^+\to W^-W^+$ diagram in \cref{fig:mumutoww-feynman-diagrams} and attaching a $Z$ boson to any charged leg.  The same considerations discussed in \cref{sec:wwa} apply to the digrams shown here, so we again simulate all HVT and SM diagrams together to allow for interference and cancellations between terms that grow with energy.

\begin{figure}
    \centering
    \includegraphics[width=0.49\linewidth]{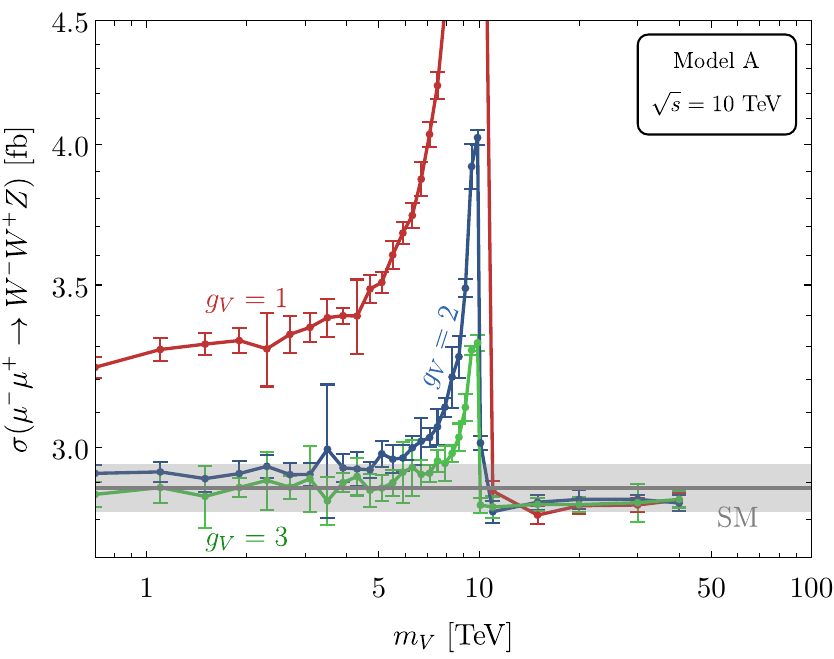}
    \includegraphics[width=0.49\linewidth]{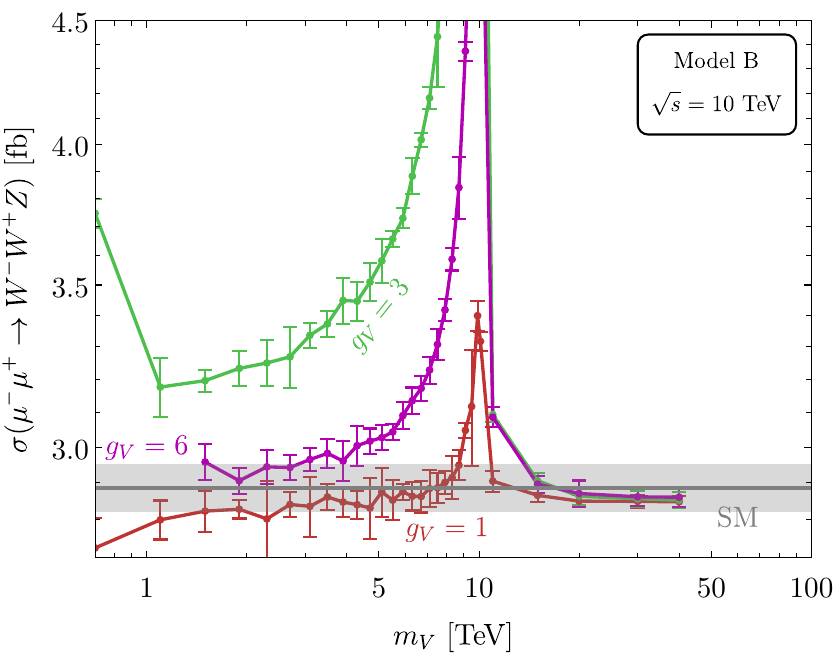}
    \caption{\small The total cross-section $\sigma(\mu^-\mu^+\to W^-W^+ Z)$ as a function of the HVT mass, $m_V$, for model A (left) and model B (right) at $\sqrt{s} = 10\,$TeV. The coloured lines show the cross-sections for different values of $g_V$. The grey shaded region shows the range of uncertainty on the SM cross-section.}
    \label{fig:wwz_cross_sections}
\end{figure}

We show the $\mu^-\mu^+\to W^-W^+Z$ cross-sections as a function of $m_V$ in \cref{fig:wwz_cross_sections}.  We see that compared to the $W^-W^+\gamma$ cross-section shown in \cref{fig:wwa_cross_sections} the SM cross-section is slightly smaller while the HVT cross-section (for the same coupling and mass) is slightly larger, due to the larger number of signal diagrams. The dependence of the cross-sections on $g_V$ is the same as discussed in \cref{sec:wwa}. 

As the Feynman diagrams suggest that the resonant mass of the SM gauge bosons produced in the HVT decay will provide a useful discriminator, we need to identify which of the two bosons came from the HVT decay.  Note that in addition to the virtual $V^\pm \to W^\pm Z$ decays shown in \cref{fig:fey_diag_wwz}, there are also diagrams where a virtual $V^0$ decays to $W^- W^+$ with the $Z$ boson coming from, e.g., an incoming muon. Since we are not performing detector simulation or smearing, we simply pick the two bosons which have an invariant mass closest to $m_V$.  That is, we take the resonant mass to be
\begin{align}
    \mres \equiv m_{ij} \,,
\end{align}
where $ij$ is the pair in $\{Z W^-, Z W^+, W^-W^+\}$ with the smallest
\begin{align}
    \left|m^2_{ij} - m^2_V\right| 
    \,.
\end{align}
In a realistic setting where the four-momenta of the $W$ or $Z$ bosons are not well reconstructed, it may be better to use the recoiling energy of the non-resonant boson, $E_\text{recoil}$.  Note that these are in one-to-one correspondence since
\begin{align}
    m_{ij}^2 = s -2\sqrt{s}E_\text{recoil} + m_\text{recoil}^2
    \,,
\end{align}
where $m_\text{recoil}$ is the mass of the non-resonant boson.  Since the cross-section is again largest at the boundaries of the $(E_Z,E_{W^+})$~plane, we enrich the Monte Carlo sampling of central bins using the bias function
\begin{align}
    f_\text{bias}
    =
    E_{W^-}\left(\frac{\sqrt{s}}{2}-E_{W^-}\right)
    +
    E_{W^+}\left(\frac{\sqrt{s}}{2}-E_{W^+}\right)
    +
    E_Z\left(\frac{\sqrt{s}}{2}-E_Z\right)
    \,.
\end{align}
This produces comparable relative errors for different values of $\mres$, analogous to that seen in \cref{fig:wwa_mres_dist}.

\begin{figure}
    \centering
    \includegraphics[width=0.49\linewidth]{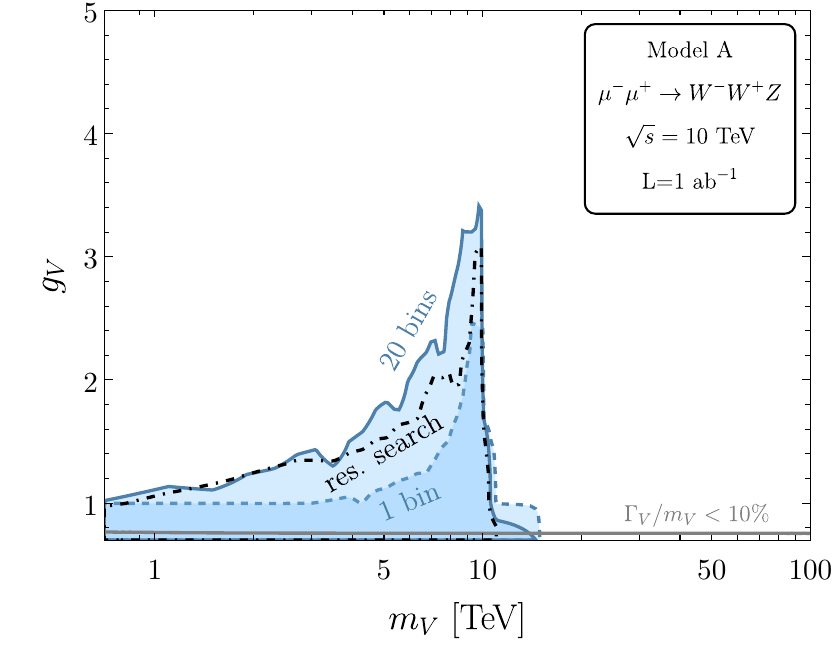} 
    \includegraphics[width=0.49\linewidth]{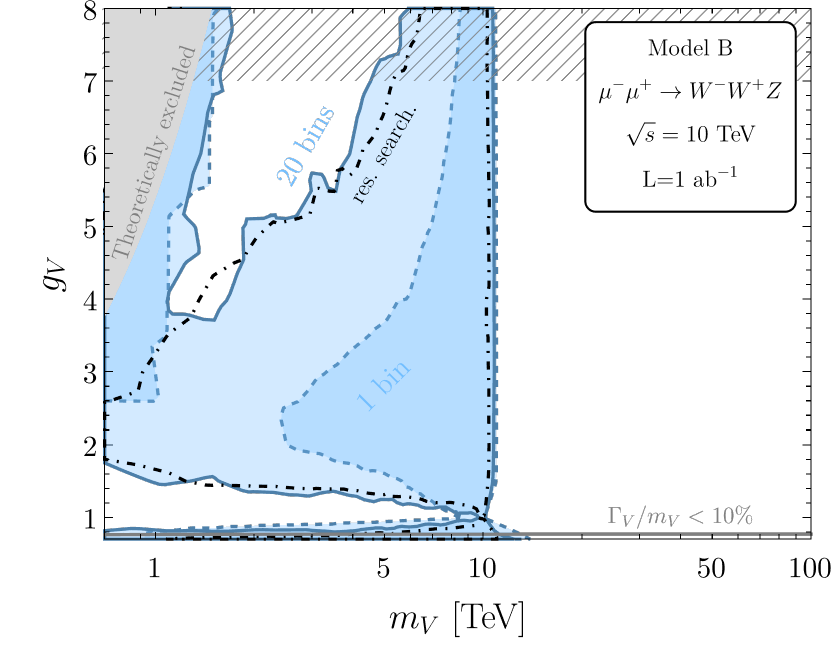} 
    \caption{\small 95\% CL sensitivity projections for three different analyses of $\mu^-\mu^+ \to W^- W^+ Z$ with hadronically decaying gauge bosons for model A (left) and model B (right) at $\sqrt{s}=10$\,TeV with 1\,ab$^{-1}$.  For model B (right), the solid grey region is theoretically excluded and the hatched region in grey indicates the loss of perturbative unitarity.}
    \label{fig:exclusion_wwz}
\end{figure}

We show the 95\% confidence limits for three different analyses in \cref{fig:exclusion_wwz}.  We take BR$(W \to \text{hadrons}) = 67\%$, BR$(Z \to \text{hadrons}) = 70\%$ and account for a $73\%$ tagging efficiency of hadronic $W$ bosons and a $64\%$ tagging efficiency of hadronic $Z$ bosons~\cite{CMS-DP-2023-065}.  As in \cref{sec:wwa} we perform a total cross-section analysis, an analysis using 20 bins equally spaced in $\mres$, and a resonant analysis requiring $\mres \in [m_V-2\Delta,m_V+2\Delta]$ with $\Delta$ defined as in \cref{eq:Delta}.  In each analysis we use the statistical procedure outlined in \cref{sec:stats-2-to-3} with a systematic error of $\epsilon_{\text{syst}} = 5\%$.  We see that the general features of the limits are the same as in $W^-W^+\gamma$ but that the exclusions are stronger, as there are more signal diagrams.  Again the fully differential analysis performs the best while the total cross-section analysis performs the worst.

Our resonant analysis shown in \cref{fig:exclusion_wwz} can be compared to Fig.~6 of Ref.~\cite{Liu:2023jta}.  However, note that while the sensitivity in Ref.~\cite{Liu:2023jta} seems to be significantly stronger than the results of our resonant analysis, there are two significant discrepancies. Firstly, the signal in Ref.~\cite{Liu:2023jta} was simulated as $\sigma(\mu^- \mu^+ \to V^\pm W^\mp) + \sigma(\mu^- \mu^+ \to V^0 Z)$ and multiplied by the branching ratios $\text{BR}(V^\pm \to W^\mp Z)$ and $\text{BR}(V^0 \to W^- W^+)$, which neglects interference with the SM diagrams. As discussed in \cref{sec:ww}, this leads to a significant overestimation of the signal (since the amplitudes grow with energy). This is the main effect that leads to the stringent exclusion reach in Ref.~\cite{Liu:2023jta}. Secondly, the simulation in Ref.~\cite{Liu:2023jta} assumes the narrow width approximation to hold for the HVT. This is not the case for several reasons \cite{Berdine:2007uv}: there is significant interference with non-resonant processes, $m_V\ll \sqrt{s}$ is not satisfied in much of the parameter space we consider, and $\Gamma_V\ll m_V$ starts to fail for $g_V > 8$ or $g_V<0.8$ (as shown in the right panel of \cref{fig:widths}).

    \begin{figure}
    \centering
    \includegraphics[width=0.49\linewidth]{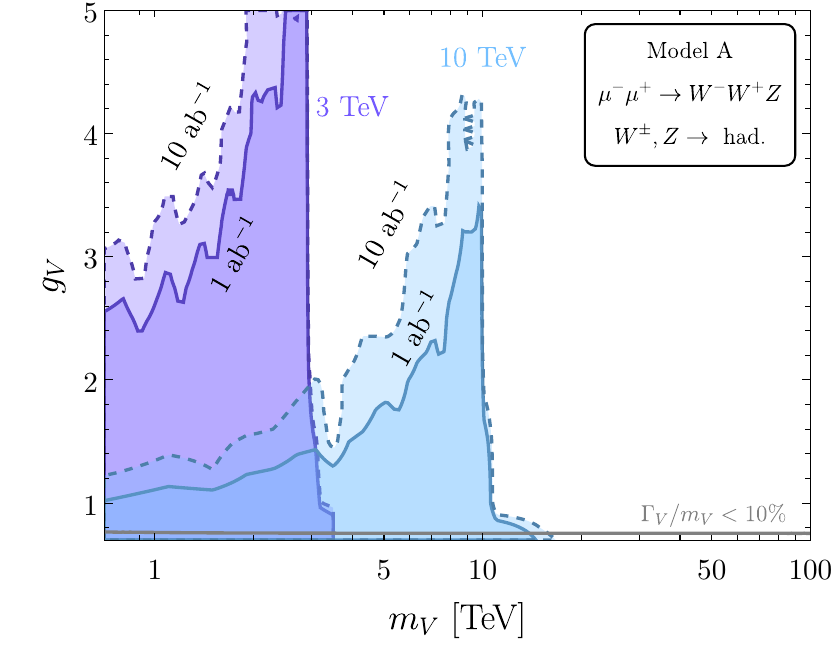} 
    \includegraphics[width=0.49\linewidth]{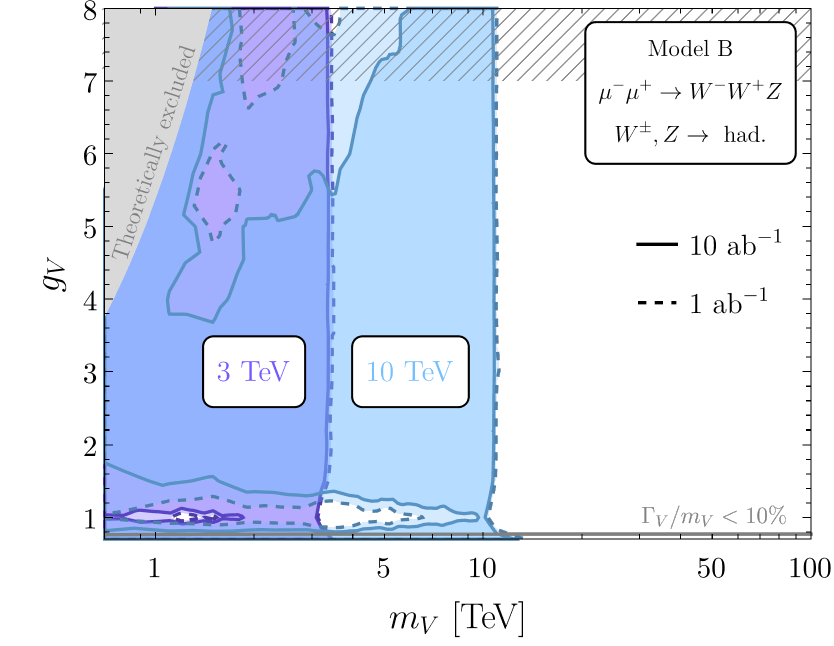} 
    \caption{\small  95\% CL sensitivity projections for the $\mu^-\mu^+ \to W^-W^+Z$ process in the $(m_V,g_V)$ plane for a 10\,TeV (blue) and a 3\,TeV (purple) muon collider with luminosities of 1 and 10\,ab$^{-1}$ in solid and dashed, respectively. Notation is the same as in previous figures.}
    \label{fig:wwz-10-vs-3}
\end{figure}

In \cref{fig:wwz-10-vs-3} we show the sensitivity projections for model A (left) and model B (right) for a $3$ and $10\,$TeV muon collider, in purple and blue, for $1$ and $10\,$ab$^{-1}$, in solid and dashed, respectively. We see that the sensitivity in model A peaks for $m_V \sim \sqrt{s}$. It declines slowly for HVT masses below the collider energy and drops rapidly for HVT masses above the collider energy. In this channel a $3\,$TeV run could probe values of $g_V$ that a $10\,$TeV collider is not sensitive to. Increasing the luminosity to $10\,$ab$^{-1}$ increases the sensitivity in $g_V$.  In model B, coupling values up to $g_V = 8$ can be probed if the HVT mass is in the vicinity of the collider energy. For this reason, a $3\,$TeV run fills in sensitivity gaps of a $10\,$TeV run. An increase in luminosity in a 10\,TeV run increases the sensitivity for large $g_V$ and HVT masses between $2$ and $6\,$TeV. The reduced sensitivity for $g_V \sim 0.92$ can also be ameliorated with a larger integrated luminosity.

\subsection{Summary}
\label{sec:summary}

We now combine the different sensitivity estimates obtained above to compare the different analyses and draw some general conclusions.

\begin{figure}
    \centering
    \includegraphics[width=0.49\linewidth]{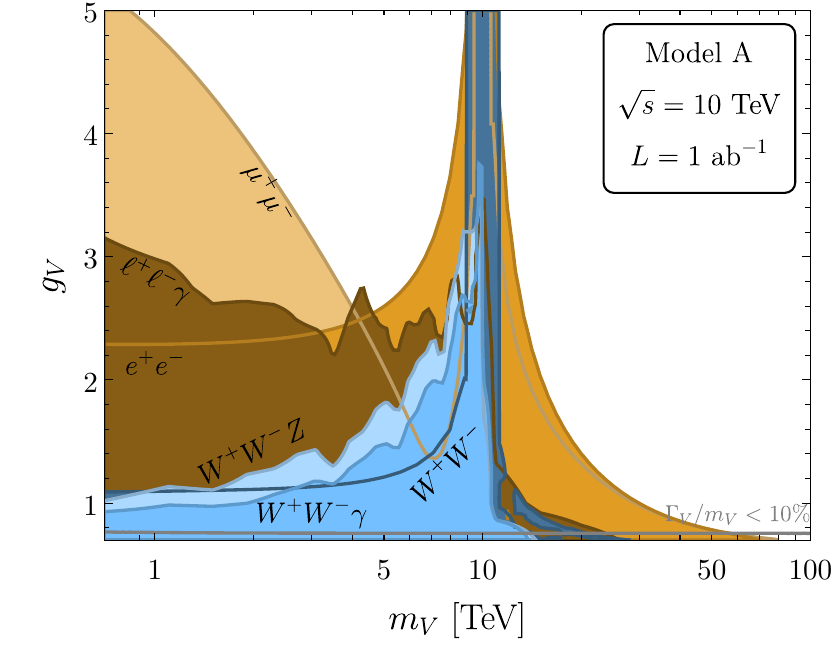}
    \includegraphics[width=0.49\linewidth]{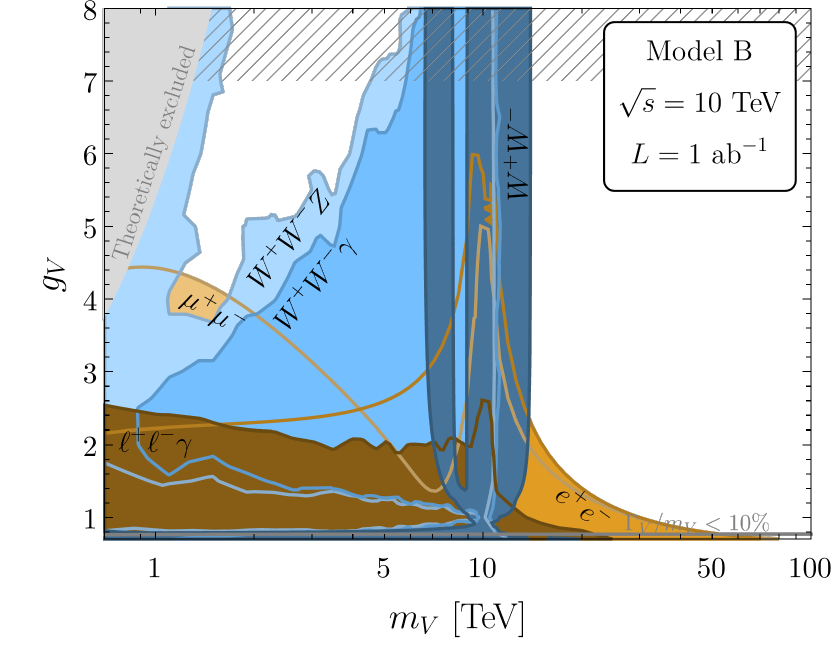}
    \\
    \includegraphics[width=0.49\linewidth]{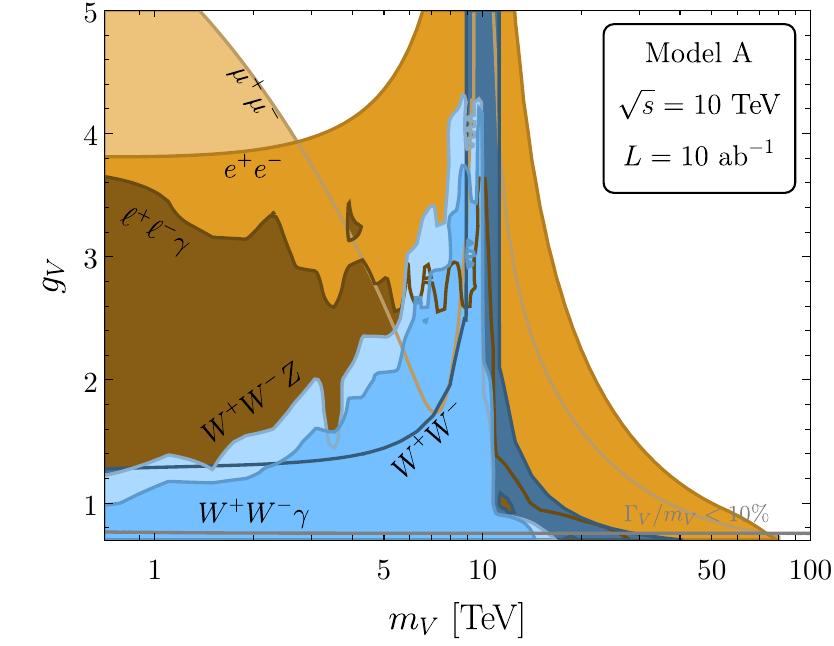}
    \includegraphics[width=0.49\linewidth]{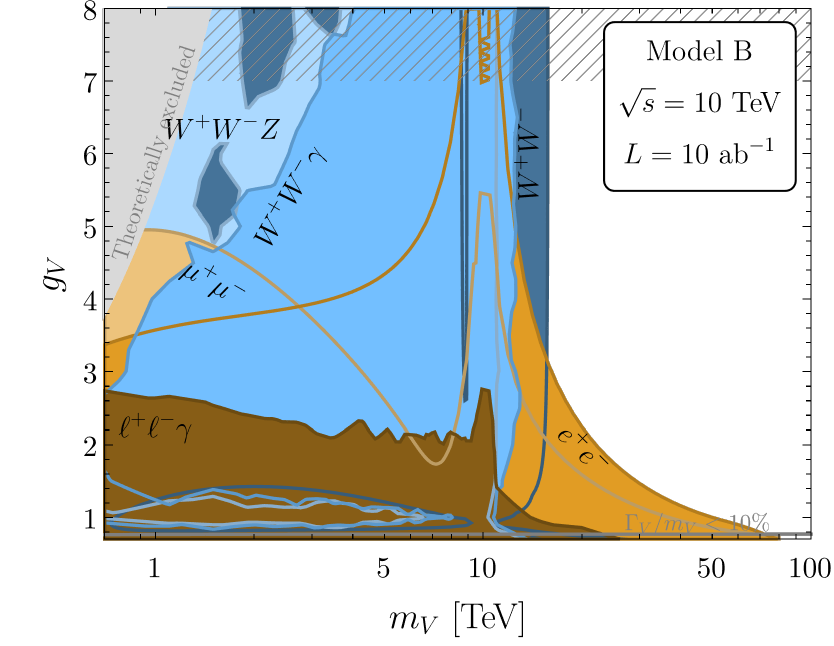}
    \caption{\small A comparison of the different channels discussed above for model A (left) and model B (right) at a 10\,TeV muon collider with 1 (top) and 10\,ab$^{-1}$ (bottom).  With $\ell$ we denote either $\mu$ or $e$.  For model B (right), the solid grey region is theoretically excluded and the hatched region in grey indicates the loss of perturbative unitarity.}
    \label{fig:summary-10-tev}
\end{figure}

In \cref{fig:summary-10-tev} we show the projected constraints on model A (left) and model B (right) at a 10\,TeV muon collider with 1 and 10\,ab$^{-1}$ (top and bottom panels, respectively).  We see that in model A the leptonic $2 \to 2$ processes provide the strongest bounds.  At $m_V \lesssim 4$\,TeV the $\mu^-\mu^+$ channel is strongest, due to the $t$-channel signal process, while at larger masses the $e^-e^+$ channel is stronger, due to negative interference in the $\mu^-\mu^+$ channel.  At $m_V > 10$\,TeV both channels provide similar sensitivity, and have a mass reach up to $m_V \approx 80$\,TeV.  This large mass reach results from a combination of large couplings, a relatively broad width ($\approx 10$\,\%) and our analysis, which goes beyond a bump hunt.  The $\ell^-\ell^+\gamma$ channel is competitive, and outperforms the $e^-e^+$ channel at $m_V \lesssim 5$\,TeV due to the fact that the heavy vector can be produced on-shell even when $m_V \not\approx \sqrt{s}$, but is ultimately weaker than $\mu^-\mu^+$.  Note, however, that the errors in our $2 \to 3$ analyses are larger than in our $2 \to 2$ analyses.  We take the systematic error $\epsilon_i$ in \cref{def:conway_beta} for $\ell^-\ell^+$ to be $1\%$, while $\epsilon_i$ in \cref{eq:2_to_3_epsilon_i} for $\ell^-\ell^+\gamma $ is $\sim 6\%$, when a systematic uncertainty of $5\%$ and the Monte Carlo uncertainty are combined.  If this error can be reduced then the $\ell^-\ell^+\gamma$ channel may fill more of the gap between the $\mu^-\mu^+$ and $e^-e^+$ channels.  In model A, the bosonic channels are not competitive with the leptonic channels.  However, we see that $W^-W^+Z$ is stronger than $W^-W^+$ (which does not have a $t$-channel signal process).  With an integrated luminosity of 10\,ab$^{-1}$ we see that a 10\,TeV muon collider could probe even more of the parameter space of model A, probing $g_V \lesssim 4$ for $m_V \lesssim 12$\,TeV.

In model B, for the leptonic channels we again see that the $\mu^-\mu^+$ channel provides the best performance at $m_V \lesssim 4$\,TeV and that $e^-e^+$ is better at higher masses.  In this case, however, we see that the $c_H$ coupling in this model leads to even stronger constraints from the bosonic channels.  With 1\,ab$^{-1}$ most of the allowed parameter space below $m_V \approx 12$\,TeV can be probed, and it is completely covered with 10\,ab$^{-1}$.  We see that with 1\,ab$^{-1}$ the $2 \to 3$ channels outperform the $2 \to 2$ channel at $m_V \lesssim \sqrt{s}$.  As the $W^-W^+$ signal process only features an $s$-channel diagram, it is suppressed away from resonance, while in the $2 \to 3$ processes the heavy vector can be on resonance at $m_V \not\approx \sqrt{s}$.  However, with 10\,ab$^{-1}$ the $W^-W^+$ channel can cover almost the whole parameter space below $m_V \approx 12$\,TeV.  At $m_V \gtrsim \sqrt{s}$ the heavy vector is always off-shell and the $W^-W^+$ channel is the best (due to the cross-section behaviour and Monte Carlo errors in the $2 \to 3$ channels). We see that the $W^-W^+Z$ search outperforms the $W^-W^+\gamma$ search.  This is because in both models the $V^+W^-\gamma$ vertex vanishes, so there are more signal diagrams that contribute to $W^-W^+Z$ than to $W^-W^+\gamma$.  Note, however, that this is not true in all HVT models, so $W^-W^+\gamma$ may be comparable to $W^-W^+Z$.  Finally, the bosonic channels feature a loss in sensitivity at $g_V \approx 0.9$, which can be covered by any of the leptonic channels.

\begin{figure}
    \centering
    \includegraphics[width=0.49\linewidth]{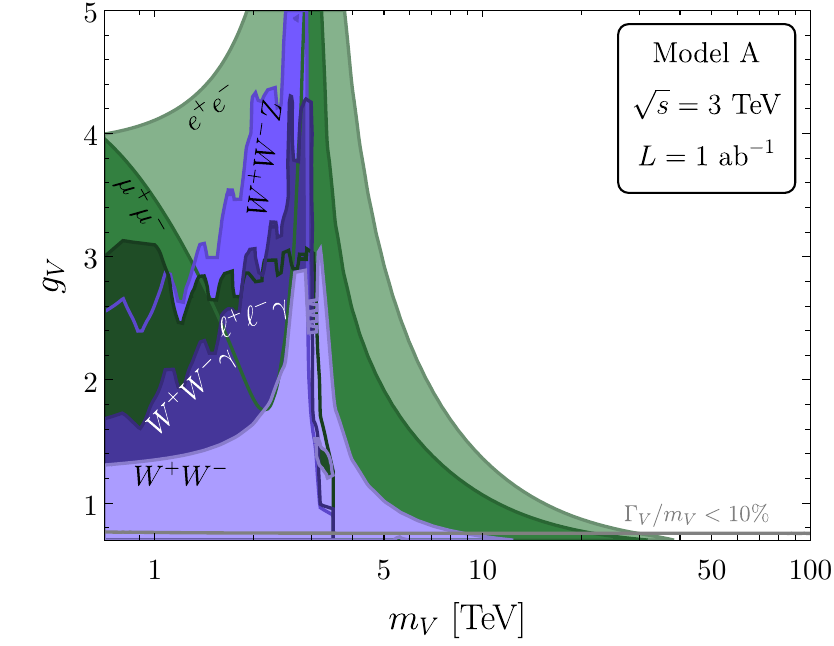}
    \includegraphics[width=0.49\linewidth]{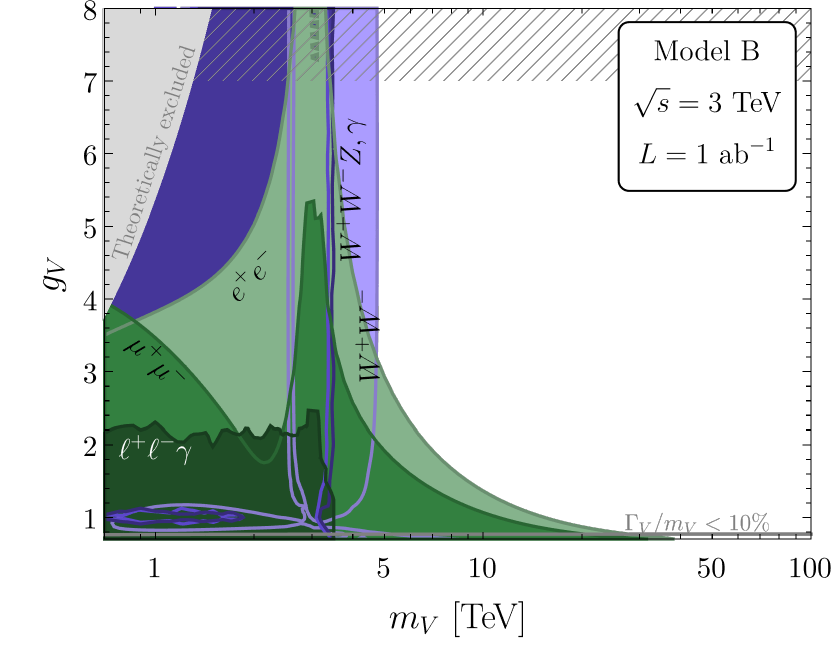}
    \\
    \includegraphics[width=0.49\linewidth]{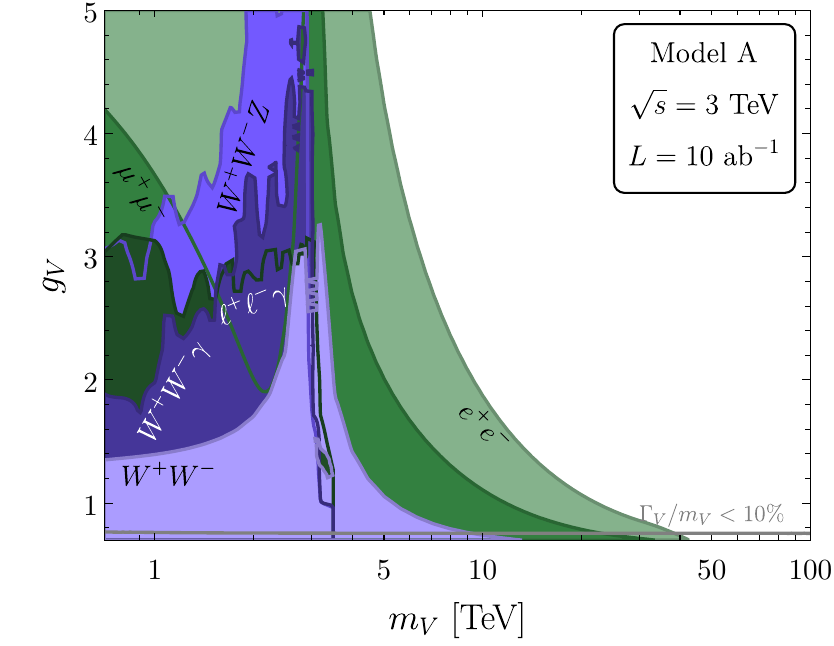}
    \includegraphics[width=0.49\linewidth]{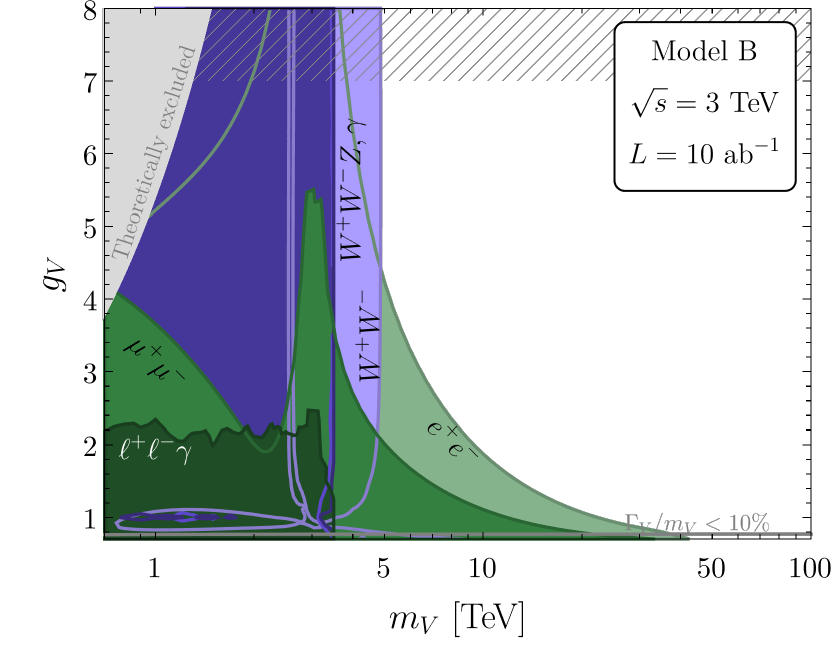}
    \caption{\small A comparison of the different channels discussed above for model A (left) and model B (right) at a 3\,TeV muon collider with 1 (top) and 10\,ab$^{-1}$ (bottom).  With $\ell$ we denote either $\mu$ or $e$.  For model B (right), the solid grey region is theoretically excluded and the hatched region in grey indicates the loss of perturbative unitarity.}
    \label{fig:summary-3-tev}
\end{figure}

In \cref{fig:summary-3-tev} we show the projected constraints on model A and model B at a 3\,TeV muon collider with 1 and 10\,ab$^{-1}$.  In model A we see that $e^-e^+$ is more sensitive than $\mu^-\mu^+$ in the whole of the parameter space.  This is because the searches are now most sensitive at $m_V \approx 3$\,TeV and while the $\mu^-\mu^+$ channel still features a loss of sensitivity due to interference below this mass, the $s$-channel remains sensitive down to below $m_V \approx 1$\,TeV.  The $\ell^-\ell^+\gamma$ search is weaker than the $2 \to 2$ searches, except in the $\mu^-\mu^+$ interference dip, but the $2 \to 3$ diboson channels are stronger than the $W^-W^+$ channel.  In model B the diboson channels are again very strong, and a combination of $W^-W^+$, $e^-e^+$ and $W^-W^+Z$ or $W^-W^+\gamma$ can probe $m_V \lesssim 5$\,TeV.  There is again a loss of sensitivity at $g_V \approx 0.9$ which can be covered by leptonic searches.  In this case, there is only marginal improvement in going to 10\,ab$^{-1}$.

Comparing the $3$ and $10\,$TeV runs highlights the power of a lower energy run. While the mass reach of a $3\,$TeV collider is reduced compared to a $10\,$TeV run, the coupling reach can outperform a higher energy collider for HVT masses that are close to the lower collider energy. For model A we see that a $3\,$TeV run with $1 (10)\,$ab$^{-1}$ can almost (entirely) close the sensitivity gap for $g_V > 4$ and $m_V = 1.5$ to $7.5\,$TeV which remains after a $10\,$TeV run with $10\,$ab$^{-1}$. In model B, a $10\,$TeV run with $10\,$ab$^{-1}$ provides the largest sensitivity. However, a $10\,$TeV run with $1\,$ab$^{-1}$ leaves significant gaps which either an increase in luminosity can close or a $3\,$TeV run with $1\,$ab$^{-1}$. We see that there is an interesting trade-off between energy and luminosity.

Beyond the sensitivity comparison between the channels, we also note that the charged component of the HVT can not be produced in a detectable $2\to 2$ process. A $2 \to 3$ channel such as $W^- W^+ Z$ would be needed to produce the charged component and to establish that it is indeed part of an $su(2)_L$ triplet.

\section{Comparison to the LHC, HL-LHC, HE-LHC and FCC-hh}
\label{sec:comparison}

In this section we compare the sensitivity projections for the muon collider to the projected sensitivity of future proton-proton colliders. We focus on the forthcoming high-luminosity LHC (HL-LHC), the proposed 27\,TeV high-energy LHC (HE-LHC) \cite{FCC:2018bvk,CidVidal:2018eel} and a 100\,TeV Future Circular Collider (FCC-hh) \cite{FCC:2018vvp}. The benchmark centre-of-mass energies and integrated luminosities are listed in \cref{tab:future-colliders}.

\begin{table} 
    \centering
    \begin{tabular}{ c | c c c c}
      	 	& Collisions 	& $ \sqrt{s}$ [TeV]	& $L$ [ab$^{-1}$]  & References \\ \hline		
            Muon Collider & $\mu^-\mu^+$ & 10 & 10 & \cite{InternationalMuonCollider:2025sys} \\
      HL-LHC 	& $pp$ 		& $14$ 			& $3$	    & \cite{ZurbanoFernandez:2020cco} \\
      HE-LHC 	& $pp$ 		& $27$ 			& $15$	    & \cite{FCC:2018bvk} \\
      FCC-hh 	& $pp$ 		& $100$ 		& $20$  & \cite{FCC:2018vvp}\\ 
    \end{tabular} \caption{\small Benchmark centre-of-mass energies and integrated luminosities for various future collider proposals used in this work. }
    \vspace{-.2cm}
    \label{tab:future-colliders}
\end{table}

To estimate the sensitivity of a future hadron collider, we follow the extrapolation procedure described in Ref.~\cite{Thamm:2015zwa}, which is based on current LHC exclusion limits. The method relies on the observation that, for a narrow invariant-mass window centred on the resonance mass, the upper limit on the number of signal events is primarily determined by the expected number of background events in that region. Assuming that the signal acceptance and efficiency remain approximately constant, the projected exclusion reach of a future hadron collider can be estimated by requiring an equivalent background yield in the corresponding signal region. The number of background events is governed by the parton luminosities, so by rescaling the parton luminosities to the desired centre-of-mass energy and accounting for the change in integrated luminosity we can map existing LHC limits to projected exclusion bounds at future hadron colliders.

\begin{figure}
    \centering
    \includegraphics[width=0.49\linewidth]{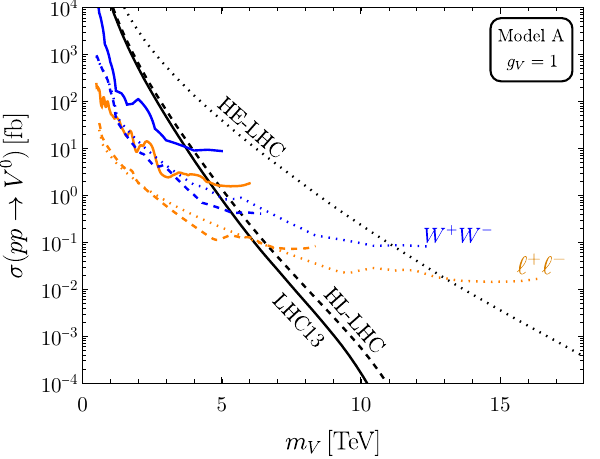}
    \includegraphics[width=0.49\linewidth]{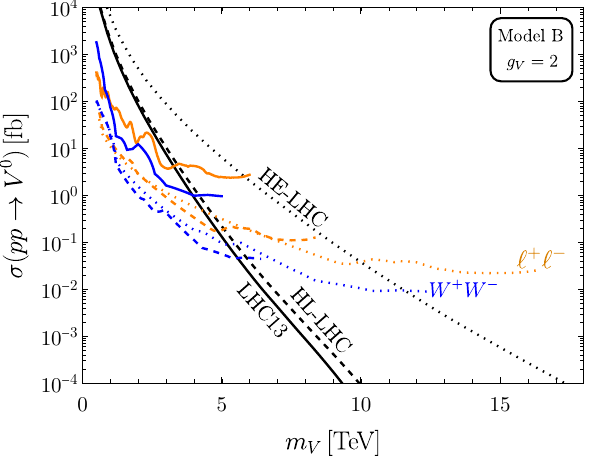} \\
    \includegraphics[width=0.49\linewidth]{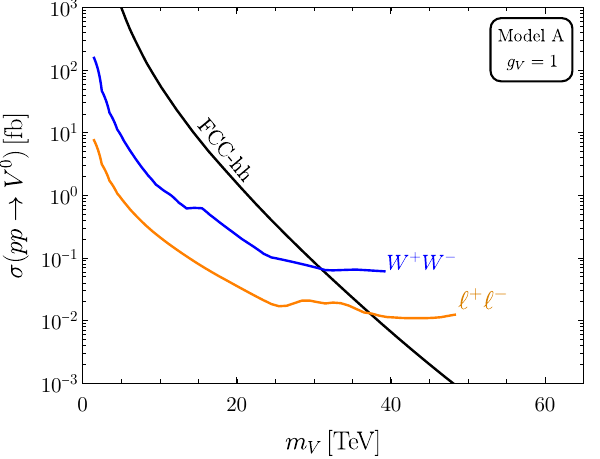}
    \includegraphics[width=0.49\linewidth]{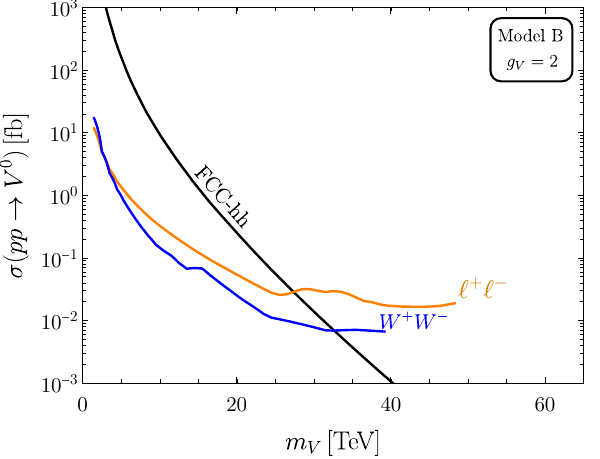}
    \caption{\small 95\% CL sensitivity projections on the production cross-section of the neutral component of the HVT for $g_V = 1$ in model A (left) and $g_V = 2$ in model B (right) at the LHC (top solid blue and orange), and projected limits for the 14\,TeV HL-LHC with $3\,$ab$^{-1}$ (top dashed), 27\,TeV HE-LHC with $15\,$ab$^{-1}$ (top dotted) and FCC-hh with $10\,$ab$^{-1}$ (bottom solid). Orange shows the di-lepton final state search of Ref.~\cite{ATLAS:2019erb} with $L=139\,$fb$^{-1}$ and blue shows the semi-leptonic di-boson final state search of Ref.~\cite{ATLAS:2020fry} with $L=139\,$fb$^{-1}$.  The production cross-sections are shown in black.}
    \label{fig:pp-colider projections}
\end{figure}

We apply the extrapolation procedure to the di-lepton final state \cite{ATLAS:2019erb}, shown in orange in \cref{fig:pp-colider projections}, and to the search for a semi-leptonically decaying di-boson final state \cite{ATLAS:2020fry}, shown in blue. For a discussion of the dominant backgrounds see Ref.~\cite{Baker:2024xwh}. The expected exclusion reach at the HL-LHC, HE-LHC and FCC-hh is shown in \cref{fig:pp-colider projections}, along with the production cross-section of a neutral HVT with $g_V = 1$ in model~A and $g_V = 2$ in model~B.

\begin{figure}
    \centering
    \includegraphics[width=0.49\linewidth]{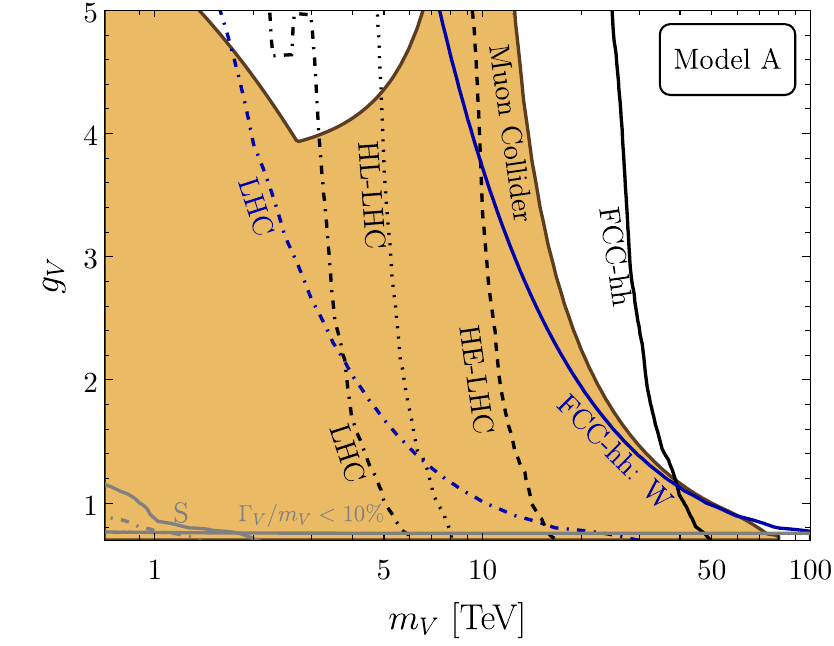}
    \includegraphics[width=0.49\linewidth]{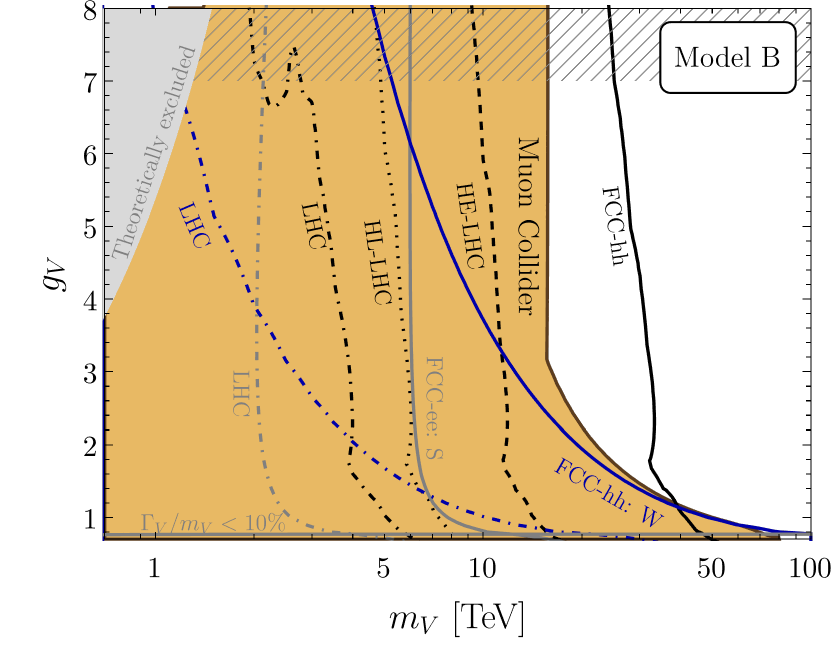}
    \caption{\small 95\% CL sensitivity projections for searches with leptonic or bosonic final states ($e^-e^+$, $\mu^-\mu^+$, $\ell^-\ell^+\gamma$, $W^-W^+$, $W^-W^+\gamma$ and $W^-W^+Z$) at a range of future colliders for Model A (left) and Model B (right). For model B (right), the solid grey region is theoretically excluded and the hatched region in grey indicates the loss of perturbative unitarity. The grey and blue curves show indirect limits from the parameters $S$ and $W$, respectively.}
    \label{fig:comparison}
\end{figure}

In \cref{fig:comparison} we show the expected reach of a 10\,TeV muon collider compared to the LHC, HL-LHC, HE-LHC and FCC-hh in the $(m_V,g_V)$ parameter space for model A (left) and B (right). We see that in model A the muon collider (solid yellow region) outperforms the LHC, HL-LHC and HE-LHC (black dot-dashed, dotted and dashed lines, respectively) except for a small region with $m_V = 2$ to $7\,$TeV and $g_V > 4$, which can be covered by a muon collider run at $3\,$TeV as shown in \cref{fig:summary-3-tev}. Our sensitivity projections for the muon collider are weaker than the expected sensitivity of the FCC-hh except for the very high mass reach of the muon collider for small $g_V$. For model A, the sensitivity at the muon collider and at the hadron machines is dominated by the sensitivity of leptonic final states. 

The conclusions are similar for model B. The muon collider clearly outperforms the HE-LHC throughout the entire parameter space and has a $\sim 5\,$TeV higher mass reach for all values of $g_V$. However, the FCC-hh is expected to have an even higher mass reach, except for $g_V \lesssim 1$ where we find sensitivity for HVT masses up to $80\,$TeV at the muon collider but only $50\,$TeV at the FCC-hh. For $g_V > 3$, the reach at the muon collider and at proton colliders is dominated by di-boson searches.

These results are particularly relevant when considering the feasibility of future collider options. While existing tunnel infrastructure could potentially be reused for the HE-LHC (e.g., the LHC tunnel) and, at least in part, for a muon collider (e.g., the LHC or Tevatron tunnels), the FCC would necessitate the construction of a new tunnel that is unprecedented in circumference, representing a substantial technical, organisational and financial effort. Our analysis highlights the strong physics potential of the FCC, while also indicating that a muon collider offers a compelling, albeit somewhat less powerful, alternative, with projected sensitivities that exceed those of the HE-LHC in the scenarios considered.

In \cref{fig:comparison}, we also include projections from indirect probes of heavy vector triplets. HVTs contribute to the oblique parameters $S$ and $W$ \cite{Pappadopulo:2014qza}, providing complementary sensitivity to direct searches. However, dedicated projections for a muon collider are not yet available. Improvements on $S,T$ and $U$ may be challenging to obtain in the absence of a $Z$-pole run at a muon collider, although it has been suggested that electroweak precision measurements could instead be extracted from vector boson fusion processes \cite{Li:2025ptq}. For comparison, we show current constraints from the $S$-parameter \cite{ParticleDataGroup:2024cfk} and expected values at the FCC-ee \cite{deBlas:2016ojx}. We also show current and expected constraints from the measurement of $W$ at the LHC and FCC-hh \cite{Farina:2016rws}. Note that we do not show the analogous projections for the HL-LHC since they are somewhat outdated and are weaker than current LHC measurements. This may mean that the projections for the electroweak fits at the FCC are also unnecessarily conservative.

We see that while the constraints from $S$, shown in grey, are very weak in model A, current and projected measurements of $W$, shown in blue, provide competitive limits. In model B, current fits of the $S$-parameter provide limits up to HVT masses of $4\,$TeV. The improvement at the FCC-ee is moderate. Current and projected measurements of the $W$-parameter are constraining in model B but not competitive with direct searches.\footnote{While we verified the $W$-parameter constraint on the HVT parameter space obtained in Ref.~\cite{Farina:2016rws}, we note that the benchmark model considered there does not coincide with our models A and B.}

\section{Conclusions}
\label{sec:conclusions}

In this work, we investigate the sensitivity of a muon collider to heavy vector triplets. Heavy vectors are well-motivated new physics candidates, arising in a broad class of models beyond the SM, including both weakly and strongly coupled frameworks. Their phenomenology is particularly well suited for assessing the reach of future collider facilities as they can be produced at both hadron and lepton machines and lead to experimentally accessible signatures in leptonic and bosonic final states.

Interest in a muon collider has steadily grown over the past decade. While significant technological challenges associated with the short muon lifetime remain to be addressed, it is a good time to assess the physics potential of a muon collider and to compare it with other proposed future collider options, such as the HE-LHC and the FCC.

In this work we study the sensitivity of the $2 \to 2$ final states $\mu^- \mu^+$, $e^- e^+$ and $W^- W^+$, and the $2 \to 3$ channels $\ell^- \ell^+ \gamma$, $W^- W^+ \gamma$ and $W^- W^+ Z$ in a weakly coupled benchmark model (model A) and a strongly coupled model (model B). We compute analytic and/or Monte Carlo cross-sections and differential cross-sections in pseudorapidity or the invariant mass. We highlight the importance of interference and demonstrate that neglecting interference can lead to significant under or overestimation of the sensitivity. The Monte Carlo generation of the background and the generation of signal and background including interference can differ due to Monte Carlo error even in regions where the HVT contribution is negligible. This leads to an irreducible contribution to the test statistic. To mitigate this, we estimate the Monte Carlo error using Monte Carlo data, employ bias functions to reduce the error in regions of small cross-section, and include the error in our statistical log-likelihood analysis. Furthermore, we ensure that the irreducible contribution to the test statistic for decoupled HVT masses remains small.

We provide sensitivity projections of the above channels at a $3$ and $10\,$TeV muon collider with $1$ and $10\,$ab$^{-1}$ and find that $\mu^- \mu^+$ and $e^- e^+$ give the strongest sensitivity in our weakly coupled benchmark model. The $\mu^- \mu^+$ analysis leverages the power of the $t$-channel diagram which provides great sensitivity for $m_V < \sqrt{s}$. The $\ell^- \ell^+ \gamma$ channel is competitive but ultimately weaker than the leptonic $2\to2$ channels. An important reason is that the errors are larger for our $2\to3$ analyses than for the $2\to2$ processes. In the weakly coupled model, bosonic channels are not competitive with the leptonic channels.

In the strongly coupled benchmark model, bosonic channels provide the largest sensitivity for large coupling values and dominate over leptonic final states for HVT masses below $12\,$TeV. For larger masses, the $e^- e^+$ channel becomes most constraining. 

We highlight the potential of a $3\,$TeV run of a future muon collider. While the mass reach is lower compared to a  $10\,$TeV run, the coupling reach can outperform a higher energy collider, especially on resonance. Gaps in the coverage of the parameter space of a $10\,$TeV collider with $1\,$ab$^{-1}$ can either be filled by collecting more luminosity or with a $3\,$TeV run at $1\,$ab$^{-1}$.

Finally, we provide a comparison of the projected sensitivity of a $10\,$TeV muon collider with limits from the LHC and expected sensitivities from the HL-LHC, HE-LHC and the FCC-hh. We find that the muon collider outperforms the LHC, HL-LHC and the HE-LHC but provides a somewhat weaker sensitivity than the FCC-hh. This highlights the strong physics potential of the FCC and, at the same time, indicates that a muon collider offers a compelling alternative.

\section*{Acknowledgements}
\label{sec:acknowledgements}

We would like to thank Riccardo Torre for collaboration and Andrea Wulzer for discussion in the early stages of this project. This work was performed in part at the Aspen Center for Physics, which is supported by National Science Foundation grant PHY-2210452. FA, MJB and AT are also grateful to the Mainz Institute for Theoretical Physics (MITP) of the Cluster of Excellence PRISMA$^{+}$ (Project ID 390831469) for its hospitality and its partial support during a critical stage of this work.

\bibliographystyle{JHEP}
\bibliography{refs}

\providecommand{\href}[2]{#2}\begingroup\raggedright\begin{thebibliography}{100}

\bibitem{Delahaye:2019omf}
J.~P. Delahaye, M.~Diemoz, K.~Long, B.~Mansouli{\'e}, N.~Pastrone, L.~Rivkin,
  D.~Schulte, A.~Skrinsky, and A.~Wulzer, {\it {Muon Colliders}},
  \href{http://arxiv.org/abs/1901.06150}{{\tt arXiv:1901.06150}}.

\bibitem{mu_smash}
H.~Al~Ali et~al., {\it {The muon Smasher{\textquoteright}s guide}},  {\em Rept.
  Prog. Phys.} {\bf 85} (2022), no.~8 084201,
  [\href{http://arxiv.org/abs/2103.14043}{{\tt arXiv:2103.14043}}].

\bibitem{toward_muc}
C.~Accettura et~al., {\it {Towards a muon collider}},  {\em Eur. Phys. J. C}
  {\bf 83} (2023), no.~9 864, [\href{http://arxiv.org/abs/2303.08533}{{\tt
  arXiv:2303.08533}}]. [Erratum: Eur.Phys.J.C 84, 36 (2024)].

\bibitem{muon_accelerator_program}
R.~B. Palmer, {\it Muon colliders},  {\em Reviews of Accelerator Science and
  Technology} {\bf 07} (2014) 137--159,
  [\href{http://arxiv.org/abs/https://doi.org/10.1142/S1793626814300072}{{\tt
  https://doi.org/10.1142/S1793626814300072}}].

\bibitem{Black:2022cth}
K.~M. Black et~al., {\it {Muon Collider Forum report}},  {\em JINST} {\bf 19}
  (2024), no.~02 T02015, [\href{http://arxiv.org/abs/2209.01318}{{\tt
  arXiv:2209.01318}}].

\bibitem{MuonCollider:2022xlm}
{\bf Muon Collider} Collaboration, J.~de~Blas et~al., {\it {The physics case of
  a 3 TeV muon collider stage}},  \href{http://arxiv.org/abs/2203.07261}{{\tt
  arXiv:2203.07261}}.

\bibitem{Narain:2022qud}
M.~Narain et~al., {\it {The Future of US Particle Physics - The Snowmass 2021
  Energy Frontier Report}},  \href{http://arxiv.org/abs/2211.11084}{{\tt
  arXiv:2211.11084}}.

\bibitem{P5:2023wyd}
{\bf P5} Collaboration, S.~Asai et~al., {\it {Exploring the Quantum Universe:
  Pathways to Innovation and Discovery in Particle Physics}},
  \href{http://arxiv.org/abs/2407.19176}{{\tt arXiv:2407.19176}}.

\bibitem{InternationalMuonCollider:2024jyv}
{\bf International Muon Collider} Collaboration, C.~Accettura et~al., {\it
  {Interim report for the International Muon Collider Collaboration (IMCC)}},
  {\em CERN Yellow Rep. Monogr.} {\bf 2/2024} (2024) 176,
  [\href{http://arxiv.org/abs/2407.12450}{{\tt arXiv:2407.12450}}].

\bibitem{InternationalMuonCollider:2025sys}
{\bf International Muon Collider} Collaboration, C.~Accettura et~al., {\it {The
  Muon Collider}},  \href{http://arxiv.org/abs/2504.21417}{{\tt
  arXiv:2504.21417}}.

\bibitem{Begel:2025ldu}
M.~Begel et~al., {\it {United States Muon Collider Community White Paper for
  the European Strategy for Particle Physics Update}},
  \href{http://arxiv.org/abs/2503.23695}{{\tt arXiv:2503.23695}}.

\bibitem{Ruhdorfer:2023uea}
M.~Ruhdorfer, E.~Salvioni, and A.~Wulzer, {\it {Invisible Higgs boson decay
  from forward muons at a muon collider}},  {\em Phys. Rev. D} {\bf 107}
  (2023), no.~9 095038, [\href{http://arxiv.org/abs/2303.14202}{{\tt
  arXiv:2303.14202}}].

\bibitem{Andreetto:2024rra}
P.~Andreetto et~al., {\it {Aspects of Higgs Physics at a $\sqrt{s}=3$ TeV Muon
  Collider with detailed detector simulation}},  {\em Eur. Phys. J. C} {\bf 85}
  (2025), no.~3 221, [\href{http://arxiv.org/abs/2405.19314}{{\tt
  arXiv:2405.19314}}].

\bibitem{Marzocca:2025inb}
D.~Marzocca, F.~Montagno, M.~Morales-Alvarado, and A.~Wulzer, {\it {Quark
  mixing from muon collider neutrinos}},
  \href{http://arxiv.org/abs/2511.23288}{{\tt arXiv:2511.23288}}.

\bibitem{Costantini:2020stv}
A.~Costantini, F.~De~Lillo, F.~Maltoni, L.~Mantani, O.~Mattelaer, R.~Ruiz, and
  X.~Zhao, {\it {Vector boson fusion at multi-TeV muon colliders}},  {\em JHEP}
  {\bf 09} (2020) 080, [\href{http://arxiv.org/abs/2005.10289}{{\tt
  arXiv:2005.10289}}].

\bibitem{Chiesa:2020awd}
M.~Chiesa, F.~Maltoni, L.~Mantani, B.~Mele, F.~Piccinini, and X.~Zhao, {\it
  {Measuring the quartic Higgs self-coupling at a multi-TeV muon collider}},
  {\em JHEP} {\bf 09} (2020) 098, [\href{http://arxiv.org/abs/2003.13628}{{\tt
  arXiv:2003.13628}}].

\bibitem{Han:2020pif}
T.~Han, D.~Liu, I.~Low, and X.~Wang, {\it {Electroweak couplings of the Higgs
  boson at a multi-TeV muon collider}},  {\em Phys. Rev. D} {\bf 103} (2021),
  no.~1 013002, [\href{http://arxiv.org/abs/2008.12204}{{\tt
  arXiv:2008.12204}}].

\bibitem{deBlas:2022ofj}
J.~de~Blas, Y.~Du, C.~Grojean, J.~Gu, V.~Miralles, M.~E. Peskin, J.~Tian,
  M.~Vos, and E.~Vryonidou, {\it {Global SMEFT Fits at Future Colliders}},  in
  {\em {Snowmass 2021}}, 6, 2022.
\newblock \href{http://arxiv.org/abs/2206.08326}{{\tt arXiv:2206.08326}}.

\bibitem{Han:2021lnp}
T.~Han, W.~Kilian, N.~Kreher, Y.~Ma, J.~Reuter, T.~Striegl, and K.~Xie, {\it
  {Precision test of the muon-Higgs coupling at a high-energy muon collider}},
  {\em JHEP} {\bf 12} (2021) 162, [\href{http://arxiv.org/abs/2108.05362}{{\tt
  arXiv:2108.05362}}].

\bibitem{Forslund:2022xjq}
M.~Forslund and P.~Meade, {\it {High precision higgs from high energy muon
  colliders}},  {\em JHEP} {\bf 08} (2022) 185,
  [\href{http://arxiv.org/abs/2203.09425}{{\tt arXiv:2203.09425}}].

\bibitem{Dawson:2022zbb}
S.~Dawson et~al., {\it {Report of the Topical Group on Higgs Physics for
  Snowmass 2021: The Case for Precision Higgs Physics}},  in {\em {Snowmass
  2021}}, 9, 2022.
\newblock \href{http://arxiv.org/abs/2209.07510}{{\tt arXiv:2209.07510}}.

\bibitem{Liu:2023yrb}
Z.~Liu, K.-F. Lyu, I.~Mahbub, and L.-T. Wang, {\it {Top Yukawa coupling
  determination at high energy muon collider}},  {\em Phys. Rev. D} {\bf 109}
  (2024), no.~3 035021, [\href{http://arxiv.org/abs/2308.06323}{{\tt
  arXiv:2308.06323}}].

\bibitem{Glioti:2025zpn}
A.~Glioti, D.~Marzocca, and A.~Wulzer, {\it {Flavor physics at high-energy muon
  colliders}},  {\em JHEP} {\bf 12} (2025) 152,
  [\href{http://arxiv.org/abs/2509.08132}{{\tt arXiv:2509.08132}}].

\bibitem{Airen:2026szh}
S.~Airen and R.~Franceschini, {\it {Top quark FCNC in Randall-Sundrum models:
  post-LHC allowed rates and searches at $e^+e^-$ and $\mu^+ \mu^-$
  colliders}},  \href{http://arxiv.org/abs/2601.14966}{{\tt arXiv:2601.14966}}.

\bibitem{Han:2020uak}
T.~Han, Z.~Liu, L.-T. Wang, and X.~Wang, {\it {WIMPs at High Energy Muon
  Colliders}},  {\em Phys. Rev. D} {\bf 103} (2021), no.~7 075004,
  [\href{http://arxiv.org/abs/2009.11287}{{\tt arXiv:2009.11287}}].

\bibitem{Ruhdorfer:2019utl}
M.~Ruhdorfer, E.~Salvioni, and A.~Weiler, {\it {A Global View of the Off-Shell
  Higgs Portal}},  {\em SciPost Phys.} {\bf 8} (2020) 027,
  [\href{http://arxiv.org/abs/1910.04170}{{\tt arXiv:1910.04170}}].

\bibitem{Li:2023tbx}
P.~Li, Z.~Liu, and K.-F. Lyu, {\it {Heavy neutral leptons at muon colliders}},
  {\em JHEP} {\bf 03} (2023) 231, [\href{http://arxiv.org/abs/2301.07117}{{\tt
  arXiv:2301.07117}}].

\bibitem{Capdevilla:2021rwo}
R.~Capdevilla, D.~Curtin, Y.~Kahn, and G.~Krnjaic, {\it {No-lose theorem for
  discovering the new physics of~$(g-2)_\mu$ at muon colliders}},  {\em Phys.
  Rev. D} {\bf 105} (2022), no.~1 015028,
  [\href{http://arxiv.org/abs/2101.10334}{{\tt arXiv:2101.10334}}].

\bibitem{Capdevilla:2021kcf}
R.~Capdevilla, D.~Curtin, Y.~Kahn, and G.~Krnjaic, {\it {Systematically testing
  singlet models for $(g-2)_{\mu}$}},  {\em JHEP} {\bf 04} (2022) 129,
  [\href{http://arxiv.org/abs/2112.08377}{{\tt arXiv:2112.08377}}].

\bibitem{Buttazzo:2020ibd}
D.~Buttazzo and P.~Paradisi, {\it {Probing the muon $g-2$ anomaly with the
  Higgs boson at a muon collider}},  {\em Phys. Rev. D} {\bf 104} (2021), no.~7
  075021, [\href{http://arxiv.org/abs/2012.02769}{{\tt arXiv:2012.02769}}].

\bibitem{Huang:2021biu}
G.-y. Huang, S.~Jana, F.~S. Queiroz, and W.~Rodejohann, {\it {Probing the RK(*)
  anomaly at a muon collider}},  {\em Phys. Rev. D} {\bf 105} (2022), no.~1
  015013, [\href{http://arxiv.org/abs/2103.01617}{{\tt arXiv:2103.01617}}].

\bibitem{Liu:2021akf}
W.~Liu, K.-P. Xie, and Z.~Yi, {\it {Testing leptogenesis at the LHC and future
  muon colliders: A Z' scenario}},  {\em Phys. Rev. D} {\bf 105} (2022), no.~9
  095034, [\href{http://arxiv.org/abs/2109.15087}{{\tt arXiv:2109.15087}}].

\bibitem{Cesarotti:2022ttv}
C.~Cesarotti, S.~Homiller, R.~K. Mishra, and M.~Reece, {\it {Probing New Gauge
  Forces with a High-Energy Muon Beam Dump}},  {\em Phys. Rev. Lett.} {\bf 130}
  (2023), no.~7 071803, [\href{http://arxiv.org/abs/2202.12302}{{\tt
  arXiv:2202.12302}}].

\bibitem{Capdevilla:2021fmj}
R.~Capdevilla, F.~Meloni, R.~Simoniello, and J.~Zurita, {\it {Hunting wino and
  higgsino dark matter at the muon collider with disappearing tracks}},  {\em
  JHEP} {\bf 06} (2021) 133, [\href{http://arxiv.org/abs/2102.11292}{{\tt
  arXiv:2102.11292}}].

\bibitem{Asadi:2021gah}
P.~Asadi, R.~Capdevilla, C.~Cesarotti, and S.~Homiller, {\it {Searching for
  leptoquarks at future muon colliders}},  {\em JHEP} {\bf 10} (2021) 182,
  [\href{http://arxiv.org/abs/2104.05720}{{\tt arXiv:2104.05720}}].

\bibitem{Gu:2020ldn}
J.~Gu, L.-T. Wang, and C.~Zhang, {\it {Unambiguously Testing Positivity at
  Lepton Colliders}},  {\em Phys. Rev. Lett.} {\bf 129} (2022), no.~1 011805,
  [\href{http://arxiv.org/abs/2011.03055}{{\tt arXiv:2011.03055}}].

\bibitem{Barger:1980ix}
V.~D. Barger, W.-Y. Keung, and E.~Ma, {\it {A Gauge Model With Light $W$ and
  $Z$ Bosons}},  {\em Phys. Rev. D} {\bf 22} (1980) 727.

\bibitem{Hewett:1988xc}
J.~L. Hewett and T.~G. Rizzo, {\it {Low-Energy Phenomenology of Superstring
  Inspired E(6) Models}},  {\em Phys. Rept.} {\bf 183} (1989) 193.

\bibitem{Cvetic:1995zs}
M.~Cvetic and S.~Godfrey, {\em {Discovery and identification of extra gauge
  bosons}}, pp.~383--415.
\newblock 3, 1995.
\newblock \href{http://arxiv.org/abs/hep-ph/9504216}{{\tt hep-ph/9504216}}.

\bibitem{Rizzo:2006wq}
T.~G. Rizzo, {\it {$Z'$ Phenomenology and the LHC}},
  \href{http://arxiv.org/abs/hep-ph/0610104}{{\tt hep-ph/0610104}}.
  [\href{http://inspirehep.net/record/728548}{Inspire}].

\bibitem{Agashe:2007hh}
K.~Agashe, H.~Davoudiasl, S.~Gopalakrishna, T.~Han, G.-Y. Huang, G.~Perez,
  Z.-G. Si, and A.~Soni, {\it {LHC Signals for Warped Electroweak Neutral Gauge
  Bosons}},  {\em Phys. Rev.} {\bf D 76} (2007) 115015,
  [\href{http://arxiv.org/abs/0709.0007}{{\tt arXiv:0709.0007}}].
  [\href{http://inspirehep.net/record/759584}{Inspire}].

\bibitem{Langacker:2008yv}
P.~Langacker, {\it {The Physics of Heavy $Z^\prime$ Gauge Bosons}},  {\em Rev.
  Mod. Phys.} {\bf 81} (2009) 1199--1228,
  [\href{http://arxiv.org/abs/0801.1345}{{\tt arXiv:0801.1345}}].

\bibitem{Salvioni:2010p2769}
E.~Salvioni, A.~Strumia, G.~Villadoro, and F.~Zwirner, {\it Non-universal
  minimal {Z}' models: present bounds and early {LHC} reach},  {\em JHEP} {\bf
  03} (2010) 010, [\href{http://arxiv.org/abs/0911.1450}{{\tt
  arXiv:0911.1450}}]. [\href{http://inspirebeta.net/record/836375}{Inspire}].

\bibitem{Accomando:2013ve}
E.~Accomando, D.~Becciolini, A.~S. Belyaev, S.~Moretti, and C.~H.
  Shepherd-Themistocleous, {\it {$Z'$ at the LHC: Interference and Finite Width
  Effects in Drell-Yan}},  {\em JHEP} {\bf 10} (2013) 153,
  [\href{http://arxiv.org/abs/1304.6700}{{\tt arXiv:1304.6700}}].
  [\href{http://inspirehep.net/record/1229792}{Inspire}].

\bibitem{Agashe:2009bj}
K.~Agashe, S.~Gopalakrishna, T.~Han, G.-Y. Huang, and A.~Soni, {\it {LHC
  signals for warped electroweak charged gauge bosons}},  {\em Phys. Rev.} {\bf
  D 80} (2009) 075007, [\href{http://arxiv.org/abs/0810.1497}{{\tt
  arXiv:0810.1497}}]. [\href{http://inspirehep.net/record/798910}{Inspire}].

\bibitem{Schmaltz:2010p2610}
M.~Schmaltz and C.~Spethmann, {\it Two {S}imple ${W}^\prime$ {M}odels for the
  {E}arly {LHC}},  {\em JHEP} {\bf 07} (2011) 046,
  [\href{http://arxiv.org/abs/1011.5918}{{\tt arXiv:1011.5918}}].
  [\href{http://inspirehep.net/record/878831}{Inspire}].

\bibitem{Grojean:2011vu}
C.~Grojean, E.~Salvioni, and R.~Torre, {\it {A weakly constrained W' at the
  early LHC}},  {\em JHEP} {\bf 07} (2011) 002,
  [\href{http://arxiv.org/abs/1103.2761}{{\tt arXiv:1103.2761}}].

\bibitem{Langacker:1989xa}
P.~Langacker and S.~U. Sankar, {\it {Bounds on the Mass of W(R) and the
  W(L)-W(R) Mixing Angle xi in General SU(2)-L x SU(2)-R x U(1) Models}},  {\em
  Phys. Rev. D} {\bf 40} (1989) 1569--1585.

\bibitem{Frank:2010p2250}
M.~Frank, A.~Hayreter, and I.~Turan, {\it Production and {D}ecays of ${W}_{R}$
  bosons at the {LHC}},  {\em Phys. Rev.} {\bf D 83} (2011) 035001,
  [\href{http://arxiv.org/abs/1010.5809}{{\tt arXiv:1010.5809}}].
  [\href{http://inspirebeta.net/record/874800}{Inspire}].

\bibitem{Accomando:2011up}
E.~Accomando, D.~Becciolini, S.~de~Curtis, D.~Dominici, L.~Fedeli, and C.~H.
  Shepherd-Themistocleous, {\it {Interference effects in heavy $W'$-boson
  searches at the LHC}},  {\em Phys. Rev.} {\bf D 85} (2011) 115017,
  [\href{http://arxiv.org/abs/1110.0713}{{\tt arXiv:1110.0713}}].
  [\href{http://inspirehep.net/record/930438}{Inspire}].

\bibitem{Accomando:2011gt}
E.~Accomando, D.~Becciolini, S.~D. Curtis, D.~Dominici, and L.~Fedeli, {\it
  {$W'$ production at the LHC in the 4-site Higgsless model}},  {\em Phys.
  Rev.} {\bf D 84} (2011) 115014, [\href{http://arxiv.org/abs/1107.4087}{{\tt
  arXiv:1107.4087}}]. [\href{http://inspirehep.net/record/919245}{Inspire}].

\bibitem{Dobrescu:2021vak}
B.~A. Dobrescu and F.~Yu, {\it {Dijet and electroweak limits on a Z' boson
  coupled to quarks}},  {\em Phys. Rev. D} {\bf 109} (2024), no.~3 035004,
  [\href{http://arxiv.org/abs/2112.05392}{{\tt arXiv:2112.05392}}].

\bibitem{Chanowitz:1993fc}
M.~S. Chanowitz and W.~Kilgore, {\it {Complementarity of Resonant and
  Nonresonant Strong $WW$ Scattering at the LHC}},  {\em Phys. Lett.} {\bf B
  322} (1993) 147--153, [\href{http://arxiv.org/abs/hep-ph/9311336}{{\tt
  hep-ph/9311336}}]. [\href{http://inspirehep.net/record/360435}{Inspire}].

\bibitem{Barbieri:2008cc}
R.~Barbieri, G.~Isidori, V.~S. Rychkov, and E.~Trincherini, {\it {Heavy Vectors
  in Higgs-less models}},  {\em Phys. Rev. D} {\bf 78} (2008) 036012,
  [\href{http://arxiv.org/abs/0806.1624}{{\tt arXiv:0806.1624}}].

\bibitem{Barbieri:2009p33}
R.~Barbieri, A.~E. {C{\'a}rcamo Hern{\'a}ndez}, G.~Corcella, R.~Torre, and
  E.~Trincherini, {\it Composite vectors at the large hadron collider},  {\em
  JHEP} {\bf 03} (2010) 068, [\href{http://arxiv.org/abs/0911.1942}{{\tt
  arXiv:0911.1942}}]. [\href{http://inspirehep.net/record/836568}{Inspire}].

\bibitem{Agashe:2009dg}
K.~Agashe and R.~Contino, {\it {Composite Higgs-Mediated FCNC}},  {\em Phys.
  Rev.} {\bf D 80} (2009) 075016, [\href{http://arxiv.org/abs/0906.1542}{{\tt
  arXiv:0906.1542}}]. [\href{http://inspirehep.net/record/822501}{Inspire}].

\bibitem{Agashe:2009ve}
K.~Agashe, A.~Azatov, T.~Han, Y.~Li, Z.-G. Si, and L.~Zhu, {\it {LHC Signals
  for Coset Electroweak Gauge Bosons in Warped/Composite PGB Higgs Models}},
  {\em Phys. Rev.} {\bf D 81} (2010) 096002,
  [\href{http://arxiv.org/abs/0911.0059}{{\tt arXiv:0911.0059}}].
  [\href{http://inspirehep.net/record/835687}{Inspire}].

\bibitem{Cata:2009iy}
O.~Cata, G.~Isidori, and J.~F. Kamenik, {\it {Drell-Yan production of Heavy
  Vectors in Higgsless models}},  {\em Nucl. Phys. B} {\bf 822} (2009)
  230--244, [\href{http://arxiv.org/abs/0905.0490}{{\tt arXiv:0905.0490}}].

\bibitem{Barbieri:2010mn}
R.~Barbieri, S.~Rychkov, and R.~Torre, {\it {Signals of composite
  electroweak-neutral Dark Matter: LHC/Direct Detection interplay}},  {\em
  Phys. Lett. B} {\bf 688} (2010) 212--215,
  [\href{http://arxiv.org/abs/1001.3149}{{\tt arXiv:1001.3149}}].

\bibitem{CarcamoHernandez:2010wpm}
A.~E. Carcamo~Hernandez, {\it {Top quark effects in composite vector pair
  production at the LHC}},  {\em Eur. Phys. J. C} {\bf 72} (2012) 2154,
  [\href{http://arxiv.org/abs/1008.1039}{{\tt arXiv:1008.1039}}].

\bibitem{CarcamoHernandez:2010qxf}
A.~E. Carcamo~Hernandez and R.~Torre, {\it {A 'Composite' scalar-vector system
  at the LHC}},  {\em Nucl. Phys. B} {\bf 841} (2010) 188--204,
  [\href{http://arxiv.org/abs/1005.3809}{{\tt arXiv:1005.3809}}].

\bibitem{Falkowski:2011ua}
A.~Falkowski, C.~Grojean, A.~Kaminska, S.~Pokorski, and A.~Weiler, {\it {If no
  Higgs then what?}},  {\em JHEP} {\bf 11} (2011) 028,
  [\href{http://arxiv.org/abs/1108.1183}{{\tt arXiv:1108.1183}}].
  [\href{http://inspirehep.net/record/922186}{Inspire}].

\bibitem{Contino:2011np}
R.~Contino, D.~Marzocca, D.~Pappadopulo, and R.~Rattazzi, {\it {On the effect
  of resonances in composite Higgs phenomenology}},  {\em JHEP} {\bf 10} (2011)
  081, [\href{http://arxiv.org/abs/1109.1570}{{\tt arXiv:1109.1570}}].

\bibitem{Chanowitz:2011ew}
M.~S. Chanowitz, {\it {A Heavy little Higgs and a light Z' under the radar}},
  {\em Phys. Rev. D} {\bf 84} (2011) 035014,
  [\href{http://arxiv.org/abs/1102.3672}{{\tt arXiv:1102.3672}}].

\bibitem{Bellazzini:2012tv}
B.~Bellazzini, C.~Csaki, J.~Hubisz, J.~Serra, and J.~Terning, {\it {Composite
  Higgs Sketch}},  {\em JHEP} {\bf 11} (2012) 003,
  [\href{http://arxiv.org/abs/1205.4032}{{\tt arXiv:1205.4032}}].

\bibitem{Accomando:2012us}
E.~Accomando, L.~Fedeli, S.~Moretti, S.~D. Curtis, and D.~Dominici, {\it
  {Charged di-boson production at the LHC in a 4-site model with a composite
  Higgs boson}},  {\em Phys.Rev.} {\bf D86} (2012) 115006,
  [\href{http://arxiv.org/abs/1208.0268}{{\tt arXiv:1208.0268}}].
  [\href{http://inspirehep.net/record/1124599}{Inspire}].

\bibitem{Greco:2014aza}
D.~Greco and D.~Liu, {\it {Hunting composite vector resonances at the LHC:
  naturalness facing data}},  {\em JHEP} {\bf 12} (2014) 126,
  [\href{http://arxiv.org/abs/1410.2883}{{\tt arXiv:1410.2883}}].

\bibitem{Low:2015uha}
M.~Low, A.~Tesi, and L.-T. Wang, {\it {Composite spin-1 resonances at the
  LHC}},  {\em Phys. Rev. D} {\bf 92} (2015), no.~8 085019,
  [\href{http://arxiv.org/abs/1507.07557}{{\tt arXiv:1507.07557}}].

\bibitem{Accomando:2016mvz}
E.~Accomando, D.~Barducci, S.~De~Curtis, J.~Fiaschi, S.~Moretti, and C.~H.
  Shepherd-Themistocleous, {\it {Drell-Yan production of multi Z$^{'}$-bosons
  at the LHC within Non-Universal ED and 4D Composite Higgs Models}},  {\em
  JHEP} {\bf 07} (2016) 068, [\href{http://arxiv.org/abs/1602.05438}{{\tt
  arXiv:1602.05438}}].

\bibitem{Liu:2018hum}
D.~Liu, L.-T. Wang, and K.-P. Xie, {\it {Prospects of searching for composite
  resonances at the LHC and beyond}},  {\em JHEP} {\bf 01} (2019) 157,
  [\href{http://arxiv.org/abs/1810.08954}{{\tt arXiv:1810.08954}}].

\bibitem{Liu:2019bua}
D.~Liu, L.-T. Wang, and K.-P. Xie, {\it {Broad composite resonances and their
  signals at the LHC}},  {\em Phys. Rev. D} {\bf 100} (2019), no.~7 075021,
  [\href{http://arxiv.org/abs/1901.01674}{{\tt arXiv:1901.01674}}].

\bibitem{DeCurtis:2021fdm}
S.~De~Curtis and D.~Dominici, {\it {Spin-1 resonances}},  {\em Eur. Phys. J.
  ST} {\bf 231} (2022), no.~7 1299--1308,
  [\href{http://arxiv.org/abs/2110.01907}{{\tt arXiv:2110.01907}}].

\bibitem{Liu:2023jta}
D.~Liu, L.-T. Wang, and K.-P. Xie, {\it {Composite resonances at a 10 TeV muon
  collider}},  {\em JHEP} {\bf 04} (2024) 084,
  [\href{http://arxiv.org/abs/2312.09117}{{\tt arXiv:2312.09117}}].

\bibitem{Pappadopulo:2014qza}
D.~Pappadopulo, A.~Thamm, R.~Torre, and A.~Wulzer, {\it {Heavy Vector Triplets:
  Bridging Theory and Data}},  {\em JHEP} {\bf 09} (2014) 060,
  [\href{http://arxiv.org/abs/1402.4431}{{\tt arXiv:1402.4431}}].

\bibitem{Baker:2024xwh}
M.~J. Baker, T.~Martonhelyi, A.~Thamm, and R.~Torre, {\it {A simplified model
  of Heavy Vector Singlets for the LHC and future colliders}},  {\em JHEP} {\bf
  06} (2025) 187, [\href{http://arxiv.org/abs/2407.11117}{{\tt
  arXiv:2407.11117}}].

\bibitem{Hosseini:2022urq}
Y.~Hosseini and M.~M. Najafabadi, {\it {Unitarity constraints and collider
  searches for dark photons}},  {\em Phys. Rev. D} {\bf 106} (2022), no.~1
  015028, [\href{http://arxiv.org/abs/2202.10058}{{\tt arXiv:2202.10058}}].

\bibitem{Dasgupta:2023zrh}
A.~Dasgupta, P.~S.~B. Dev, T.~Han, R.~Padhan, S.~Wang, and K.~Xie, {\it
  {Searching for heavy leptophilic Z': from lepton colliders to gravitational
  waves}},  {\em JHEP} {\bf 12} (2023) 011,
  [\href{http://arxiv.org/abs/2308.12804}{{\tt arXiv:2308.12804}}].

\bibitem{Cheung:2025uaz}
K.~Cheung, J.~Kim, S.~Lee, P.~Sanyal, and J.~Song, {\it {Probing a heavy dark Z
  boson at multi-TeV muon colliders: Leveraging the optimized recoil mass
  technique}},  {\em Phys. Rev. D} {\bf 112} (2025), no.~9 095010,
  [\href{http://arxiv.org/abs/2501.02224}{{\tt arXiv:2501.02224}}].

\bibitem{Baker:2022zxv}
M.~J. Baker, T.~Martonhelyi, A.~Thamm, and R.~Torre, {\it {The role of vector
  boson fusion in the production of heavy vector triplets at the LHC and
  HL-LHC}},  {\em JHEP} {\bf 11} (2022) 066,
  [\href{http://arxiv.org/abs/2207.05091}{{\tt arXiv:2207.05091}}].

\bibitem{Contino:2013gna}
R.~Contino, C.~Grojean, D.~Pappadopulo, R.~Rattazzi, and A.~Thamm, {\it {Strong
  Higgs Interactions at a Linear Collider}},  {\em JHEP} {\bf 02} (2014) 006,
  [\href{http://arxiv.org/abs/1309.7038}{{\tt arXiv:1309.7038}}].

\bibitem{ZurbanoFernandez:2020cco}
I.~Zurbano~Fernandez et~al., {\it {High-Luminosity Large Hadron Collider
  (HL-LHC): Technical design report}}, .

\bibitem{FCC:2018bvk}
{\bf FCC} Collaboration, A.~Abada et~al., {\it {HE-LHC: The High-Energy Large
  Hadron Collider}: {Future Circular Collider Conceptual Design Report Volume
  4}},  {\em Eur. Phys. J. ST} {\bf 228} (2019), no.~5 1109--1382.

\bibitem{CidVidal:2018eel}
X.~Cid~Vidal et~al., {\it {Report from Working Group 3}: {Beyond the Standard
  Model physics at the HL-LHC and HE-LHC}},  {\em CERN Yellow Rep. Monogr.}
  {\bf 7} (2019) 585--865, [\href{http://arxiv.org/abs/1812.07831}{{\tt
  arXiv:1812.07831}}].

\bibitem{FCC:2018vvp}
{\bf FCC} Collaboration, A.~Abada et~al., {\it {FCC-hh: The Hadron Collider}:
  {Future Circular Collider Conceptual Design Report Volume 3}},  {\em Eur.
  Phys. J. ST} {\bf 228} (2019), no.~4 755--1107.

\bibitem{Thamm:2015zwa}
A.~Thamm, R.~Torre, and A.~Wulzer, {\it {Future tests of Higgs compositeness:
  direct vs indirect}},  {\em JHEP} {\bf 07} (2015) 100,
  [\href{http://arxiv.org/abs/1502.01701}{{\tt arXiv:1502.01701}}].

\bibitem{Han:2020uid}
T.~Han, Y.~Ma, and K.~Xie, {\it {High energy leptonic collisions and
  electroweak parton distribution functions}},  {\em Phys. Rev. D} {\bf 103}
  (2021), no.~3 L031301, [\href{http://arxiv.org/abs/2007.14300}{{\tt
  arXiv:2007.14300}}].

\bibitem{Buttazzo:2020uzc}
D.~Buttazzo, R.~Franceschini, and A.~Wulzer, {\it {Two Paths Towards Precision
  at a Very High Energy Lepton Collider}},  {\em JHEP} {\bf 05} (2021) 219,
  [\href{http://arxiv.org/abs/2012.11555}{{\tt arXiv:2012.11555}}].

\bibitem{Han:2022edd}
T.~Han, S.~Li, S.~Su, W.~Su, and Y.~Wu, {\it {BSM Higgs Production at a Muon
  Collider}},  in {\em {Snowmass 2021}}, 5, 2022.
\newblock \href{http://arxiv.org/abs/2205.11730}{{\tt arXiv:2205.11730}}.

\bibitem{Korshynska:2024suh}
K.~Korshynska, M.~L{\"o}schner, M.~Marinichenko,
  K.~M\c{e}ka\textnormal{l}\llap{/}a, and J.~Reuter, {\it {Z{\textquoteright}
  boson mass reach and model discrimination at muon colliders}},  {\em Eur.
  Phys. J. C} {\bf 84} (2024), no.~6 568,
  [\href{http://arxiv.org/abs/2402.18460}{{\tt arXiv:2402.18460}}].

\bibitem{Chen:2022msz}
S.~Chen, A.~Glioti, R.~Rattazzi, L.~Ricci, and A.~Wulzer, {\it {Learning from
  radiation at a very high energy lepton collider}},  {\em JHEP} {\bf 05}
  (2022) 180, [\href{http://arxiv.org/abs/2202.10509}{{\tt arXiv:2202.10509}}].

\bibitem{MuCoL:2025quu}
{\bf MuCoL} Collaboration, A.~Chanc{\'e} et~al., {\it {MuCol Milestone Report
  No. 7: Consolidated Parameters}},
  \href{http://arxiv.org/abs/2510.27437}{{\tt arXiv:2510.27437}}.

\bibitem{MICE:2023vpa}
{\bf MICE} Collaboration, M.~Bogomilov et~al., {\it {Transverse emittance
  reduction in muon beams by ionization cooling}},  {\em Nature Phys.} {\bf 20}
  (2024), no.~10 1558--1563, [\href{http://arxiv.org/abs/2310.05669}{{\tt
  arXiv:2310.05669}}]. [Erratum: Nature Phys. 20, 1687 (2024)].

\bibitem{MuonCollider:2022ded}
{\bf Muon Collider} Collaboration, N.~Bartosik et~al., {\it {Simulated Detector
  Performance at the Muon Collider}},
  \href{http://arxiv.org/abs/2203.07964}{{\tt arXiv:2203.07964}}.

\bibitem{Ally:2022rgk}
D.~Ally, L.~Carpenter, T.~Holmes, L.~Lee, and P.~Wagenknecht, {\it {Strategies
  for Beam-Induced Background Reduction at Muon Colliders}},  in {\em {Snowmass
  2021}}, 3, 2022.
\newblock \href{http://arxiv.org/abs/2203.06773}{{\tt arXiv:2203.06773}}.

\bibitem{MAIA:2025hzm}
{\bf MAIA} Collaboration, C.~Bell et~al., {\it {MAIA: A new detector concept
  for a 10 TeV muon collider}},  \href{http://arxiv.org/abs/2502.00181}{{\tt
  arXiv:2502.00181}}.

\bibitem{Andreetto:2025mrd}
P.~Andreetto et~al., {\it {MUSIC: A Multi-Purpose Detector Concept for Physics
  at the 10 TeV Muon Collider}},  \href{http://arxiv.org/abs/2511.23273}{{\tt
  arXiv:2511.23273}}.

\bibitem{Ruhdorfer:2024dgz}
M.~Ruhdorfer, E.~Salvioni, and A.~Wulzer, {\it {Building the case for forward
  muon detection at a muon collider}},  {\em Phys. Rev. D} {\bf 111} (2025),
  no.~5 053010, [\href{http://arxiv.org/abs/2411.00096}{{\tt
  arXiv:2411.00096}}].

\bibitem{Bartosik:2019dzq}
N.~Bartosik et~al., {\it {Preliminary Report on the Study of Beam-Induced
  Background Effects at a Muon Collider}},
  \href{http://arxiv.org/abs/1905.03725}{{\tt arXiv:1905.03725}}.

\bibitem{Bredt:2022dmm}
P.~M. Bredt, W.~Kilian, J.~Reuter, and P.~Stienemeier, {\it {NLO electroweak
  corrections to multi-boson processes at a muon collider}},  {\em JHEP} {\bf
  12} (2022) 138, [\href{http://arxiv.org/abs/2208.09438}{{\tt
  arXiv:2208.09438}}].

\bibitem{Ma:2024ayr}
Y.~Ma, D.~Pagani, and M.~Zaro, {\it {EW corrections and heavy boson radiation
  at a high-energy muon collider}},  {\em Phys. Rev. D} {\bf 111} (2025), no.~5
  053002, [\href{http://arxiv.org/abs/2409.09129}{{\tt arXiv:2409.09129}}].

\bibitem{Frixione:2025guf}
S.~Frixione, F.~Maltoni, D.~Pagani, and M.~Zaro, {\it {Precision phenomenology
  at multi-TeV muon colliders}},  {\em JHEP} {\bf 09} (2025) 036,
  [\href{http://arxiv.org/abs/2506.10733}{{\tt arXiv:2506.10733}}].

\bibitem{Buttazzo:2018qqp}
D.~Buttazzo, D.~Redigolo, F.~Sala, and A.~Tesi, {\it {Fusing Vectors into
  Scalars at High Energy Lepton Colliders}},  {\em JHEP} {\bf 11} (2018) 144,
  [\href{http://arxiv.org/abs/1807.04743}{{\tt arXiv:1807.04743}}].

\bibitem{Liu:2021jyc}
W.~Liu and K.-P. Xie, {\it {Probing electroweak phase transition with multi-TeV
  muon colliders and gravitational waves}},  {\em JHEP} {\bf 04} (2021) 015,
  [\href{http://arxiv.org/abs/2101.10469}{{\tt arXiv:2101.10469}}].

\bibitem{Frixione:2023gmf}
S.~Frixione and G.~Stagnitto, {\it {The muon parton distribution functions}},
  {\em JHEP} {\bf 12} (2023) 170, [\href{http://arxiv.org/abs/2309.07516}{{\tt
  arXiv:2309.07516}}].

\bibitem{Asadi:2026kpt}
P.~Asadi, A.~Batz, S.~Homiller, and T.-T. Yu, {\it {On the Run from the Dark
  Side of the Muon}},  \href{http://arxiv.org/abs/2602.16771}{{\tt
  arXiv:2602.16771}}.

\bibitem{Berdine:2007uv}
D.~Berdine, N.~Kauer, and D.~Rainwater, {\it {Breakdown of the Narrow Width
  Approximation for New Physics}},  {\em Phys. Rev. Lett.} {\bf 99} (2007)
  111601, [\href{http://arxiv.org/abs/hep-ph/0703058}{{\tt hep-ph/0703058}}].

\bibitem{deLima:2025ctj}
C.~H. de~Lima, {\it {Boosting vector boson fusion reconstruction at muon
  colliders}},  {\em Phys. Rev. D} {\bf 113} (2026), no.~5 L051703,
  [\href{http://arxiv.org/abs/2507.22108}{{\tt arXiv:2507.22108}}].

\bibitem{Cowan:2010js}
G.~Cowan, K.~Cranmer, E.~Gross, and O.~Vitells, {\it {Asymptotic formulae for
  likelihood-based tests of new physics}},  {\em Eur. Phys. J. C} {\bf 71}
  (2011) 1554, [\href{http://arxiv.org/abs/1007.1727}{{\tt arXiv:1007.1727}}].
  [Erratum: Eur.Phys.J.C 73, 2501 (2013)].

\bibitem{ParticleDataGroup:2024cfk}
{\bf Particle Data Group} Collaboration, S.~Navas et~al., {\it {Review of
  particle physics}},  {\em Phys. Rev. D} {\bf 110} (2024), no.~3 030001.

\bibitem{Conway:2011in}
J.~S. Conway, {\it {Incorporating Nuisance Parameters in Likelihoods for
  Multisource Spectra}},  in {\em {PHYSTAT 2011}}, pp.~115--120, 2011.
\newblock \href{http://arxiv.org/abs/1103.0354}{{\tt arXiv:1103.0354}}.

\bibitem{Alwall:2014hca}
J.~Alwall, R.~Frederix, S.~Frixione, V.~Hirschi, F.~Maltoni, O.~Mattelaer,
  H.~S. Shao, T.~Stelzer, P.~Torrielli, and M.~Zaro, {\it {The automated
  computation of tree-level and next-to-leading order differential cross
  sections, and their matching to parton shower simulations}},  {\em JHEP} {\bf
  07} (2014) 079, [\href{http://arxiv.org/abs/1405.0301}{{\tt
  arXiv:1405.0301}}].

\bibitem{comb_exp_results}
T.~P. Hill and J.~Miller, {\it How to combine independent data sets for the
  same quantity},  {\em Chaos: An Interdisciplinary Journal of Nonlinear
  Science} {\bf 21} (07, 2011) 033102.

\bibitem{hepmdb}
M.~Bondarenko, A.~Belyaev, J.~Blandford, L.~Basso, E.~Boos, V.~Bunichev,
  et~al., {\it {High Energy Physics Model Database : Towards decoding of the
  underlying theory (within Les Houches 2011: Physics at TeV Colliders New
  Physics Working Group Report)}},  \href{http://arxiv.org/abs/1203.1488}{{\tt
  arXiv:1203.1488}}.

\bibitem{HVTGitHub}
R.~Torre, {\it {HVT Tools}},  2022.
\newblock \href{httpshttps://github.com/riccardotorre/HVT_tools}{GitHub}.

\bibitem{Carena:2016npr}
M.~Carena and Z.~Liu, {\it {Challenges and opportunities for heavy scalar
  searches in the $ t\overline{t} $ channel at the LHC}},  {\em JHEP} {\bf 11}
  (2016) 159, [\href{http://arxiv.org/abs/1608.07282}{{\tt arXiv:1608.07282}}].

\bibitem{Accomando:2011eu}
E.~Accomando, D.~Becciolini, S.~De~Curtis, D.~Dominici, L.~Fedeli, and
  C.~Shepherd-Themistocleous, {\it {Interference effects in heavy W'-boson
  searches at the LHC}},  {\em Phys. Rev. D} {\bf 85} (2012) 115017,
  [\href{http://arxiv.org/abs/1110.0713}{{\tt arXiv:1110.0713}}].

\bibitem{Choudhury:2011cg}
D.~Choudhury, R.~M. Godbole, and P.~Saha, {\it {Dijet resonances, widths and
  all that}},  {\em JHEP} {\bf 01} (2012) 155,
  [\href{http://arxiv.org/abs/1111.1054}{{\tt arXiv:1111.1054}}].

\bibitem{Accomando:2013sfa}
E.~Accomando, D.~Becciolini, A.~Belyaev, S.~Moretti, and
  C.~Shepherd-Themistocleous, {\it {Z' at the LHC: Interference and Finite
  Width Effects in Drell-Yan}},  {\em JHEP} {\bf 10} (2013) 153,
  [\href{http://arxiv.org/abs/1304.6700}{{\tt arXiv:1304.6700}}].

\bibitem{CMS-DP-2023-065}
{\bf CMS} Collaboration, {\it {Jet Tagging with the Boosted Event Shape Tagger
  at CMS}}, .

\bibitem{Byers:1964ryc}
N.~Byers and C.~N. Yang, {\it {Physical Regions in Invariant Variables for n
  Particles and the Phase-Space Volume Element}},  {\em Rev. Mod. Phys.} {\bf
  36} (1964), no.~2 595--609.

\bibitem{Kajantie:1968jtt}
K.~Kajantie and P.~Lindblom, {\it {Physical region on the plane of two
  invariant momentum transfers for a reaction with three particles in the final
  state}},  {\em Phys. Rev.} {\bf 175} (1968) 2203--2213.

\bibitem{Burns:2009zi}
M.~Burns, K.~T. Matchev, and M.~Park, {\it {Using kinematic boundary lines for
  particle mass measurements and disambiguation in SUSY-like events with
  missing energy}},  {\em JHEP} {\bf 05} (2009) 094,
  [\href{http://arxiv.org/abs/0903.4371}{{\tt arXiv:0903.4371}}].

\bibitem{ATLAS:2019erb}
{\bf ATLAS} Collaboration, G.~Aad et~al., {\it {Search for high-mass dilepton
  resonances using 139 fb$^{-1}$ of $pp$ collision data collected at
  $\sqrt{s}=$13 TeV with the ATLAS detector}},  {\em Phys. Lett. B} {\bf 796}
  (2019) 68--87, [\href{http://arxiv.org/abs/1903.06248}{{\tt
  arXiv:1903.06248}}].

\bibitem{ATLAS:2020fry}
{\bf ATLAS} Collaboration, G.~Aad et~al., {\it {Search for heavy diboson
  resonances in semileptonic final states in pp collisions at $\sqrt{s}=13$ TeV
  with the ATLAS detector}},  {\em Eur. Phys. J. C} {\bf 80} (2020), no.~12
  1165, [\href{http://arxiv.org/abs/2004.14636}{{\tt arXiv:2004.14636}}].

\bibitem{Li:2025ptq}
H.-Q. Li, H.-N. Yan, J.~Gu, and X.-Z. Tan, {\it {Probing Z/W pole physics at
  high-energy muon colliders via vector-boson-fusion processes}},  {\em Chin.
  Phys. C} {\bf 49} (2025), no.~10 103102,
  [\href{http://arxiv.org/abs/2503.19073}{{\tt arXiv:2503.19073}}].

\bibitem{deBlas:2016ojx}
J.~de~Blas, M.~Ciuchini, E.~Franco, S.~Mishima, M.~Pierini, L.~Reina, and
  L.~Silvestrini, {\it {Electroweak precision observables and Higgs-boson
  signal strengths in the Standard Model and beyond: present and future}},
  {\em JHEP} {\bf 12} (2016) 135, [\href{http://arxiv.org/abs/1608.01509}{{\tt
  arXiv:1608.01509}}].

\bibitem{Farina:2016rws}
M.~Farina, G.~Panico, D.~Pappadopulo, J.~T. Ruderman, R.~Torre, and A.~Wulzer,
  {\it {Energy helps accuracy: electroweak precision tests at hadron
  colliders}},  {\em Phys. Lett. B} {\bf 772} (2017) 210--215,
  [\href{http://arxiv.org/abs/1609.08157}{{\tt arXiv:1609.08157}}].

\end{thebibliography}\endgroup

\end{document}